\documentclass{article}%
\usepackage{amsfonts}
\usepackage{amsmath}
\usepackage{fancybox}
\usepackage{appendix}%
\usepackage{amssymb}%
\usepackage[dvips]{graphicx}
\usepackage{graphicx}
\usepackage{psfrag}

\providecommand{\U}[1]{\protect\rule{.1in}{.1in}}
\newtheorem{theorem}{Theorem}

\newtheorem{definition}[theorem]{Definition}

\begin{document}

\title{The Klein-Gordon-Fock equation in the curved spacetime of the Kerr-Newman (anti) de
Sitter black hole}
\author{G. V. Kraniotis \footnote{email: gkraniot@cc.uoi.gr}\\
University of Ioannina, Physics Department \\ Section of
Theoretical Physics, GR- 451 10, Greece \\
}

 \maketitle

\begin{abstract}
Exact solutions of the Klein-Gordon-Fock (KGF) general relativistic equation that describe the dynamics of a massive, electrically charged scalar particle in the curved spacetime geometry of an electrically charged, rotating Kerr-Newman-(anti) de Sitter black hole are investigated. In the general case of a rotating, charged, cosmological black hole the solution of the KGF equation with the method of separation of variables results in Fuchsian differential equations for the radial and angular parts which for most of the parameter space contain more than three finite singularities and thereby generalise the Heun differential equations. For particular values of the physical parameters (i.e mass of the scalar particle)  these Fuchsian equations reduce to the case of Heun equation and the closed form analytic solutions we derive are expressed in terms of Heun functions. For other values of the parameters some of the extra singular points are false singular points. We derive the conditions on the coefficients of the generalised Fuchsian equation such that a singular point is a false point. In such a case the exact solution of the Fuchsian equation can in principle be simplified and expressed in terms of Heun functions. This is the generalisation of the case of a Heun equation with a false singular point in which the exact solution of Heun's differential equation is expressed in terms of Gau$\ss$  hypergeometric function.
We also derive the exact solutions of the radial and angular equations for a charged massive scalar particle in the Kerr-Newman spacetime. The analytic solutions are expressed in terms of confluent Heun functions. Moreover, we derived the constraints on the parameters of the theory such that the solution simplifies and expressed in terms of confluent Kummer hypergeometric functions.
We also investigate the radial solutions in the KN case in the regions near the event horizon and far from the black hole.
Finally, we construct several expansions of the solutions of the Heun equation in terms of generalised hypergeometric functions of Lauricella-Appell.

\end{abstract}

\section{Introduction}

The investigation of the interaction of a scalar particle with the gravitational field is of importance in the attempts to construct quantum theories on curved spacetime backgrounds \cite{KLEIN},\cite{GORDON},\cite{FOCK}. The general relativistic form of the so called Klein-Gordon-Fock (KGF) wave equation were obtained independently in \cite{KLEIN} and \cite{FOCK}.

On the other hand, black holes are intensively studied both at the experimental \cite{GhezA}, \cite{GenzelETAL} as well as at the theoretical level \cite{GRGKRANIOTIS},\cite{CQGKraniotis},\cite{GeorgeVKraniotis} and this fruitful research can lead to tests of General Relativity at the strong field regime in particular for the supermassive Galactic Centre SgrA* black hole. This can be achieved by identifying the type of the black hole that resides at the Sagittarius A* region and confirming the relativistic predictions for the periastron precession and frame-dragging effects for the orbits of S-stars \cite{GeorgeVKraniotis}, the gravitational lensing effects near the event horizon \cite{GRGKRANIOTIS},\cite{CQGKraniotis},\cite{SHADOW} time delays \cite{CQGKraniotis},\cite{LINHE} and gravitational redshift \cite{HERRERA},\cite{PRETOSAHA}.
Also the discovery of a Higgs-like scalar field \cite{PHIGGS} at Cern in the mass region of $126$GeV \cite{CMS},\cite{ATLAS}, provides additional impetus for probing the interaction of a scalar particle with the strong gravitational field of a black hole.

The recent spectacular observations of gravitational waves predicted by the theory of General Relativity from the binary black hole mergers GW150914 \cite{GW150914} and GW151226 \cite{GW151226} enhanced our knowledge of the spacetime physical structure and motivates further studies of the interaction of the spacetime with particles such as the scalar degrees of freedom.

Potentially interesting applications of the theory we develop in our  work based on the exact solutions of the general relativistic KFG equation, include the gravitational radiation from a hypothetical axion cloud around a black hole \cite{DIMO},\cite{AXIONCLOUD}\footnote{Ultralight axion fields are ubiquitous  in Calabi-Yau compactifications of string theory \cite{diameaxion}.}. We note that an axion field of mass $m_A= 10^{-10}{\rm eV}$ has a Compton wavelength $\frac{h}{m_A c}=12417{\rm m}$ which corresponds to the size of a black hole with a mass $m_{\rm BH}\sim 10M_{\odot}$ while for an axion mass $m_A=10^{-16}{\rm eV}$ its length is comparable to the length $\frac{GM_{BH}}{c^2}$ of a supermassive black hole $M_{\rm BH}=4.04\times 10^6M_{\odot}$ such as the SgrA*.

One of the most fundamental exact non-vacuum solutions of the
gravitational field equations of general relativity is the
Kerr-Newman black hole \cite{Newman}. The Kerr-Newman (KN) exact
solution describes the curved spacetime geometry surrounding a
charged, rotating black hole and it solves the coupled system of
differential equations for the gravitational and electromagnetic
fields \cite{Newman} (see also \cite{HansOhanian}).

 The KN exact solution generalised the Kerr
solution \cite{KerrR}, which describes the curved spacetime
geometry around a rotating black hole, to include a net electric
charge carried by the black hole.

Taking into account the contribution from the cosmological
constant $\Lambda,$ the generalisation of the Kerr-Newman solution
is described by the Kerr-Newman de Sitter $($KNdS$)$ metric
element which in Boyer-Lindquist (BL) coordinates is given by
\cite{Stuchlik1},\cite{BCAR},\cite{GrifPod},\cite{ZdeStu} (in units where $G=1$ and $c=1$):
\begin{align}
\mathrm{d}s^{2}  & =\frac{\Delta_{r}^{KN}}{\Xi^{2}\rho^{2}}(\mathrm{d}%
t-a\sin^{2}\theta\mathrm{d}\phi)^{2}-\frac{\rho^{2}}{\Delta_{r}^{KN}%
}\mathrm{d}r^{2}-\frac{\rho^{2}}{\Delta_{\theta}}\mathrm{d}\theta
^{2}\nonumber \\ &-\frac{\Delta_{\theta}\sin^{2}\theta}{\Xi^{2}\rho^{2}}(a\mathrm{d}%
t-(r^{2}+a^{2})\mathrm{d}\phi)^{2}%
\label{KNADSelement}
\end{align}%
\begin{equation}
\Delta_{\theta}:=1+\frac{a^{2}\Lambda}{3}\cos^{2}\theta,
\;\Xi:=1+\frac {a^{2}\Lambda}{3},
\end{equation}

\begin{equation}
\Delta_{r}^{KN}:=\left(  1-\frac{\Lambda}{3}r^{2}\right)  \left(  r^{2}
+a^{2}\right)  -2Mr+e^{2},
\label{DiscrimiL}
\end{equation}

\begin{equation}
\rho^{2}=r^{2}+a^{2}\cos^{2}\theta,
\end{equation}
where $a,M,e,$ denote the Kerr parameter, mass and electric charge
of the black hole, respectively.
The KN(a)dS metric is the most general exact stationary black hole solution of the Einstein-Maxwell system of differential equations.
This
is accompanied by a non-zero electromagnetic field
$F=\mathrm{d}A,$ where the vector potential is
\cite{ZST},\cite{GrifPod}:
\begin{equation}
A=-\frac{er}{\Xi(r^{2}+a^{2}\cos^{2}\theta)}(\mathrm{d}t-a\sin^{2}\theta
\mathrm{d}\phi).
\end{equation}

For the surrounding spacetime to represent a black hole, i.e.
the singularity surrounded by the horizon, the electric charge and
angular
momentum $J$ must be restricted by the relation \cite{Bicak}:
\begin{equation}
\fbox{$M\geq\left[  \left(  \dfrac{J}{M}\right)  ^{2}%
+e^{2}\right]  ^{1/2}\Leftrightarrow M^2\geq a^2+e^2.$}
\end{equation}%
Exact solutions of the null geodesics in the Kerr-Newman and the Kerr-Newman-(anti) de Sitter black hole spacetimes have been recently obtained in \cite{GRGKRANIOTIS} in terms of the Weiersta\ss elliptic functions and the generalised hypergeometic functions of Appell-Lauricella \cite{Appell}. Gravitational lensing and frame dragging of light has been studied intensively in those spacetimes in \cite{GRGKRANIOTIS},\cite{CQGKraniotis}. For the case of
charged particle geodesic orbits in the KN spacetime, we refer the reader
to the works of \cite{Ruffini} and \cite{EVAh} (see also
\cite{RufII},\cite{Calvani}).
We also mention at this point the possibility of Kerr naked singularity spacetime as the source of strong gravity. In this scenario, both accretion phenomena \cite{ZSTUCH} and collissional processes \cite{NakedZSTUCH} give signatures clearly distinct from those related to black holes.

The investigation of the separability of the general relativistic wave equations started with the work Chandrasekhar in the context of the Kerr metric (rotating uncharged black hole) see for instance \cite{CHANDRA}, see also the work of Teukolsky \cite{STEUKOLSKY}. In the context of the Newman-Penrose formalism the Dirac equation for an electron around a Kerr-Newman black hole was separated into decoupled ordinary differential equations in \cite{page}. However, in these works no attempt was made to solve the resulting ordinary differential equations.
Attempts in solving the resulting differential equations with a different degree of accuracy, in the Kerr (K) and Kerr-Newman (KN) spacetimes can be traced in the works \cite{ROWANSTEPHEN}(investigation of the radial equation for a massive particle in KN spacetime ),\cite{Blandin} (investigation of the angular equation in K-spacetime for a massless particle),\cite{WU},\cite{Batic}. More recently solutions of the radial equation resulting from the separation of the KFG equation in the K and KN-spacetimes have been investigated in \cite{HODs},\cite{BEZERRA}. In the former case, it was pointed out that stationary spinning black holes can develope 'hair' in the presence of massive bosonic fields.
Solutions of wave equations and eigenfrequencies in the KdS and KNdS black hole spacetimes have been investigated in \cite{Suzuki},\cite{RK}, however for the case of massless particles.

Thus the most general case of finding the exact solutions of the KFG equation in the curved spacetime of a KNdS black hole for a \textit{massive} and \textit{electrically charged} scalar particle has not been studied.
Our work aims to fill this important gap in the literature.
It is thus pleasing that the theory produced in this work is a \textit{complete} theory for the massive KGF equation in the field of rotating, charged cosmological black holes: \textit{all} of its fundamental parameters enter the analysis of the exact solutions derived.
The Klein-Gordon-Fock equation for a scalar field $\Phi$ that describes the dynamics of a massive scalar electrically charged particle of charge $q$, in a curved spacetime is described by the equation:
\begin{equation}
\Box \Phi+\mu^2 \Phi=0
\label{KleinGordonFock1}
\end{equation}
where,
\begin{equation}
\Box \Phi=\frac{1}{\sqrt{-g}}D_{\nu}(\sqrt{-g}g^{\mu\nu}D_{\mu}\Phi)
\label{DALEMBERT}
\end{equation}
The generalised D'Alembertian involves the inverse spacetime metric $g^{\mu\nu}$ which in our case of investigation is computed from (\ref{KNADSelement}). Also in our case it involves the gauge differential operator:
\begin{equation}
D_{\mu}=\partial_{\mu}-iqA_{\mu}
\end{equation}

The material of this work is organised as follows. In section \ref{variabseparation}, we derive for the first time,  using a  separation of variables ansatz for the KFG equation of a charged massive scalar particle, in the KNdS black hole spacetime, the resulting ordinary radial and angular differential equations, eqns.(\ref{FORTIORADIAL}),(\ref{FORTIOANGULAR}) respectively. In section \ref{HeunMun} we provide a brief mathematical background of Heun's differential equation, its properties, local solutions and Heun functions. As we shall show in the main body of the paper for particular values of the scalar mass and for a non-zero cosmological constant the Fuchsian ordinary differential equations reduce to the case of Heun equations and the exact solution can be expressed in terms of local Heun solutions or functions. Merging two of the four regular singularities of Heun's differential equation yields the confluent Heun equation, the subject of section \ref{confluentheunfuchs}. We note that in the particular limit of the general theory for vanishing cosmological constant the exact solutions of the radial and angular ordinary differential equations are expressed in terms of confluent Heun functions.
In section \ref{ANGULARLAMBDAMASS} and for a particular value of the scalar mass in terms of the cosmological constant $\Lambda$ we derive the exact solution of the angular differential equation by determining the Heun differential equation that the angular variable obeys, see Eqn.(\ref{HeunMunchen}). The parameters of this equation have been determined in terms of the physical parameters of the theory, see Eqns(\ref{ALPHABETAH}),(\ref{auxialiaryfield}).
Moreover, we obtained its elliptic function representation  in both the Jacobian and the Weierstra$\ss$ approach.
In section \ref{APPARENTSINGULARFUCHS} we discuss the theory of \textit{false} or \textit{apparent} singularities of Fuchsian equations \cite{Yoshida}. We focus in this section in the case of a false singularity with exponents $(0,2)$. We derive a condition that guarantees the absence of a logarithmic singularity with exponents $(0,2)$, see eqns. (\ref{apparantfalse}),(\ref{nolog}). This is of importance since as we show in subsection \ref{WICHTIGFALSE}, the angular equation for generic values of the physical parameters is a Fuchsian equation with five singularities, see eqn.(\ref{preheunB}). For particular values of the scalar mass in terms of the cosmological constant we derive the conditions on the physical parameters of the theory that guarantee that the fifth singular point of the angular equation is a false singular point of type $(0,2)$, see Eqns.(\ref{physicalapparency1}),(\ref{physicappara02}),(\ref{physicalfalsesingular}).

In section \ref{LIMITLAMBDAANGULAR} by applying the confluence limit of Heun's equation-subsection \ref{confluentheunfuchs}-we derive the exact solution of the angular differential equation,  under the assumption of a vanishing cosmological constant. The novel closed form analytic solutions are given in terms of confluent Heun functions, see (\ref{CONFLUENTHEUNC}) ,(\ref{conflugwnia1})-(\ref{conflugwnia22}). In
subsection \ref{KUMMERCONFLUENCE}, by applying a recent mathematical theory developed in \cite{ISHKHANYAN}, we derive the conditions for expanding the confluence Heun function solution of the angular equation for a massive scalar particle in KN spacetime in terms of the Kummer confluent hypergeometric functions.
In subsection \ref{aktinamazikoubatmotoul0}, assuming vanishing cosmological constant, we derive the exact solutions of the radial equation first for a massive neutral scalar particle-see eqns (\ref{GVKKGFEXACTKNA}),(\ref{exactnessKGF}),(\ref{alphagammadeltaexactKFG})-(\ref{wexactKFG}) and then generalise them for the case of a charged massive scalar particle Eqns. (\ref{GVKKGFEXACTKNCr}) and (\ref{FORTIORADEXACTHCG}). The novel solutions for the charged massive scalar field, as in the angular case, are given in terms of confluent Heun functions. Again in both cases we derive the conditions on the physical parameters of the theory such that the solutions are expressed in terms of the Kummer confluent hypergeometric functions.
In addition, in subsections \ref{farmakriaeventhorizon} and \ref{kontastonorizonta}, we investigated the asymptotic behaviour of the radial solutions for a charged massive particle in KN spacetime and derived the solutions in regions far from the event horizon and near the horizon. In the former case, our results are expressed in Eqn. (\ref{MKGFCFHORIZONTHOME}) and (\ref{CMFarhorizonWhitt1}),(\ref{CMFarhorizonWhit2}). In the latter case, our results are expressed in (\ref{EventHorizonPOWERSERIES}),(\ref{OrizontasGegonotonWHIT}).
In Appendix \ref{LauricellaDivision} following recent work in the mathematical literature \cite{TAYLORAPPELL} and starting from the equation satisfied by the derivative we constructed several expansions of the solutions of the general Heun equation in terms of the Lauricella $F_D$ and the Appell $F_1$ generalised hypergeometric functions of three and two variables respectively.
In section \ref{RADEQNMPKNDS} we derive exact solutions of the radial part of the KGF equation in KNdS spacetime. For the particular value of scalar mass: $\mu=\sqrt{\frac{2\Lambda}{3}}$ the analytic solution of the radial Fuchsian equation can be given in terms of general Heun functions, see Eqn.(\ref{BHCRCHEUNRADIAL}). We also derived in this case, the Jacobian elliptic form of the resulting Heun equation  which is given in equation (\ref{FORMJACOBIHEUN}). In a series of appendices we collect some of our formal calculations. In Appendix A we discuss in detail the elliptic  function representation of Heun's equation, while in Appendix B we prove in detail that the solution of a Heun equation with a false singular point simplifies and it is given by the Gau$\ss$ hypergeometric function.

\section{The Klein-Gordon-Fock equation for a massive charged particle in KNdS spacetime}\label{variabseparation}

By calculating the D'Alembertian of the Klein-Gordon-Fock equation in the Kerr-Newman-de Sitter spacetime  and initially for the case of a massive neutral particle, we obtain the KGF equation:

\begin{align}
&
\frac{\Xi^2}{\rho^2}\left[\frac{(r^2+a^2)^2}{\Delta_r^{KN}}-\frac{
a^2 \sin^2\theta}{\Delta_{\theta}}\right]\frac{\partial^2\Phi}{\partial t^2}-\frac{1}{\rho^2}\frac{\partial }{\partial r}\left(\Delta_r^{KN}\frac{\partial \Phi}{\partial r}\right) -\frac{1}{\rho^2}\frac{1}{\sin\theta}\frac{\partial}{\partial \theta}
\left(\sin\theta\Delta_{\theta}\frac{\partial \Phi}{\partial \theta}\right)\nonumber \\
&+
2\frac{a\Xi^2}{\rho^2}\Biggl\{-\frac{1}{\Delta_{\theta}}+
\frac{r^2+a^2}{\Delta_r^{KN}}\Biggr\}\frac{\partial^2 \Phi}{\partial t \partial \phi}-\frac{\Xi^2}{\rho^2\sin^2\theta}\{\frac{1}{\Delta_{\theta}}-\frac{a^2\sin^2\theta}
{\Delta^{KN}_r}\}\frac{\partial^2\Phi}{\partial \phi^2}+\mu^2 \Phi=0.
\label{kleinEqnFock}
\end{align}
To solve (\ref{kleinEqnFock}) we assume the ansatz:
\begin{equation}
\Phi=\Phi(\vec{r},t)=R(r)S(\theta)e^{i m\varphi}e^{-i\omega t}
\label{ansatzSEP}
\end{equation}
Substituting  (\ref{ansatzSEP}) into (\ref{kleinEqnFock}) we obtain:
\begin{align}
&\frac{1}{R(r)}\frac{{\rm d}}{{\rm d}r}\left(\Delta_r^{KN}\frac{{\rm d}R}{{\rm d}r}\right)-\Xi^2\left[\frac{(r^2+a^2)^2}{\Delta_r^{KN}}-\frac{
a^2 \sin^2\theta}{\Delta_{\theta}}\right](-\omega^2) \nonumber \\
&+\frac{1}{S(\theta)}\frac{1}{\sin\theta}\frac{{\rm d}}{{\rm d}\theta}
\left(\sin\theta \Delta_{\theta}\frac{{\rm d}S(\theta)}{{\rm d}\theta}\right)+
\frac{\Xi^2}{\sin^2\theta}\left\{\frac{1}{\Delta_{\theta}}-\frac{a^2\sin^2\theta}
{\Delta^{KN}_r}\right\}(-m^2) \nonumber \\
&-2a\Xi^2\Biggl\{-\frac{1}{\Delta_{\theta}}+
\frac{r^2+a^2}{\Delta_r^{KN}}\Biggr\}m\omega-\rho^2\mu^2=0,
\end{align}
Subsequently, separating radial from polar angle parts yields the differential equations:
\begin{align}
&\frac{1}{\sin\theta}\frac{{\rm d}}{{\rm d}\theta}
\left(\sin\theta\Delta_{\theta}\frac{{\rm d}S(\theta)}{{\rm d}\theta}\right)\nonumber \\
&+S(\theta)\left[-\frac{m^2\Xi^2}{\sin^2\theta}\frac{1}{\Delta_{\theta}}
+\frac{2a\Xi^2}{\Delta_{\theta}}m\omega-\frac{\Xi^2a^2\sin^2\theta \omega^2}{\Delta_{\theta}}-\mu^2 a^2 \cos^2\theta+K_{lm}\right]=0,
\label{FORTIOANGULAR}\\
&\frac{{\rm d}}{{\rm d}r}\left(\Delta_r^{KN}\frac{{\rm d}R}{{\rm d}r}\right)
+\frac{R(r)}{\Delta_r^{KN}}[\Xi^2K^2-r^2\mu^2\Delta_r^{KN}-K_{lm}\Delta_r^{KN}]=0,
\label{NOCHARGERADIAL}
\end{align}
where
\begin{equation}
K(r):=\omega(r^2+a^2)-am
\end{equation}
In eqns. (\ref{FORTIOANGULAR}),(\ref{NOCHARGERADIAL}) the constant $K_{lm}$ denotes the separability constant.
Now including the contribution from the electric charge of the scalar particle we calculate first:
\begin{align}
A^{\rho}A_{\rho}&=g^{00}A_0A_0+g^{03}A_3A_0+g^{30}A_3A_0+g^{33}A_3A_3
=-\frac{q^2 e^2 r^2}{\rho^2 \Delta_r^{KN}}, \\
-2iqA^{\mu}\partial_{\mu}&=-2iqA^{0}\partial_0-2iq A^3\partial_3=\frac{2iq\Xi}{\rho^2 \Delta_r^{KN}}\left[(r^2+a^2)\frac{\partial}{\partial t}+a\frac{\partial }{\partial \phi}\right]
\end{align}
Then the radial ordinary equation that results from separation will take the form:
\begin{equation}
\frac{{\rm d}}{{\rm d}r}\left(\Delta_r^{KN}\frac{{\rm d}R}{{\rm d}r}\right)
+\frac{R(r)}{\Delta_r^{KN}}\left[\Xi^2\left(K-\frac{eqr}{\Xi}\right)^2-r^2\mu^2\Delta_r^{KN}-K_{lm}\Delta_r^{KN}\right]=0
\label{FORTIORADIAL}
\end{equation}
while the angular equation remains unaltered.

\section{The solutions of the angular equations}\label{ANGULARLAMBDAMASS}
As we saw in section \ref{variabseparation}, separation of variables with the ansatz (\ref{ansatzSEP}), yielded the angular differential equation (\ref{FORTIOANGULAR}).

By defining the variable $x:=\cos\theta$,  and setting
$\mu=\sqrt{\frac{2\Lambda}{3}}, \Lambda>0$, equation (\ref{FORTIOANGULAR}) becomes:
\begin{align}
& \left[\left(1+\frac{a^2\Lambda}{3}x^2\right)(1-x^2)\frac{\mathrm{d}^2}{\mathrm{d}x^2}+
2\frac{a^2\Lambda}{3}x(1-x^2)\frac{\mathrm{d}}{\mathrm{d}x}-2\left(1+\frac{a^2\Lambda}{3}x^2\right)x\frac{\mathrm{d}}{\mathrm{d}x}
\right]S \nonumber \\
&+\left[-\frac{\Xi^2 a^2 \omega^2 (1-x^2)}{1+\frac{a^2\Lambda}{3}x^2}+\frac{2a\omega m\Xi^2}{1+\frac{a^2\Lambda}{3}x^2}
-\frac{m^2\Xi^2}{(1+\frac{a^2\Lambda}{3}x^2)(1-x^2)}\right]S \nonumber \\
&+\left[-2\frac{a^2\Lambda}{3}x^2+K_{lm}\right]S=0
\label{gwnia2}
\end{align}
Since as we mention the automorphism group of the parameter space of Heun's equation has recently been determined, we apply first to equation (\ref{gwnia2}) the homographic transformation of the independent variable \footnote{The angular differential equation (\ref{gwnia2}) has four regular singularities at the points $\pm 1,\pm \frac{i}{\sqrt{\alpha_{\Lambda}}}$, which we denote with the tuple $(a_1,a_2,a_3,a_4)=(-1,1,-\frac{i}{\sqrt{\alpha_{\Lambda}}},\frac{i}{\sqrt{\alpha_{\Lambda}}}).$ The fourth singularity $a_3\overset{(\ref{homograph})}\rightarrow z_3=\frac{a_3-a_1}{a_3-a_4}\frac{a_2-a_4}{a_2-a_1}. $}:
\begin{equation}
z=\frac{a_2-a_4}{a_2-a_1}\frac{x-a_1}{x-a_4}=\frac{1-\frac{i}{\sqrt{\alpha_{\lambda}}}}{2}\frac{x+1}{x-\frac{i}{\sqrt{\alpha_{\Lambda}}}},\;
\;\alpha_{\Lambda}:=\frac{a^2\Lambda}{3},
\label{homograph}
\end{equation}
where such a transformation is designed to map the three singularities $a_1,a_2,a_4$ into $0,1,\infty$. With this transformation we have:
\begin{equation}
(1+\alpha_{\Lambda}x^2)(1-x^2)=\frac{\alpha_{\Lambda}16i \Xi^2}{\sqrt{\alpha_{\Lambda}}}\frac{z(z-1)(z-z_3)}{[2z\sqrt{\alpha_{\Lambda}}-\sqrt{\alpha_{\Lambda}}+i]^4},
\end{equation}
where
\begin{equation}
z_3=-\frac{1}{2}\left(-1+\frac{\alpha_{\Lambda}-1}{2i\sqrt{\alpha_{\Lambda}}}\right).
\end{equation}
Equation (\ref{gwnia2}) with the aid of (\ref{homograph}) becomes:
\begin{align}
& \Biggl\{ \frac{\mathrm{d}^2}{\mathrm{d}z^2}+\left[\frac{1}{z}+\frac{1}{z-1}
+\frac{1}{z-z_3}-\frac{2}{z-z_{\infty}}\right]\frac{\mathrm{d}}{\mathrm{d}z}\nonumber \\
&-\frac{m^2}{4}\frac{1}{z^2}-\frac{m^2}{4}\frac{1}{(z-1)^2}+\left(\frac{\Xi a \omega}{2\sqrt{\alpha_{\Lambda}}}-\frac{m \sqrt{\alpha_{\Lambda}}}{2}\right)^2 \frac{1}{(z-z_3)^2}+
\frac{2}{(z-z_{\infty})^2}+ \nonumber \\
&\frac{1}{z}\left[\frac{m^2(1+2i\sqrt{\alpha_{\Lambda}}+3\alpha_{\Lambda})}{2(-i+\sqrt{\alpha_{\Lambda}})^2}
+\frac{2m\Xi \xi}{(1+i\sqrt{\alpha_{\Lambda}})^2}-\frac{2\alpha_{\Lambda}}{(1+i\sqrt{\alpha_{\Lambda}})^2}+
\frac{K_{lm}}{(1+i\sqrt{\alpha_{\Lambda}})^2}\right] \nonumber \\
&+\frac{1}{z-1}\left[\frac{-m^2(1-2i\sqrt{\alpha_{\Lambda}}+3\alpha_{\Lambda})}{2(i+\sqrt{\alpha_{\Lambda}})^2}
+\frac{-2m\xi \Xi}{(1-i\sqrt{\alpha_{\Lambda}})^2}+\frac{2\alpha_{\Lambda}}{(1-i\sqrt{\alpha_{\Lambda}})^2}
-\frac{K_{lm}}{(1-i\sqrt{\alpha_{\Lambda}})^2}\right] \nonumber \\
&+\frac{1}{z-z_3}\left[\frac{-8i m^2 \alpha_{\Lambda}\sqrt{\alpha_{\Lambda}}}{\Xi^2}+\frac{8i m \sqrt{\alpha_{\Lambda}}\xi}{\Xi}+\frac{8i\sqrt{\alpha_{\Lambda}}}{\Xi^2}+\frac{4i\sqrt{\alpha_{\Lambda}}K_{lm}}{\Xi^2} \right] \nonumber \\
&+\frac{1}{z-z_{\infty}}\frac{-8i\sqrt{\alpha_{\Lambda}}}{\Xi} \Biggr\}S(z)=0,
\label{preheun}
\end{align}
where $z_{\infty}=-\frac{-i(1+\sqrt{\alpha_{\Lambda}}i)}{2\sqrt{\alpha_{\Lambda}}}$ and $\xi:=a\omega$.
The four singularities $z=0,1,z_3,z_{\infty}$ have exponents $\{\frac{|m|}{2},-\frac{|m|}{2}\},\{\frac{|m|}{2},-\frac{|m|}{2}\},\{\frac{i}{2}\left(\frac{\Xi\xi}{\sqrt{\alpha_{\Lambda}}}-m\sqrt{\alpha_{\Lambda}}\right),
-\frac{i}{2}\left(\frac{\Xi\xi}{\sqrt{\alpha_{\Lambda}}}-m\sqrt{\alpha_{\Lambda}}\right)
\},\{2,1\}$.
Thus equation (\ref{preheun}) is not of a Heun type. The \textit{F-homotopic transformation} or \textit{index transformation} of the dependent variable $S$:
\begin{equation}
S(z)=z^{\alpha_1}(z-1)^{\alpha_2}(z-z_3)^{\alpha_3}(z-z_{\infty})^{\alpha_4}\bar{S}(z)
\label{fhomotopy}
\end{equation}
where $\alpha_1=\alpha_2=\frac{|m|}{2}, \alpha_3=\pm\frac{i}{2}\left(\frac{\Xi\xi}{\sqrt{\alpha_{\Lambda}}}-m\sqrt{\alpha_{\Lambda}}\right),\alpha_4=1$ is designed to reduce one of the exponents of the finite singularities $0,1,z_3$ to zero and to eliminate the finite $z_{\infty}$ singularity. In other words transforms  (\ref{preheun}) into the Heun form (\ref{KarlHeunI}).
Indeed application of (\ref{fhomotopy}) into (\ref{preheun}) yields:
\begin{equation}
\Biggl\{\frac{\mathrm{d}^2}{\mathrm{d}z^2}+\left[\frac{2\alpha_1+1}{z}+
\frac{2\alpha_2+1}{z-1}+\frac{2\alpha_3+1}{z-z_3}\right]\frac{\mathrm{d}}{\mathrm{d}z}+
\frac{{\alpha}{\beta}z-q}{z(z-1)(z-z_3)}\Biggr\}\bar{S}(z)=0,
\label{HeunMunchen}
\end{equation}
where the auxiliary parameter $q$ is calculated in terms of the cosmological constant, spin of the black hole,the parameters $m,\omega$ and is given by the expression
\begin{align}
q&=\frac{i}{4\sqrt{\alpha_{\Lambda}}}\Biggl\{-(1+i\sqrt{\alpha_{\Lambda}})^2
[2\alpha_1\alpha_2+\alpha_2+\alpha_1]-4\sqrt{\alpha_{\Lambda}}i[2\alpha_1\alpha_3+\alpha_3+\alpha_1]\nonumber \\
&-\frac{m^2}{2}((1+i\sqrt{\alpha_{\Lambda}})^2+4\alpha_{\Lambda})+K_{lm}-2i\sqrt{\alpha_{\Lambda}}+2\Xi m\xi\Biggr\}
\label{auxialiaryfield}
\end{align}

The parameters $\alpha,\beta$ are given in terms of the physical parameters by the expression \footnote{The parameter $\mathcal{B}$ is the total coefficient of the term $\frac{1}{z-1}$ that results after the application of the F-homotopic transformation (\ref{fhomotopy}) in (\ref{preheun}).}
\begin{align}
\alpha\beta &=q-(z_3-1)\times {\mathcal{B}} \nonumber \\
&=\frac{i}{4\sqrt{\alpha_{\Lambda}}}\Biggl\{-(1+i\sqrt{\alpha_{\Lambda}})^2
[2\alpha_1\alpha_2+\alpha_2+\alpha_1]-4\sqrt{\alpha_{\Lambda}}i[2\alpha_1\alpha_3+\alpha_3+\alpha_1]\nonumber \\
&-\frac{m^2}{2}((1+i\sqrt{\alpha_{\Lambda}})^2+4\alpha_{\Lambda})+K_{lm}-2i\sqrt{\alpha_{\Lambda}}+2\Xi m\xi\Biggr\} \nonumber \\
&+ \frac{i}{4\sqrt{\alpha_{\Lambda}}}\Biggl\{\frac{m^2}{2}\left((1-i\sqrt{\alpha_{\Lambda}})^2+4\alpha_{\Lambda}\right)
-2m\xi\Xi-K_{lm}-2\sqrt{\alpha_{\Lambda}}i \nonumber \\
&+(1-i\sqrt{\alpha_{\Lambda}})^2[2\alpha_1\alpha_2+\alpha_2+\alpha_1]
+i4\sqrt{\alpha_{\Lambda}}(-2\alpha_2\alpha_3-\alpha_3-\alpha_2)\Biggr\}
\label{ALPHABETAH}
\end{align}

\subsubsection{Exact solution of the Heun angular equation in KNdS spacetime in hypergeometric polynomials}\label{augmentedSYGLISI}
In this section we derive an analytic solution of the angular Heun equation in terms of hypergeometric polynomials-Jacobi polynomials.  For $u(z)$ a function that satisfies Heun's equation, we make the following ansatz:
\begin{equation}
u(z)=\sum_{\nu=0}^{\infty}c_{\nu}y_{\nu}(z)
\label{JacobiP}
\end{equation}
where
\begin{equation}
y_{\nu}(z)=F(-\nu,\nu+\omega,\gamma,z)=\frac{\nu!\Gamma(\gamma)}{\Gamma(\nu+\gamma)}P_{\nu}^{(\gamma-1,\omega-\gamma)}(1-2z)
\label{PolyCGCJACOBI}
\end{equation}
In eqn.(\ref{PolyCGCJACOBI}), $\omega=\delta+\gamma-1$ and should not be confused with the angular frequency that appears in the separation ansatz.
The polynomials satisfy the differential equation \footnote{The type of solution of Heun's differential equation expressed as an infinite series of hypergeometric polynomials was first investigated by Svartholm \cite{Svartholm}. See also \cite{Kristensson}.}
\begin{equation}
y^{\prime\prime}_{\nu}(z)+\left[\frac{\gamma}{z}+\frac{\delta}{z-1}\right]y^{\prime}_{\nu}(z)-
\frac{\nu(\nu+\omega)}{z(z-1)}y_{\nu}(z)=0
\end{equation}
and the recursion relations $\nu\in \mathbb{Z}^{+}$:
\begin{align}
& zy_{\nu}(z)=P_{\nu}y_{\nu+1}(z)+Q_{\nu}y_{\nu}(z)+R_{\nu}y_{\nu-1}(z),\\
&z(z-1)\frac{{\rm d}}{{\rm d}z}y_{\nu}(z)=P^{\prime}_{\nu}y_{\nu+1}(z)+Q^{\prime}_{\nu}y_{\nu}(z)+R^{\prime}_{\nu}y_{\nu-1}(z)
\end{align}
where ($\nu\in \mathbb{Z}^{+}$)
\[\left\{\begin{array}{l}
P_{\nu}=-\frac{(\nu+\omega)(\nu+\gamma)}{(2\nu+\omega)(2\nu+\omega+1)}\\
Q_{\nu}=\frac{(\omega-1)(\gamma-\delta)}{2(2\nu+\omega+1)(2\nu+\omega-1)}+\frac{1}{2}\\
R_{\nu}=-\frac{\nu(\nu+\delta-1)}{(2\nu+\omega)(2\nu+\omega-1)}\end{array}\right.
\left\{\begin{array}{l}
P^{\prime}_{\nu}=-\frac{\nu(\nu+\omega)(\nu+\gamma)}{(2\nu+\omega)(2\nu+\omega+1)}\\
Q^{\prime}_{\nu}=\frac{\nu(\nu+\omega)(\gamma-\delta)}{(2\nu+\omega+1)(2\nu+\omega-1)}\\
R^{\prime}_{\nu}=\frac{\nu(\nu+\omega)(\nu+\delta-1)}{(2\nu+\omega)(2\nu+\omega-1)}\end{array}\right.
\]
with initialising values ($\nu=0$)
\[\left\{\begin{array}{l}
P_0=-\frac{\gamma}{\omega+1}\\
Q_0=\frac{\gamma}{\omega+1}\\
R_0=0\end{array}\right.
\left\{\begin{array}{l}
P^{\prime}_0=0\\
Q^{\prime}_0=0\\
R^{\prime}_0=0\end{array}\right.\]

Inserting (\ref{JacobiP}) into Heun's equation (\ref{KarlHeunI}) we find that if the series (\ref{JacobiP}) is a solution of the Heun's differential equation the coefficients in the series of hypergeometric polynomials, $c_{\nu}$, satisfy:
\begin{equation}
c_{\nu+1}=E_{\nu}c_{\nu}+F_{\nu}c_{\nu-1},
\label{recursionAUGCONVERGENCE}
\end{equation}
where
\[\left\{\begin{array}{l}
E_{\nu}=-\frac{[\nu(\nu+\omega)+\alpha\beta]Q_{\nu}+\varepsilon Q^{\prime}_{\nu}-a\nu(\nu+\omega)-q}{\varepsilon R_{\nu+1}^{\prime}+[(\nu+1)(\nu+1+\omega)+\alpha\beta]R_{\nu+1}}\\
F_{\nu}=-\frac{\varepsilon P^{\prime}_{\nu-1}+((\nu-1)(\nu-1+\omega)+\alpha\beta)P_{\nu-1}}{\varepsilon R_{\nu+1}^{\prime}+[(\nu+1)(\nu+1+\omega)+\alpha\beta]R_{\nu+1}}\end{array}\right.\]
The sequence is initialised by
\begin{equation}
c_1=-\frac{c_0(\alpha\beta\gamma-q(\omega+1))(2+\omega)}{\delta((1+\omega)(\varepsilon-1)-\alpha\beta)}
\end{equation}
Using the first of the recursion relations and the well defined limits of the coefficients $P_{\nu},Q_{\nu},R_{\nu}$ as $\nu\rightarrow \infty$:
\[\left\{\begin{array}{l}
P_{\nu}=-\frac{1}{4}+\frac{1-2\gamma}{8\nu}+\mathcal{O}\left(\frac{1}{\nu^2}\right)\\
Q_{\nu}=\frac{1}{2}+\mathcal{O}\left(\frac{1}{\nu^2}\right)\;\;\;{\rm as}\; \nu\rightarrow \infty \\
R_{\nu}=-\frac{1}{4}-\frac{1-2\gamma}{8\nu}+\mathcal{O}\left(\frac{1}{\nu^2}\right)\end{array}\right.\]
one can show that \footnote{The exceptional case $\lim_{\nu\rightarrow \infty}\frac{y_{\nu+1}(z)}{y_{\nu}(z)}=s_1(z)=\frac{1}{s(z)}$ does not occur \cite{Erdelyi}.}
\begin{equation}
\lim_{\nu\rightarrow \infty}\frac{y_{\nu+1}(z)}{y_{\nu}(z)}=s(z)\equiv s_2(z)=\frac{(1-z^{-1})^{1/2}+1}{(1-z^{-1})^{1/2}-1}\equiv \frac{Z+1}{Z-1}
\end{equation}
exists where the branch of the square root satisfies $\Re(1-z^{-1})^{1/2}>0$ (assuming $z\not \in[0,1])$. Moreover, $\frac{y_{\nu+1}(z)}{y_{\nu}(z)}=s(z)+\frac{\sigma(z)}{\nu}+\mathcal{O}(1/\nu^2),\;{\rm as} \;\nu\rightarrow \infty$, where $\sigma(z)=s(z)\frac{1-2\gamma}{2}$.
On the other hand, the asymptotic behaviour of the coefficients $c_{\nu}$ is given as follows:
The limit
\begin{equation}
\lim_{\nu\rightarrow \infty}\frac{c_{\nu+1}}{c_{\nu}}=t_2=\frac{(1-a^{-1})^{1/2}+1}{(1-a^{-1})^{1/2}-1}\equiv \frac{A+1}{A-1},
\end{equation}
exists, where the branch of the square root satisfies $\Re A=\Re(1-a^{-1})^{1/2}>0$ (under the assumption $|a|>1$), and $|t_2|>1$. In the exceptional case, the limit is
\begin{equation}
\lim_{\nu\rightarrow \infty}\frac{c_{\nu+1}}{c_{\nu}}=t_1=\frac{(1-a^{-1})^{1/2}-1}{(1-a^{-1})^{1/2}+1}\equiv\frac{A-1}{A+1}
\end{equation}
and $|t_1|<1$.
Moreover, if we write $\frac{c_{\nu+1}}{c_{\nu}}=t_n+\frac{\tau}{\nu}+\mathcal{O}(1/\nu^2),\; {\rm as}\;\nu\rightarrow \infty\; (n=1,2)$ then $\tau$ satisfies: $\tau=t_n(2\gamma+2\varepsilon-5)\frac{(1-2a)t_n-1}{t_n^2-1},n=1,2$.
By the D'Alembert's ratio test absolute convergence of the series (\ref{JacobiP}) is guaranteed provided
\begin{equation}
\lim_{\nu\rightarrow \infty}\left|\frac{c_{\nu+1}y_{\nu+1}(z)}{c_{\nu}y_{\nu}(z)}\right|=|t_{n}s_2(z)|<1, \;n=1,2
\end{equation}
and diverges if $|t_{n}s_2(z)|>1, \;n=1,2$. In general $t_2$ is the proper root for the asymptotic limit of the coefficients. The boundary of the domain of convergence, $|t_2 s_2(z)|=1$, consists of all $z\in \mathbb{C}$ satisfying:
\begin{equation}
\left|\frac{Z+1}{Z-1}\right|=|s_2(z)|=\frac{1}{|t_2|}=\left|\frac{A-1}{A+1}\right|<1
\end{equation}
The solution set is void since $|Z+1|<|Z-1|\equiv \Re Z<0$ which contradicts the assumption $\Re Z>0$.
The other root, $t_1$, gives a domain of convergence determined \footnote{The exceptional limit to the root $t_1$ is also known in the literature as the phenomenon of augmented convergence.} by
\begin{equation}
\left|\frac{Z+1}{Z-1}\right|=|s_2(z)|<\frac{1}{|t_1|}=\left|\frac{A+1}{A-1}\right|
\end{equation}
which defines the interior of an ellipse in the complex $z$-plane, with foci at $z=0,1$ and passing through $z=a$. On the ellipse $t_1 s_2(z)=1$. Raabe's test guarantees absolute convergence, if there exists $c>0$ such that:
\begin{equation}
\lim_{\nu\rightarrow \infty}\nu \Re\left(\frac{c_{\nu+1}y_{\nu+1}(z)}{c_{\nu}y_{\nu}(z)}-1\right)=-1-c
\end{equation}
From the results above
\begin{equation}
\lim_{\nu\rightarrow \infty}\nu \Re\left(\frac{c_{\nu+1}y_{\nu+1}(z)}{c_{\nu}y_{\nu}(z)}-1\right)=\Re \varepsilon -2.
\end{equation}
Thus by Raabe's test the series converges absolutely on the ellipse if $\Re \varepsilon<1$.
The formal procedure can be applied to angular Heun equation with the ansatz: $\bar{S}(z)=\sum_{\nu=0}^{\infty}c_{\nu}y_{\nu}(z)$ where $y_{\nu}(z)$ are given in (\ref{PolyCGCJACOBI}) and $a=z_3$. The parameters of angular Heun's equation are given in (\ref{HeunMunchen}),(\ref{auxialiaryfield}), $\alpha=\sum_{i=1}^{3}\alpha_i+\alpha_3^{*}+1$,
$\beta=\sum_{i=1}^{3}\alpha_i-\alpha_3^{*}+1$.

Now we will describe how the separability constant $K_{lm}$ can be determined from the recurrence relation (\ref{recursionAUGCONVERGENCE}) compatible with the augmented convergence of the series expansion (\ref{JacobiP}) solution. For this we can rewrite  (\ref{recursionAUGCONVERGENCE}) in the form:
\begin{equation}
\mathcal{D}_{\nu}c_{\nu+1}+\mathcal{E}_{\nu}c_{\nu}+\mathcal{F}_{\nu}c_{\nu-1}=0,
\label{anadromsigklisi}
\end{equation}
\begin{align}
\mathcal{F}_{\nu}&=-\frac{(\nu-1+\omega)(\nu-1+\gamma)(\nu-1+\alpha)(\nu-1+\beta)}
{(2\nu+\omega-2)(2\nu+\omega-1)} \\
\mathcal{D}_{\nu}&=-\frac{(\nu+\delta)(\nu+1)(\nu+1+\omega-\alpha)(\nu+1+\omega-\beta)}
{(2\nu+\omega+2)(2\nu+\omega+1)}\\
\mathcal{E}_{\nu}&=\frac{J_{\nu}}{(2\nu+\omega+1)(2\nu+\omega-1)}-z_3\nu(\nu+\omega)-q,
\label{SEPCONSTANTQUANTUM}
\end{align}
\begin{equation}
J_{\nu}=[\nu(\nu+\omega)+\alpha\beta][2\nu(\nu+\omega)+\gamma(\omega-1)]+\varepsilon\nu
(\nu+\omega)(\gamma-\delta)
\end{equation}
We note that the separability constant $K_{lm}$ enters only in the expression for $\mathcal{E}_{\nu}$, Eqn.(\ref{SEPCONSTANTQUANTUM}) through $q$. Thus we can write $\mathcal{E}_{\nu}=\mathcal{Q}_{\nu}-q$ where the quantities $\mathcal{D}_{\nu},\mathcal{F}_{\nu},\mathcal{Q}_{\nu}$ do not contain $K_{lm}$. Now if we define $v_{\nu}=\frac{c_{\nu+1}}{c_{\nu}}$ we can obtain from (\ref{anadromsigklisi}) the equation:
\begin{equation}
v_{\nu-1}=\frac{-\mathcal{F}_{\nu}}{\mathcal{D}_{\nu}v_{\nu}+\mathcal{E}_{\nu}}
\label{contifraction}
\end{equation}
Likewise we can obtain the equation $u_{\nu+1}=\frac{-\mathcal{D}_{\nu}}{\mathcal{E}_{\nu}+\mathcal{F}_{\nu}u_{\nu}}$. Now applying twice (\ref{contifraction}) we obtain the continued fraction
\begin{equation}
v_{\nu-1}=\frac{-\mathcal{F}_{\nu}}{\mathcal{Q}_{\nu}-q-}\frac{\mathcal{D}_{\nu}\mathcal{F}_{\nu+1}}
{\mathcal{Q}_{\nu+1}-q+\mathcal{D}_{\nu+1}v_{\nu+1}}
\end{equation}
Following this idea $v_{\nu-1}$ can be expressed as an infinite continued fraction:
\begin{equation}
\mathcal{E}_{0}=\frac{\mathcal{D}_{0}\mathcal{F}_{1}}{\mathcal{Q}_1-q-}
\frac{\mathcal{D}_{1}\mathcal{F}_{2}}{\mathcal{Q}_2-q-}\cdots
\end{equation}

In a more compact form we can summarise the procedure of determining the eigenparameter $K_{lm}$ for the angular equation. The constant of separation in the angular equation that appears in $q$ in (\ref{auxialiaryfield}) is determined from  the following transcendental equation compatible with the convergence of the series:
\begin{equation}
R_{\nu}L_{\nu-1}=1,
\label{continuedfracTRANSCE}
\end{equation}
where now the continued fractions are defined as follows:
\begin{equation}
R_{\nu}=\frac{c_{\nu}}{c_{\nu-1}},\;\;\;L_{\nu}=\frac{c_{\nu}}{c_{\nu+1}}.
\end{equation}
A detailed account of the series solution in hypergeometric polynomials of the angular Heun equation will necessarily be a subject of a separate publication.

\subsection{Heun equation elliptic functions and Inozemtsev system}
There is a deep connection between the Heun equation and quantum mechanical system such as the Inozemtsev model \cite{INOZEMTSEV}. The connection stems from the fact that the Heun equation admits an expression in terms of elliptic functions and this expression is closely related to the $BC_1$ Inozemtsev model \cite{RONVEAUX} and in particular with the Darboux transformation, see Appendix A for some details.
Let us explain this.

Let $f(x)$ denotes an eigenfunction of the Hamiltonian $H$ with eigenvalue $E$:
\begin{equation}
\left(-\frac{{\rm d}^2}{{\rm d}x^2}+\sum_{i=0}^3l_i (l_i+1)\wp(x+\omega_i)-E\right)f(x)=0,
\label{Darboux}
\end{equation}
where $\wp(x)$ is the Weierstra\ss elliptic function (which is also a Jacobi modular form of weight $2$) with periods ($1,\tau$), $\omega_0=0,\omega_1=1/2,\omega_2= \frac{1+\tau}{2},\omega_3=\tau/2$ are half-periods and $l_i (i=0,1,2,3)$ are coupling constants. Also assuming $\Im(\tau)>0$ we have $e_i=\wp(\omega_i), (i=1,2,3)$.
 Applying the transformation
\begin{equation}
w=\frac{e_1-e_3}{\wp(x)-e_3},t=\frac{e_1-e_3}{e_2-e_3},
\end{equation}
\begin{equation}
\tilde{\Phi}(w)=w^{\frac{l_0+1}{2}}(w-1)^{\frac{l_1+1}{2}}(w-t)^{\frac{l_2+1}{2}}
\label{JACOBIDARBOUX}
\end{equation}
reduces the differential equation (\ref{Darboux}) into a Heun's differential equation
\begin{equation}
\left(\left(\frac{{\rm d}}{{\rm d}w}\right)^2+\left(\frac{l_0+\frac{3}{2}}{w}+\frac{l_1+\frac{3}{2}}{w-1}
+\frac{l_2+\frac{3}{2}}{w-t}\right)\frac{\rm d}{{\rm d}w}+\frac{\left(\frac{\sum_{i=0}^3l_i+4}{2}\right)(\frac{3+\sum_{i=0}^2l_i-l_3}{2})w-q}{w(w-1)(w-t)}
\right)\tilde{f}(w)=0
\label{HeDarb}
\end{equation}
where the auxiliary parameter is given by
\begin{equation}
q=-\frac{t}{4}\left(\frac{E}{e_1-e_3}+\left(\frac{t+1}{3t}\right)\sum_{i=0}^3l_i(l_i+1)-
\frac{1}{t}(l_0+l_2+2)^2-(l_0+l_1+2)^2\right)
\end{equation}
Also we have
\begin{equation}
f(x)=\tilde{f}\left(\frac{e_1-e_3}{\wp(x)-e_3}\right)\tilde{\Phi}\left(\frac{e_1-e_3}{\wp(x)-e_3}\right)
=\tilde{f}(w)\tilde{\Phi}(w)
\end{equation}
We note the very useful relationships involved in the transformation (\ref{JACOBIDARBOUX})
\begin{align}
\left(\frac{{\rm d}w}{{\rm d}x}\right)^2&=4(e_2-e_3)w(w-1)(w-t),\\
\frac{{\rm d}^2 w}{{\rm d}x^2}&=\frac{1}{2}\left(\frac{{\rm d}w}{{\rm d}x}\right)^2\left[\frac{1}{w}+\frac{1}{w-1}+\frac{1}{w-t}\right],\;\frac{{\rm d}w}{{\rm d}x}=-\frac{(e_1-e_3)\wp^{\prime}(x)}{(\wp(x)-e_3)^2}
\end{align}
as well as
\begin{align}
&\wp(x+\omega_1)=\frac{(e_2-e_1)w}{w-1}+e_1,\;\wp(x+\omega_2)=-\frac{(e_2-e_1)w}{w-t}+e_2, \nonumber \\
&\wp(x+\omega_3)=e_3-w(e_3-e_2)
\end{align}
We also note that
\begin{equation}
\gamma+\delta+\varepsilon=l_0+l_1+l_2+\frac{9}{2}=\alpha+\beta+1,
\label{Lfuchs}
\end{equation}
and thus the condition (\ref{Fuchs}) is satisfied, therefore equation (\ref{HeDarb}) is indeed a Heun equation.

Conversely, if a differential equation of second order with four regular singular points on a Riemann sphere is given, we can transform it into a Heun equation as equation (\ref{HeDarb}) with the Fuchsian condition (\ref{Lfuchs}) with suitable $l_i,(i=0,1,2,3)$ and $q$ by using a change of variables through a homographic transformation of the independent variable $w\rightarrow \frac{a^{\prime} w+b^{\prime}}{c^{\prime}w+d^{\prime}}$ followed by a transformation of the dependent variable $f\rightarrow w^{\alpha_1}(w-1)^{\alpha_2}(w-t)^{\alpha_3}f$. If $t^{-1}\not =0,1$ than there exists a solution $\tau$ to the equation $t^{-1}=\frac{e_2-e_3}{e_1-e_3}=k^2$,
the roots $e_i, (i=1,2,3)$ depend on $\tau$.
Indeed (see also Appendix A, equation (\ref{KarlWeierstrass})) the \textit{real half-period} is the value of $u$ for $x=e_1$ and we have:
\begin{equation}
\omega_1=\int_{e_1}^{\infty}\frac{{\rm d}x}{\sqrt{X}}=\int_{e_3}^{e_2}\frac{{\rm d}x}{\sqrt{X}}=\frac{K}{\sqrt{e_1-e_3}}.
\end{equation}
For values of $x$ between $e_1$ and $e_2$ or between $e_3$ and $-\infty$, $\sqrt{X}$ is imaginary; however, the value of $\int {\rm d}x/\sqrt{X}$ between the limits $e_3$ and $-\infty$ is denoted by $\omega_3$, and called the \textit{imaginary half period}, so that
\begin{equation}
\omega_3=\int_{e_2}^{e_1}\frac{{\rm d}x}{\sqrt{X}}=\int_{-\infty}^{e_3}\frac{{\rm d}x}{\sqrt{X}}=\frac{iK^{\prime}}{\sqrt{e_1-e_3}}
\end{equation}
Thus the parameter $\tau$ is determined. This in turn, determines the values of the cubic roots $e_1,e_2,e_3$ and $E$. Thus we obtain a Hamiltonian of $BC_1$ Inozemtsev model with an eigenvalue $E$ starting from a differential equation of second order with four regular points on a Riemann sphere.

In terms of Jacobian elliptic functions (see Appendix A) the fuchsian equation (\ref{HeunMunchen}) acquires the following elliptic representation:
\begin{align}
&\frac{{\rm d}^2 \bar{S}}{{\rm d}u^2}+\left[(4\alpha_1+1)\frac{{\rm cn}u{\rm dn}u}{{\rm sn}u}-(4\alpha_2+1)\frac{{\rm sn}u{\rm dn}u}{{\rm cn}u}-k^2(4\alpha_3+1)\frac{{\rm sn}u{\rm cn}u}{{\rm dn}u}\right]\frac{{\rm d}\bar{S}}{{\rm d}u} \nonumber \\
&+\left(4\alpha\beta k^2{\rm sn}^2 u-4k^2q\right)\bar{S}=0
\end{align}
where the auxiliary parameter $q$ is given by equation (\ref{auxialiaryfield}), the parameters $\alpha,\beta$ are determined by equation (\ref{ALPHABETAH}) while $k^{-2}=z_3= -\frac{1}{2}\left(-1+\frac{\alpha_{\Lambda}-1}{2i\sqrt{\alpha_{\Lambda}}}\right)$, $z={\rm sn}^2(u,k)$.

\section{False singular points and exact solution of the angular KGF equation}
\label{APPARENTSINGULARFUCHS}

In this section we shall see that for special values of the scalar field mass the fourth finite singularity $z_{\infty}$ can be of special character namely that of a \textit{false} singularity. In this case, as in the case of Heun's equation, a skillfull change of variables can transform the Fuchsian equation into a \textit{finite-gap} elliptic Schr\"{o}dinger equation. In fact Smirnov in \cite{SMIRNOV} has shown the theorem: \textit{A Fuchsian equation with M+4 singular points is `finite-gap' if and only if, for all $k,k=1,\cdots,M$, $z=b_k$ is a false singular point}.

Let us define in what follows the concept of a false singular point.
An arbitrary Fuchsian equation of second order can be written in the form:
\begin{equation}
\frac{{\rm d}^2 Y}{{\rm d}z^2}=f(z)\frac{{\rm d}Y}{{\rm d}z}+g(z)Y
\label{FuchsA}
\end{equation}
where $f(z)$ and $g(z)$ are known rational functions and we recall that Fuchsian equations have only regular singular points: irregular singular points occur in confluent cases when two regular points coalesce in a particular limiting process. We assume that equation (\ref{FuchsA}) has regular singular points (i.e. the poles of the coefficients $f$ and $g$) $a_i, i=1,\ldots,v$, and that a local expansion at each singular point yields a pair of exponents $\{\alpha_i,\beta_i\}$ that characterise the local behaviour there.

We call a singular point $a_i$ \textit{false} if both exponents $\alpha_i$ and $\beta_i$ are non-negative integers and there are no logarithmic terms in the local expansion near the singular point.
It is known that such logarithmic terms generally appear in the case when the difference of the exponents is an integer,so a specific restriction must be imposed on the coefficients of the equation.

We shall discuss briefly these restrictions on the coefficients of the equation (\ref{FuchsA}) so that the singular point $a_j$ is false. Considering the simplest false point with the exponents equal to $0$ and $2$, then, from the general theory of Fuchsian equations, the exponents follow from a characteristic equation, and local to $a_j$:
\begin{equation}
f(z)=\frac{1}{z-a_j}+f_0+O(z-a_j), g(z)=\frac{g_{-1}}{z-a_j}+g_0+O(z-a_j),
\end{equation}
for some constants $f_0,g_{-1},g_0$. The solution corresponding to the exponent zero can be written in the form
\begin{equation}
Y(z)=\sum_{m=0}^{\infty}c_m(z-a_j)^m=c_0+c_1(z-a_j)+c_2(z-a_j)^2+O((z-a_j)^3),
\label{zeroexpo}
\end{equation}
for some constants $c_0,c_1,c_2$. Substituting (\ref{zeroexpo}) into the equation (\ref{FuchsA}) we obtain recursive equations for the coefficients at different orders of $(z-a_j)$ and this yields:
\begin{equation}
\fbox{$\displaystyle
g_{-1} f_0-g_0+(g_{-1})^2=0,
$}
\label{nolog}
\end{equation}
and as a consequence to the absence of logarithmic terms local to $a_j$.
Indeed, considering the operator
\begin{equation}
L=x^2\frac{{\rm d}^2}{{\rm d}x^2}+xp(x)\frac{{\rm d}}{{\rm d}x}+q(x),
\end{equation}
where $p(x)$ and $q(x)$ are analytic functions at $x=0$ with the power series expansions:
\begin{equation}
p(x)=\sum_{j=0}^{\infty} p_j x^j,\;q(x)=\sum_{j=0}^{\infty}q_jx^{j}
\end{equation}
Put
\begin{equation}
y(x)=x^r\sum_{m=0}^{\infty}c_m x^m, (c_0\not=0, {\rm we\;can\;choose\;} c_0=1)
\end{equation}
we can calculate $Ly$
\begin{equation}
Ly=\sum_{m\geq 0}\{((m+r)(m+r-1)+(m+r)p_0+q_0)c_m+R_m \}x^{m+r},
\end{equation}
where
\begin{equation}
R_0=0,R_m=\sum_{i+j=m,i\not =m}\{(i+r)c_ip_j+c_iq_j\}=
\sum_{k=0}^{m-1}[(k+r)p_{m-k}+q_{m-k}]c_k,m>0
\end{equation}
If we set
\begin{equation}
F(r)=r(r-1)+rp_0+q_0,
\end{equation}
then we see that $Ly=0$ if, and only if:
\begin{equation}
F(r+m)c_m+R_m=0,m=0,1,2,\ldots
\end{equation}
The second order algebraic equation $F(r)=0$ is called the \textit{characteristic equation} or \textit{indicial equation} and the roots of the equation are called the \textit{characteristic exponents}. If the exponents are denoted by $r_1,r_2$, their difference is an integer $r_2-r_1=m\in \mathbb{Z}$, $m\not =0$  and $R_m=0$, the singularity is called non-logarithmic because no logarithm appears in the expansions for the two solutions.
Now returning to our case of a false singularity with exponents $(0,2)$ we have $p_0=-1,q_0=0,p_1=-f_0,q_1=-g_{-1},q_2=-g_0$ we get
\begin{align}
R_2&=[-rp_2+q_2]c_0+[(1+r)p_1+q_1]c_1=q_2 c_0+(p_1+q_1)c_1 \nonumber \\
&=-g_0c_0+(-f_0-g_{-1})(-g_{-1}c_0)=-g_0c_0+f_0g_{-1}c_0+g_{-1}^2 c_0,
\label{apparantfalse}
\end{align}
which since $c_0\not=0$ leads to the condition of absence of a logarithmic singularity, Eq.(\ref{nolog}).

\subsection{Conditions on the coefficients of the Fuchsian angular Klein-Gordon-Fock equation and the scalar mass such that the fifth singular point is a false point}
\label{WICHTIGFALSE}

Equation (\ref{FORTIOANGULAR}) with the aid of (\ref{homograph}) becomes:
\begin{align}
& \Biggl\{ \frac{\mathrm{d}^2}{\mathrm{d}z^2}+\left[\frac{1}{z}+\frac{1}{z-1}
+\frac{1}{z-z_3}-\frac{2}{z-z_{\infty}}\right]\frac{\mathrm{d}}{\mathrm{d}z}\nonumber \\
&-\frac{m^2}{4}\frac{1}{z^2}-\frac{m^2}{4}\frac{1}{(z-1)^2}+\left(\frac{\Xi a \omega}{2\sqrt{\alpha_{\Lambda}}}-\frac{m \sqrt{\alpha_{\Lambda}}}{2}\right)^2 \frac{1}{(z-z_3)^2}+
\frac{a^2\mu^2}{\alpha_{\Lambda}(z-z_{\infty})^2}+ \nonumber \\
&\frac{1}{z}\left[\frac{m^2(1+2i\sqrt{\alpha_{\Lambda}}+3\alpha_{\Lambda})}{2(-i+\sqrt{\alpha_{\Lambda}})^2}
+\frac{2m\Xi \xi}{(1+i\sqrt{\alpha_{\Lambda}})^2}+\frac{a^2\mu^2}{(-i+\sqrt{\alpha_{\Lambda}})^2}+
\frac{K_{lm}}{(1+i\sqrt{\alpha_{\Lambda}})^2}\right] \nonumber \\
&+\frac{1}{z-1}\left[\frac{-m^2(1-2i\sqrt{\alpha_{\Lambda}}+3\alpha_{\Lambda})}{2(i+\sqrt{\alpha_{\Lambda}})^2}
-\frac{-2m\xi \Xi}{(1-i\sqrt{\alpha_{\Lambda}})^2}-\frac{a^2\mu^2}{(i+\sqrt{\alpha_{\Lambda}})^2}
-\frac{K_{lm}}{(1-i\sqrt{\alpha_{\Lambda}})^2}\right] \nonumber \\
&+\frac{1}{z-z_3}\left[\frac{-8i m^2 \alpha_{\Lambda}\sqrt{\alpha_{\Lambda}}}{\Xi^2}+\frac{8i m \sqrt{\alpha_{\Lambda}}\xi}{\Xi}+\frac{4ia^2\mu^2}{\sqrt{\alpha_{\Lambda}}\Xi^2}+\frac{4i\sqrt{\alpha_{\Lambda}}K_{lm}}{\Xi^2} \right] \nonumber \\
&+\frac{1}{z-z_{\infty}}\frac{-4i a^2\mu^2}{\sqrt{\alpha_{\Lambda}}\Xi} \Biggr\}S(z)=0,
\label{preheunB}
\end{align}
We have five singular points. The exponentials at the singular point $z_{\infty}$ are obtained by solving the indicial equation:
\begin{equation}
F(r)=r(r-1)+p_0 r+q_0=0
\end{equation}
where $p_0=\lim_{z\rightarrow z_{\infty}}(z-z_{\infty})\frac{-2}{z-z_{\infty}}=-2$ and $q_0=\lim_{z\rightarrow z_{\infty}}(z-z_{\infty})^2Q(z)=\frac{a^2\mu^2}{\alpha_{\Lambda}}$. Thus we obtain $r_{1,2}(\mu)=\frac{3\pm \sqrt{9-4\frac{a^2\mu^2}{\alpha_{\Lambda}}}}{2}$. Now choosing $\frac{5}{4}=\frac{a^2\mu^2}{\alpha_{\Lambda}}$, and performing the homotopy transformation (\ref{fhomotopy}) for the dependent variable but now with $\alpha_4=\frac{1}{2}$ one transforms (\ref{preheunB}) into an equation with the same singularities however the exponents of the singular point $z_{\infty}$ will be now $\{0,2\}$, i.e. non-negative integers. Thus, for this choice of scalar mass we can arrange matters so that the singularity $z_{\infty}$ becomes false. However in order for this to be true, also the condition (\ref{nolog},)that guarantees the absence of logarithmic terms needs to be satisfied.
The terms appearing in (\ref{nolog}) are calculated to be:
\begin{align}
& g_{-1}=\frac{-i\sqrt{\alpha_{\Lambda}}}{\Xi},
\label{physicalapparency1} \\
& f_0=\frac{2\alpha_1+1}{z_{\infty}}+\frac{2\alpha_2+1}{z_{\infty}-1}+\frac{2\alpha_3+1}{z_{\infty}-z_3}, \label{physicappara02}\\
& g_0=\left[\frac{m^2(1+2i\sqrt{\alpha_{\Lambda}}+3\alpha_{\Lambda})}{2(-i+\sqrt{\alpha_{\Lambda}})^2}
+\frac{2m \xi\Xi}{(1+i\sqrt{\alpha_{\Lambda}})^2}+\frac{-2\alpha_{\Lambda}}{(1+i\sqrt{\alpha_{\Lambda}})^2}
+\frac{K_{lm}}{(1+i\sqrt{\alpha_{\Lambda}})^2}\right]\frac{1}{z_{\infty}}\nonumber  \\
&+\left[\frac{m^2}{2}\left(1+\frac{4\alpha_{\Lambda}}{(1-i\sqrt{\alpha_{\Lambda}})^2}
-\right)\frac{2m \xi\Xi}{(1-i\sqrt{\alpha_{\Lambda}})^2}+\frac{2\alpha_{\Lambda}}{(1-i\sqrt{\alpha_{\Lambda}})^2}
-\frac{K_{lm}}{(1-i\sqrt{\alpha_{\Lambda}})^2}\right]\frac{1}{z_{\infty}-1} \nonumber \\
&+ \left[\frac{-8im^2 \alpha_{\Lambda}\sqrt{\alpha_{\Lambda}}}{\Xi^2}+
\frac{8im\sqrt{\alpha_{\Lambda}}\xi}{\Xi}+\frac{8i\sqrt{\alpha_{\Lambda}}}{\Xi^2}
+\frac{4i\sqrt{\alpha_{\Lambda}}K_{lm}}{\Xi^2}\right]\frac{1}{z_{\infty}-z_3}
\label{physicalfalsesingular}
\end{align}

\section{The solutions of massive radial equation for a charged particle in the KNdS black hole spacetime \label{RADEQNMPKNDS}}

We write the quantity $\Delta_r^{KN}$ in terms of the radii of the event and Cauchy horizons $r_+,r_{-}$ and the cosmological horizon $r_{\Lambda}^+$ for positive cosmological constant:
\begin{equation}
\Delta_r^{KN}=-\frac{\Lambda}{3}(r-r_+)(r-r_{-})(r-r_{\Lambda}^{+})(r-r_{\Lambda}^{-})
\label{cosmochargrot}
\end{equation}
There are five regular singularities in (\ref{FORTIORADIAL}), at the points $r_{\pm},r_{\Lambda}^{\pm},\infty$. Applying the homographic substitution
\begin{equation}
z=\left(\frac{r_+-r_{\Lambda}^{-}}{r_+-r_{-}}\right)\left(\frac{r-r_{-}}{r-r_{\Lambda}^{-}}\right)
\label{VARIABLEZ}
\end{equation}
Equation (\ref{cosmochargrot}) in terms of the new variable is written:
\begin{equation}
\Delta_r^{KN}=-\frac{\Lambda}{3}\frac{H z_{\infty}^3 z (z-1)(z-z_r)}{(z_{\infty}-z)^4},
\end{equation}
where $H:=\frac{(r_{-}-r_{\Lambda}^{-})^2 (r_+-r_{-})(r_{\Lambda}^+-r_{-})}{z_r}$.
Also we have the following relations and definitions:
\begin{equation}
r=\frac{r_{-}z_{\infty}-r_{\Lambda}^{-} z}{z_{\infty}-z},\;z_{\infty}:=\frac{r_{+}-r_{\Lambda}^{-}}{r_{+}-r_{-}},\;z_r:=z_{\infty}\left(
\frac{r_{\Lambda}^+-r_{-}}{r_{\Lambda}^+-r_{\Lambda}^{-}}\right),
\end{equation}
\begin{align}
\frac{{\rm d}z}{{\rm d}r}&=\frac{z_{\infty}(r_{-}-r_{\Lambda}^{-})}{(r-r_{\Lambda}^-)^2}
=\frac{1}{z_{\infty}}\frac{1}{r_{-}-r_{\Lambda}^-}(z_{\infty}-z)^2=
\frac{r_{+}-r_{-}}{r_{+}-r_{\Lambda}^-}\frac{1}{r_{-}-r_{\Lambda}^{-}}
(z_{\infty}-z)^2 \\
\frac{{\rm d}^2z}{{\rm d}r^2}&=\frac{-2z_{\infty}(r_{-}-r_{\Lambda}^-)}{(r-r_{\Lambda}^{-})^3},\;\;\frac{\frac{{\rm d}^2z}{{\rm d}r^2}}{\left(\frac{{\rm d}z}{{\rm d}r}\right)^2}=\frac{-2}{z_{\infty}-z}.
\end{align}

Applying the homographic transformation (\ref{VARIABLEZ}) in the radial equation for a massive charged particle (\ref{FORTIORADIAL}) we obtain:
\begin{align}
&\frac{{\rm d}^2R}{{\rm d}z^2}+\frac{1}{\left(\frac{{\rm d}z}{{\rm d}r}\right)^2}\frac{1}{\Delta_r^{KN}}\frac{{\rm d}\Delta_r^{KN}}{{\rm d}r}\frac{{\rm d}R}{{\rm d}r}+\frac{\frac{{\rm d}^2z}{{\rm d}r^2}}{\left(\frac{{\rm d}z}{{\rm d}r}\right)}\frac{{\rm d}R}{{\rm d}z}+\frac{\Xi^2(K(r)-\frac{eqr}{\Xi})^2}
{(\Delta_r^{KN})^2\left(\frac{{\rm d}z}{{\rm d}r}\right)^2}+\frac{-r^2\mu^2 R}{\left(\frac{{\rm d}z}{{\rm d}r}\right)^2 \Delta_r^{KN}}-\frac{K_{lm}R}{\Delta_r^{KN}\left(\frac{{\rm d}z}{{\rm d}r}\right)^2}\nonumber \\
&=\frac{{\rm d}^2R}{{\rm d}z^2}+\left\{\frac{1}{z}+\frac{1}{z-1}+
\frac{1}{z-z_r}-\frac{2}{z-z_{\infty}}\right\}\frac{{\rm d}R}{{\rm d}z}+ \left[\frac{A^{\prime}}{z^2}+\frac{B^{\prime}}{z}+\frac{C^{\prime}}{(z-1)^2}
+\frac{D^{\prime}}{z-1}+\frac{E^{\prime}}{(z-z_r)^2}+\frac{H^{\prime}}{z-z_r}\right]R\nonumber \\
&+\left[\frac{A}{(z_{\infty}-z)^2}+\frac{B}{z_{\infty}-z}+\frac{C}{z}+
\frac{D}{z-1}+\frac{F}{z-z_r}\right]R+\left[\frac{\mathcal{B}_{K_{lm}}}{z}
+\frac{\mathcal{D}_{K_{lm}}}{z-1}+
+\frac{\mathcal{H}_{K_{lm}}}{z-z_r}\right]R=0.
\end{align}

The coefficients of the partial fraction expansion for the term $\frac{-r^2\mu^2 R}{\left(\frac{{\rm d}z}{{\rm d}r}\right)^2 \Delta_r^{KN}}$, are given as follows:
\begin{align}
A&=\frac{3\mu^2}{\Lambda}, \\
B&=\frac{3\mu^2}{\Lambda}\frac{1}{r_{-}-r_{\Lambda}^-}\left[
\frac{(r_{\Lambda}^{-}+r_{-})z_r-2r_{-}z_{\infty}-2r_{-}z_r z_{\infty}-(r_{\Lambda}^{-}-3r_{-})z_{\infty}^2}{(1-z_{\infty})(z_r-z_{\infty})z_{\infty}}\right], \\
C&=\frac{3\mu^2}{\Lambda}\frac{1}{r_{+}-r_{-}}\frac{1}{r_{\Lambda}^{+}-r_{-}}\frac{r_{-}^2}{z_{\infty}}, \\
D&=-\frac{3\mu^2}{\Lambda}\frac{z_r}{r_{+}-r_{-}}\frac{1}{r_{\Lambda}^{+}-r_{-}}\frac{1}{z_{\infty}}
\frac{[r_{\Lambda}^{-}-r_{-}z_{\infty}]^2}{(z_r-1)(z_{\infty}-1)},\\
F&=\frac{3\mu^2}{\Lambda}\frac{1}{r_{+}-r_{-}}\frac{1}{r_{\Lambda}^{+}-r_{-}}\frac{1}{z_{\infty}}
\frac{(r_{\Lambda}^{-}z_r-r_{-}z_{\infty})^2}{(z_r-1)(z_r-z_{\infty})^2}
\end{align}
while the expansion coefficients $A^{\prime},C^{\prime},E^{\prime}$ are computed to be:
\begin{align}
A^{\prime}&=\frac{a^4}{\alpha_{\Lambda}^2}\frac{[\Xi K(r_{-})-eqr_{-}]^2}
{(r_{-}-r_{\Lambda}^-)^2(r_{+}-r_{-})^2(r_{\Lambda}^{+}-r_{-})^2}
\\
C^{\prime}&=\frac{a^4}{\alpha_{\Lambda}^2}\frac{[\Xi K(r_{+})-eqr_{+}]^2}
{(r_{+}-r_{\Lambda}^-)^2(r_{+}-r_{\Lambda}^+)^2(r_{+}-r_{-})^2}  \\
E^{\prime}&=\frac{a^4}{\alpha_{\Lambda}^2}\frac{[\Xi K(r_{\Lambda}^+)-eqr_{\Lambda}^{+}]^2}{(r_{\Lambda}^{+}-r_{-})^2(r_{\Lambda}^{+}-
r_{\Lambda}^-)^2(r_{+}-r_{\Lambda}^+)^2}
\end{align}
Let us calculate the exponents of the singularity at $z_{\infty}$. The indicial equation takes the form:
\begin{equation}
F(r)=r(r-1)-2r+\frac{3\mu^2}{\Lambda}=0,
\end{equation}
and the exponents are computed to be:
\begin{equation}
r^{1,2}_{\mu z_{\infty}}=\frac{3}{2}\pm\frac{1}{2}\sqrt{9-\frac{12\mu^2}{\Lambda}}.
\end{equation}
Subsequently we compute the exponents for the regular singularities $z=0,z=1,z=z_r$.
Indeed the indicial equation for the $z=1$ singularity takes the form:
\begin{equation}
F(r)=r(r-1)+r+\frac{a^4}{\alpha_{\Lambda}^2}\frac{[\Xi K(r_{+})-eqr_{+}]^2}{(r_{+}-r_{\Lambda}^{-})^2(r_{+}-r_{\Lambda}^{+})^2(r_{+}-r_{-})^2}=0
\end{equation}
Thus the roots are calculated to be:
\begin{equation}
r^{1,2}_{z=1}\equiv \mu_2=\pm \frac{i a^2}{\alpha_{\Lambda}}\frac{\Xi K(r_{+})-eqr_{+}}{(r_{\Lambda}^{-}-r_{+})(r_{-}-r_{+})(r_{\Lambda}^{+}-r_{+})}
\end{equation}
Likewise we compute the exponents of the other two singularities:
\begin{align}
r^{1,2}_{z=0}&\equiv \mu_1=\pm \frac{ia^2}{\alpha_{\Lambda}}\frac{\Xi K(r_{-})-eqr_{-}}{(r_{-}-r_{\Lambda}^{-})(r_{+}-r_{-})(r_{\lambda}^{+}-r_{-})},\\
r^{1,2}_{z=z_r}&\equiv \mu_3=\pm \frac{ia^2}{\alpha_{\Lambda}}\frac{[\Xi K(r_{\Lambda}^{+})-eqr_{\Lambda}^{+}]}{(r_{\Lambda}^{-}-r_{\Lambda}^{+})(r_{+}-r_{\Lambda}^{+})(r_{-}-r_{\Lambda}^{+})}.
\end{align}
Thus we see that in general  the radial Fuchsian KGF equation for a massive charged particle in the curved spacetime of a cosmological rotating charged black hole possess five singularites including the infinity. We investigate as in the case of the angular equation the possibility of deriving exact solutions in terms of Heun functions, i.e. eliminating one of the regular finite singularities. Indeed choosing a value of the scalar mass in terms of the cosmological constant as $\mu^2=\frac{2}{3}\Lambda$ the exponents of the $z_{\infty}$ singularity become $r^{1,2}_{z_{\infty},\mu^2=\frac{2}{3}\Lambda}=2,1$. Thus applying the  $F$-homotopic transformation of the dependent variable $R$
\begin{equation}
R(z)=z^{\mu_1}(z-1)^{\mu_2}(z-z_r)^{\mu_3}(z-z_{\infty})^{r^2_{z_{\infty}}}\bar{R}(z)
\label{factorisation}
\end{equation}
we eliminate the $z_{\infty}$ singularity and   reduce one of the exponents of the three finite singularities $z=0,1,z_r$ to zero. Consequently for this value  for the scalar mass the radial part of the KGF Fuchsian equation in the curved spacetime of the KNdS black hole becomes a Heun differential equation:
\begin{align}
&\Biggl\{ \frac{{\rm d}^2}{{\rm d}z^2}+\left[\frac{2\mu_1+1}{z}+\frac{2\mu_2+1}{z-1}+\frac{2\mu_3+1}{z-z_r}\right]
\frac{{\rm d}}{{\rm d}z}+\frac{\alpha\beta z-q}{z(z-1)(z-z_r)}\Biggr\}\bar{R}(z)=0.
\label{BHCRCHEUNRADIAL}
\end{align}

The $F-$ homotopic transformation (\ref{factorisation}) factors out the $z_{\infty}$ singularity because it eliminates both terms $\propto \frac{1}{(z-z_{\infty})^2}$ and $\propto\frac{1}{z-z_{\infty}}$ respectively. Indeed the last term vanishes:
\begin{align}
&\frac{1}{z-z_{\infty}}
\left(\frac{1}{z_{\infty}}-\frac{1}{1-z_{\infty}}-\frac{1}{z_r-z_{\infty}}\right)
-\frac{B}{z-z_{\infty}}\nonumber \\
&=\frac{1}{z-z_{\infty}}
\frac{(r_{-}-r_{+})(r_{\Lambda}^{-}+r_{\Lambda}^{+}+r_{-}+r_{+})}
{(r_{\Lambda}^{-}-r_{-})(r_{\Lambda}^{-}-r_{+})}=0,
\end{align}
due to Vieta's relations, i.e. $r_{\Lambda}^{-}+r_{\Lambda}^{+}+r_{-}+r_{+}=0$.
The four roots $r_{\Lambda}^{-},r_{\Lambda}^{+},r_{-},r_{+}$ of the quartic polynomial $\Delta_r^{KN}$ can be given in closed analytic form by applying the theory developed in \cite{GRGKRANIOTIS}. Indeed they are given by the following formulae in terms of the elliptic functions $\wp,\wp^{\prime}$:
\begin{align}
\alpha & =\frac{1}{2}\frac{\wp^{\prime}(-x_{1}/2+\omega)-\wp^{\prime}(x_{1}%
)}{\wp(-x_{1}/2+\omega)-\wp(x_{1})},\label{maxweierstrass}\\
\beta & =\frac{1}{2}\frac{\wp^{\prime}(-x_{1}/2+\omega+\omega^{\prime}%
)-\wp^{\prime}(x_{1})}{\wp(-x_{1}/2+\omega+\omega^{\prime})-\wp(x_{1})},\label{weirg3}\\
\gamma & =\frac{1}{2}\frac{\wp^{\prime}(-x_{1}/2+\omega^{\prime})-\wp^{\prime
}(x_{1})}{\wp(-x_{1}/2+\omega^{\prime})-\wp(x_{1})}\label{weirstathec},\\
\delta & =\frac{1}{2}\frac{\wp^{\prime}(-x_{1}/2)-\wp^{\prime}(x_{1})}%
{\wp(-x_{1}/2)-\wp(x_{1})}\label{fourth}.
\end{align}
The point $x_1$ is defined by the equation:
\begin{equation}
-6\wp(x_1)=-\frac{3}{\Lambda}+a^2,
\end{equation}
while $\omega,\omega^{\prime}$ denote the half-periods of the Weierstra$\ss$ elliptic function $\wp$. The equations:
\begin{equation}
4\wp^{\prime}(x_1)=\frac{6}{\Lambda},\;-3\wp^2(x_1)+g_2=-\frac{3}{\Lambda}(a^2+e^2),
\end{equation}
determine the Weierstra$\ss$ invariants ($g_2,g_3$) with the result:
\begin{align}
g_2&=\frac{1}{12}\left(-\frac{3}{\Lambda}+a^2\right)^2-\frac{3}{\Lambda}(a^2+e^2),\\
g_3&=-\frac{1}{216}\left(-\frac{3}{\Lambda}+a^2\right)^3-
\frac{3}{\Lambda}\frac{1}{6}(a^2+e^2)\left(-\frac{3}{\Lambda}+a^2\right)-\frac{9}{4\Lambda^2}.
\end{align}

In terms of Jacobian elliptic functions the Heun-Fuchsian differential equation (\ref{BHCRCHEUNRADIAL}) acquires the form:
\begin{align}
&\frac{{\rm d}^2 \bar{R}}{{\rm d}u^2}+\left[(4\mu_1+1)\frac{{\rm cn}u{\rm dn}u}{{\rm sn}u}-(4\mu_2+1)\frac{{\rm sn}u{\rm dn}u}{{\rm cn}u}-k^2(4\mu_3+1)\frac{{\rm sn}u{\rm cn}u}{{\rm dn}u}\right]\frac{{\rm d}\bar{R}}{{\rm d}u} \nonumber \\
&+\left(4\alpha\beta k^2{\rm sn}^2 u-4k^2q\right)\bar{R}=0,
\label{FORMJACOBIHEUN}
\end{align}
where the Jacobi modulus satisfies the equation:
\begin{equation}
k^{-2}=z_r=\frac{r_{+}-r_{\Lambda}^{-}}{r_{+}-r_{-}}\frac{r_{\Lambda}^{+}-r_{-}}{r_{\Lambda}^{+}-r_{\Lambda}^{-}},
\end{equation}
with $z={\rm sn}^2(u,k)$.
Thus we have proved the following:
\begin{theorem}
 For the  value of the scalar mass parameter: $\mu=\sqrt{\frac{2\Lambda}{3}}$ both radial and angular Fuchsian differential equations that result from separation of variables of the KGF equation in KNdS spacetime, are transformed into Heun's equations by eliminating the singularity at $z_{\infty}$.
\end{theorem}
Therefore both equations can be solved in closed analytic form in terms of \textit{general Heun functions}. Radial $\bar{R}(z)$ and angular parts $\bar{S}(z)$ can be expressed locally in terms of Heun functions: $Hl(a_i,q_i;\alpha_i,\beta_i,\gamma_i,\delta_i;z),\;\;i=\bar{R},\bar{S}$.
As we saw in section \ref{augmentedSYGLISI} using the concept of augmented convergence the angular equation can be solved in terms of an infinite series of Jacobi polynomials which converges inside the ellipse with foci at $z=0$ and $z=1$ passing through the point $z_3$ with possible exception of the line connecting the two foci. We also mentioned how the separability constant $K_{lm}$ can in principle be determined in a compatible way with augmented convergence.

\subsection{Solution of the massive radial equation in KNdS spacetime in the ellipse with foci at the event and Cauchy horizons}
We will apply the theory of augmented convergence of section \ref{augmentedSYGLISI} to the massive radial equation for a charged scalar in KNdS spacetime with  $\mu=\sqrt{\frac{2\Lambda}{3}}$.
The solution convergent in the ellipse with foci at $z=0,1$ which correspond to $r=r_{-},r_{+}$, respectively is given by
\begin{align}
\bar{R}_{\nu}(z)&=\sum_{\varrho=-\infty}^{+\infty}c_{\varrho}^{\nu}u_{\nu+\varrho}(z), \\
u_{\nu}(z)&=F(-\nu,\nu+2(\mu_1+\mu_2)+1,2\mu_1+1,z)
\end{align}
The expansion coefficients $c_{\varrho}^{\nu}$ are determined by  the recurrence relation (\ref{recursionAUGCONVERGENCE}) written as follows:
\begin{equation}
D_{\varrho}^{\nu}c_{\varrho+1}^{\nu}+E_{\varrho}^{\nu}c_{\varrho}^{\nu}+F_{\varrho}^{\nu}
c_{\varrho-1}^{\nu}=0,
\end{equation}
where
\begin{align}
D_{\varrho}^{\nu}&=-\frac{(\nu+\varrho+\delta)(\nu+\varrho+1)(\nu+\varrho+1+\omega-\alpha)(\nu+\varrho+1+\omega-\beta)}
{(2\nu+2\varrho+\omega+2)(2\nu+2\varrho+\omega+1)},\\
F_{\varrho}^{\nu}&=-\frac{(\nu+\varrho-1+\omega)(\nu+\varrho-1+\gamma)(\nu+\varrho-1+\alpha)(\nu+\varrho-1+\beta)}
{(2\nu+2\varrho+\omega-2)(2\nu+2\varrho+\omega-1)},\\
E_{\varrho}^{\nu}&=\frac{J_{\varrho}^{\nu}}{(2\nu+2\varrho+\omega+1)(2\nu+2\varrho+\omega-1)}
-z_r(\nu+\varrho)(\nu+\varrho+\omega)-q,
\end{align}
\begin{align}
J_{\varrho}^{\nu}&=[(\nu+\varrho)(\nu+\varrho+\omega)+\alpha\beta](2(\nu+\varrho)(\nu+\varrho+\omega)
+\gamma(\omega-1))\nonumber \\
&+\varepsilon(\nu+\varrho)(\nu+\varrho+\omega)(\gamma-\delta)
\end{align}
The radial solution is expressed as a series, where $\varrho$ runs from $-\infty$ to $\infty$, because $\nu\notin \mathbb{Z}$, since the separation constant $K_{lm}$ is fixed from the angular solution and the parameter $\nu$ is determined by the corresponding in the radial case transcendental equation (\ref{continuedfracTRANSCE}).
A more detailed account of the solution will be given elsewhere.

An interesting research path will be to construct radial solutions for the massive charged scalar particle which are valid not only inside an ellipse with foci at $z=0$ and 1, but also at the cosmological horizon. Such solutions will be valid in all regions of $r$ and they should satisfy certain boundary conditions. The construction of such radial solutions is further motivated for the following reason. As has been argued by the authors in  \cite{STATICRADIUS},\cite{RADIUSBEFOREL} there is a clear possibility to distinguish in the KdS spacetime the region corresponding to gravitational binding and the region of cosmic repulsion where gravitational binding is not possible. These regions are separated by the so called static radius where the KdS spacetime is very close to the flat spacetime \cite{STATICRADIUS},\cite{RADIUSBEFOREL}. For instance for Milky Way this static radius was estimated to be $\sim 11$Kpc \cite{RADIUSBEFOREL}. By constructing such radial solutions valid in all regions of $r$ we will be able to look for possible signature of the static radius in the behaviour of the scalar massive field in the KdS and KNdS spacetimes. The investigation of this very interesting issue is beyond the scope of the present paper and it will be a subject of a future publication.

\subsection{Exact solutions of angular and radial equations for a massive charged particle in the Kerr-Newman spacetime}
\label{LIMITLAMBDAANGULAR}
\subsubsection{Solution of angular KGF equation for a massive charged  particle in the KN spacetime}

Assuming vanishing cosmological constant the separated part due to angular differential equation is given by Eqn.(\ref{FORTIOANGULAR}) setting $\Lambda=0$. The angular differential
equation becomes:
\begin{equation}
\left[(1-x^2)\frac{{\rm d}^2}{{\rm d}x^2}-2x\frac{{\rm d}}{{\rm d}x}+
(\tau^2-\mu^2 a^2)x^2+\frac{-m^2}{1-x^2}+E\right]S=0,
\label{massivespinangular1}
\end{equation}
where $\tau=a\omega$.
Applying the s-homotopic transformation for the dependent variable:
\begin{equation}
S(z)=z^{\frac{m}{2}}(1-z)^{\frac{m}{2}}\exp(2\tau^{\prime}z)w(z),
\end{equation}
yields the non-symmetrical canonical form of the confluent Heun
differential equation (CHE):
\begin{equation}
w^{\prime\prime}(z)+\left(4p+\frac{\gamma}{z}+\frac{\delta}{z-1}\right)w^{\prime}(z)
+\frac{4p\alpha z-\sigma}{z(z-1)}w(z)=0,
\label{CONFLUENTHEUNC}
\end{equation}
where $\tau^{\prime}=\sqrt{\tau^2-\mu^2 a^2}=\sqrt{a^2\omega^2-\mu^2 a^2}$ and the parameters of the confluent differential equation of Heun are given by
\begin{align}
&\sigma:=E+\tau^2-\mu^2a^2-m(m+1)+2\tau^{\prime}(m+1), \;\; p=\tau^{\prime}
\label{conflugwnia1}\\
&\alpha=m+1,  \\ &\gamma=m+1,\;\;\delta=m+1
\label{conflugwnia22}
\end{align}

Thus the exact analytic solution we obtained will involve the confluent Heun functions $Hc(p,\alpha,\gamma,\delta,\sigma;z)$ in the notation of \cite{RONVEAUX}. Several other solutions with confluent Heun  functions with appropriate parameters obtained by transformations which preserve the canonical form of the CHE can in principle be constructed \cite{RONVEAUX}.
\subsubsection{Exact solutions of the angular equation for a massive spin-$\frac{1}{2}$ particle around a Kerr-Newman black hole}
If one wants to include a spin parameter in the massive case this has to be done separately for each value of spin for the massive particle. So the cases of the massive Dirac equation and the massive spin 1 particles in the curved backgrounds of the KNdS and KN black holes have to be studied on an individual basis by solving the corresponding differential equations. Although this will be a subject of a future publication \cite{ProfDrKRANIOTIS} we would like to report in what follows the first exact analytic solution of the angular differential
in the KN spacetime for a massive spin half particle. Indeed in this case the angular equation acquires the form \cite{page}:
\begin{align}
&\Biggl[\frac{1}{\sin\theta}\frac{\rm d}{\rm d\theta}\left(\sin\theta\frac{\rm d}{\rm d\theta}\right)+\frac{a\mu\sin\theta}{\lambda+a\mu\cos\theta}\frac{\rm d}{\rm d\theta}+
\left(\frac{1}{2}+a\omega\cos\theta\right)^2-\left(\frac{m-\frac{1}{2}\cos\theta}{\sin\theta}\right)^2
-\frac{3}{4} \nonumber \\
&+2a\omega m-a^2\omega^2+\frac{a\mu(\frac{1}{2}\cos\theta+a\omega\sin^2\theta-m)}{\lambda+a\mu\cos\theta}-a^2\mu^2\cos^2\theta+\lambda^2
\Biggr]S^{(-)}(\theta)=0.
\end{align}
Using the variable $x:=\cos\theta$ the previous differential equation is written:
\begin{align}
&\Biggl\{(1-x^2)\frac{{\rm d}^2}{{\rm d}x^2}-2x\frac{\rm d}{{\rm d}x}-\frac{a\mu(1-x^2)}{\lambda+a\mu x}\frac{\rm d}{{\rm d}x}\nonumber \\
&+\frac{a\mu(\frac{x}{2}+a\omega (1-x^2)-m)}{\lambda+a\mu x}+a^2(\omega^2-\mu^2)x^2+a\omega x-\frac{1}{4} \nonumber \\
&+\lambda^2+2am\omega-a^2\omega^2+\frac{-m^2+mx-\frac{1}{4}}{1-x^2}\Biggr\}S^{-}=0.
\label{DiracMas}
\end{align}
Equation (\ref{DiracMas}) possess three finite singularities at the points $x=\pm 1, x=-\lambda/a\mu$ which we denote using the triple:$(a_1,a_2,a_3)=(-1,+1,-\lambda/a\mu)$. Applying the transformation of the independent variable:
\begin{equation}
z=\frac{x-a_1}{a_2-a_1}=\frac{x+1}{2},
\end{equation}
results in transforming $a_1,a_2$ into $0,1$ while the remaining singularity $a_3$ is transformed to $z=z_3$:
\begin{equation}
z_3=\frac{a_3-a_1}{a_2-a_1}=\frac{-\lambda/a\mu+1}{2}.
\end{equation}
In terms of the new variable $z$ the Fuchsian differential equation (\ref{DiracMas}) becomes:
\begin{align}
& \Biggl\{ \frac{\mathrm{d}^2}{\mathrm{d}z^2}+\left[\frac{1}{z}+\frac{1}{z-1}
-\frac{1}{z-z_3}\right]\frac{\mathrm{d}}{\mathrm{d}z}+4a^2(\mu^2-\omega^2)+\frac{a^2(\mu^2-\omega^2)}{z-1}-\frac{a^2(\mu^2-\omega^2)}{z} \nonumber \\
&+\frac{1}{16}\frac{-4m^2-4m-1}{z^2}+\frac{1}{8}\frac{4m^2+1}{z-1}+\frac{1}{16}\frac{-4m^2+4m-1}{(z-1)^2}
+\frac{1}{8}\frac{-4m^2-1}{z} \nonumber \\
&+\frac{1}{4}\frac{8a\omega z_3^2-8a\omega z_3+2m-2z_3+1}{z_3(z_3-1)(z-z_3)}+\frac{-2m+1}{(z-1)(-4+4z_3)}+\frac{1}{4}\frac{2m+1}{z_3 z}\nonumber \\
&+\frac{1}{4}\frac{4a^2\omega^2 -8am\omega-4a\omega-4\lambda^2+1}{z-1}+
\frac{1}{4}\frac{-4a^2\omega^2+8am\omega-4a\omega+4\lambda^2-1}{z}
\Biggr\}S(z)=0.
\end{align}
Let us calculate the exponents of the singularities. The indicial equation for the $z=0$ singularity is:
\begin{equation}
F(r)=r(r-1)+r-\frac{1}{4}\left(m+\frac{1}{2}\right)^2=0,
\end{equation}
with roots: $r_{1,2}^{z=0}=\pm\frac{1}{2}|m+1/2|$. Likewise the exponents at the singularities $z=1,z=z_3$ are computed to be: $\{\frac{|m-\frac{1}{2}|}{2},-\frac{|m-\frac{1}{2}|}{2}\}, \{0,2\}$ respectively. Applying the index transformation for the dependent variable $S$:
\begin{equation}
S(z)=z^{\alpha_1}(z-1)^{\alpha_2}(z-z_3)^{\alpha_3}\bar{S}(z),
\label{fhomotopyDIRAC}
\end{equation}
where $\alpha_1=\frac{1}{2}|m+1/2|,\alpha_2=\frac{1}{2}|m-1/2|$ and $\alpha_3=0$ yields the Heun equation:
\begin{equation}
\Biggl\{\frac{\mathrm{d}^2}{\mathrm{d}z^2}+\left[\frac{2\alpha_1+1}{z}+
\frac{2\alpha_2+1}{z-1}+\frac{-1}{z-z_3}\right]\frac{\mathrm{d}}{\mathrm{d}z}+
\frac{{\alpha}{\beta}z-q}{z(z-1)(z-z_3)}\Biggr\}\bar{S}(z)=0,
\label{HeunMunchenPDIRACmas}
\end{equation}
where for instance the auxiliary parameter $q$ is computed in terms of the physical parameters and the separation constants to be:
\begin{align}
&q=-z_3\Biggl\{-a^2(\mu^2-\omega^2)+\frac{1}{8}(-4m^2-1)+\frac{2m+1}{4z_3} \nonumber \\
&+\frac{-4a^2\omega^2+8ma\omega-4a\omega+4\lambda^2-1}{4} -\alpha_2-\frac{\alpha_3}{z_3}-\alpha_1+\frac{\alpha_1}{z_3}-2\alpha_1\alpha_2-\frac{2\alpha_1\alpha_3}{z_3}\Biggr\}
\end{align}
Equation (\ref{HeunMunchenPDIRACmas}) is a Heun equation with a singular point $z_3$ with exponents $(0,2)$. Thus if the condition (\ref{nolog}) for the absence of logarithmic terms in the local expansion around the singular point is satisfied it follows that the point $z_3$ is an apparent singularity and the theory of false singularity of section \ref{APPARENTSINGULARFUCHS} applies. Then the exact solution of (\ref{HeunMunchenPDIRACmas}) simplifies and it is expressed in terms of Gau$\ss$ hypergeometric function see appendix B, equation (\ref{GAUSSHEUNFEXACT}). Of course if the conditions for a false singular point are not all satisfied an independent solution will be given locally in terms of general Heun functions of the form in Eqn(\ref{TOPIKESHEUNSYNARTISEIS}).
A more detailed account of the exact solutions of the angular and radial Fuchsian wave equations for a massive spin half particle in rotating charged black hole backgrounds will be reported elsewhere \cite{ProfDrKRANIOTIS}.

\subsubsection{Conditions for expanding the confluence Heun function solution of the angular equation KN spacetime in terms of the Kummer confluent hypergeometric functions}
\label{KUMMERCONFLUENCE}

Following the work in \cite{ISHKHANYAN} we would like to obtain conditions on the parameters of the exact solution for the angular equation for a massive particle in KN spacetime in terms of the confluent Heun function $H_C(p,\alpha,\gamma,\delta,\sigma;z)$, obtained in the previous section, such that the solution of the confluent Heun equation (\ref{CONFLUENTHEUNC}), can be expanded in terms of the Kummer confluent hypergeometric functions $F(\alpha,\gamma,z):=\sum_{\nu=0}^{\infty}\frac{(\alpha)_{\nu}z^{\nu}}{(\gamma)_{\nu}\nu!}$. In other words we are interested in obtaining expansions of the form
\begin{equation}
w=\sum_{\mu}a_{\mu}w_{\mu},\;\;\; w_{\mu}=F(\alpha_{\mu},\gamma_{\mu},s_0 z)
\label{edwardKUMMERC}
\end{equation}
This will be useful also for the exact solution of the radial part of the Klein-Gordon-Fock equation for a massive scalar particle. As we shall see in the next section in that case the exact solution will be given by the confluent Heun function which satisfies a slightly more general differential equation than (\ref{CONFLUENTHEUNC}). Namely:
\begin{equation}
\fbox{$\displaystyle w^{\prime\prime}+\left(\frac{\gamma}{z}+\frac{\delta}{z-1}+\varepsilon\right)w^{\prime}
+\frac{\alpha z-q}{z(z-1)}w=0.
\label{raymassiveKGF}$}
\end{equation}
The difference is that the parameters $\varepsilon,\alpha$ are independent.
Thus the results of this section will also be useful for the exact solution of the radial equation for a massive scalar particle in the Kerr-Newman spacetime which is derived in section \ref{AKTINAKFGMASSIVE}.

The Kummer confluent hypergeometric function $F(\alpha,\gamma,t)$ satisfies the differential equation
\begin{equation}
t\frac{{\rm d}^2y}{{\rm d}t^2}+(\gamma-t)\frac{{\rm d}y}{{\rm d}t}-\alpha y=0
\end{equation}
which for a independent variable $t=s_0 z$ and dependent $y=w_{\mu}$ can be written
\begin{equation}
w^{\prime\prime}_{\mu}+\left(\frac{\gamma_{\mu}}{z}-s_0\right)w^{\prime}_{\mu}
-\frac{\alpha_{\mu}s_0}{z}w_{\mu}=0
\label{kummersyrreon}
\end{equation}
We shall investigate the case in which the parameters of the confluent Kummer hypergeometric function are given as follows \cite{ISHKHANYAN}:
\begin{equation}
\alpha_{\mu}=\alpha_0+\mu,\gamma_{\mu}=\gamma_0=constant,\;\;\mu\in\mathbb{Z},
\end{equation}
in other words the expansion Kummer functions are of the form $w_{\mu}=F(\alpha_0+\mu,\gamma_0,s_0z)$.
Now the expansion functions obey some recurrence relations:
Using the expansions:
\begin{align}
\alpha_{\mu}w_{\mu+1}&=\alpha_{\mu}\left[1+\frac{(\alpha_0+\mu+1)z s_0}{\gamma_0 1!}+\frac{(\alpha_0+\mu+1)(\alpha_0+\mu +2)z^2 s_0^2}{(\gamma_0+1)\gamma_0 2!}+\cdots\right] \nonumber \\
(\alpha_{\mu}-\gamma_0)w_{\mu-1}&=\alpha_{\mu}\left[1+\frac{\alpha_0+\mu-1}{\gamma_0}zs_0+
\frac{(\alpha_0+\mu-1)(\alpha_0+\mu)z^2s_0^2}{\gamma_0(\gamma_0+1)2!}+\cdots\right]
\nonumber \\
(\gamma_0-2\alpha_{\mu})w_{\mu}&=\left[1+\frac{(\alpha_0+\mu)z s_0}{\gamma_0 1!}+\frac{(\alpha_0+\mu)(\alpha_0+\mu+1)z^2 s_0^2}{\gamma_0 (\gamma_0+1)2!}+\cdots\right]
\end{align}
one obtains the recurrence relations
\begin{align}
\fbox{$\displaystyle s_0 z w_{\mu}=(\alpha_{\mu}-\gamma_0)w_{\mu-1}+
(\gamma_0-2\alpha_{\mu})w_{\mu}+\alpha_{\mu}w_{\mu+1},$}
\label{anadrom1}
\end{align}
\begin{equation}
\fbox{$\displaystyle z w_{\mu}^{\prime}=\alpha_{\mu}(w_{\mu+1}-w_{\mu})
$}
\label{anadrom2two}
\end{equation}

Combining the last two equations one obtains:
\begin{align}
s_0 z^2 w_{\mu}^{\prime}&=\alpha_{\mu}[(\alpha_{\mu}+1)w_{\mu+2}
+w_{\mu+1}(\gamma_0-3\alpha_{\mu}-2) \nonumber \\
&+ w_{\mu}(3\alpha_{\mu}+1-2\gamma_0)+(\gamma_0-\alpha_{\mu})w_{\mu-1}].
\label{paragorecurrence}
\end{align}
Substituting the expansions (\ref{edwardKUMMERC}) into the confluence Heun differential equation (\ref{raymassiveKGF}) we obtain
\begin{align}
&\sum_{\mu}a_{\mu}\{[z^2(\varepsilon+s_0)+z(-\varepsilon-s_0+\gamma+\delta-\gamma_0)+\gamma_0-\gamma]w_{\mu}^{\prime}
\nonumber \\
&+[(\alpha_{\mu}s_0+\alpha)z-(q+\alpha_{\mu}s_0)]w_{\mu}\}=0.
\label{recuKFGKNADS}
\end{align}
Since $w^{\prime}_{\mu}$ is not expressed as a linear combination of the functions $w_{\mu}$ it is demanded that $\gamma_0-\gamma=0$ \cite{ISHKHANYAN}.
Substituting Eqs.(\ref{anadrom1})-(\ref{paragorecurrence}) into Eq.(\ref{recuKFGKNADS}) we obtain:
\begin{align}
& \sum_{\mu}a_{\mu}\{z\varepsilon\alpha_{\mu}(w_{\mu+1}-w_{\mu})+\alpha_{\mu}(\alpha_{\mu}+1)
w_{\mu+2}\nonumber \\
&-\alpha_{\mu}(2\alpha_{\mu}+2-\gamma_0)w_{\mu+1}+\alpha_{\mu}(\alpha_{\mu}-\gamma_0+1)w_{\mu}\nonumber \\
& +(\delta-\epsilon)\alpha_{\mu}(w_{\mu+1}-w_{\mu})-s_0\alpha_{\mu}(w_{\mu+1}-w_{\mu})+\alpha z w_{\mu}-q w_{\mu}-\alpha_{\mu}s_0w_{\mu}\}=0
\end{align}
where the coefficients are calculated to be
\begin{align}
{\rm coefficient\; of\; } w_{\mu+2}:&\;S_{\mu}:=\frac{(s_0+\varepsilon)\alpha_{\mu}(\alpha_{\mu}+1)}{s_0}  \\
{\rm coefficient\; of\; } w_{\mu+1}:&\;P_{\mu}:=\frac{\alpha_{\mu}}{s_0}\left[-(\varepsilon+s_0)(-\gamma+2\alpha_{\mu}+2-\delta-\varepsilon)-(\varepsilon+s_0)^2+\alpha-\varepsilon(\alpha_{\mu}+\delta)\right]
 \\
{\rm coefficient\; of\; } w_{\mu}:&\;Q_{\mu}:=
\frac{(\gamma-2\alpha_{\mu})(\alpha-\varepsilon\alpha_{\mu})+\alpha_{\mu}(\alpha_{\mu}-\gamma+1)(\varepsilon+s_0)+s_0(\alpha_{\mu}(\varepsilon-\delta)-q)}{s_0}
\\
{\rm coefficient\; of\; } w_{\mu-1}:&\;R_{\mu}=\frac{(\alpha_{\mu}-\gamma)(\alpha-\varepsilon\alpha_{\mu})}{s_0}
\end{align}
Thus we end up with a four-term recurrence relation for the coefficients $a_{\mu}$:
\begin{equation}
R_{\nu}a_{\nu}+Q_{\nu-1}a_{\nu-1}+P_{\nu-2}a_{\nu-2}+S_{\nu-3}a_{\nu-3}=0.
\end{equation}

If we set $s_0=-\varepsilon$-removing in this way the $z^2$ dependence in the coefficient of $w^{\prime}_{\mu}$ in Eq.(\ref{recuKFGKNADS})-then the four-term recurrence relation becomes a three-term:
\begin{equation}
R_{\nu}a_{\nu}+Q_{\nu-1}a_{\nu-1}+P_{\nu-2}a_{\nu-2}=0,
\end{equation}
where
\begin{align}
R_{\nu}&=(\alpha_{\nu}-\gamma)(\alpha_{\nu}-\frac{\alpha}{\varepsilon}), \\
Q_{\nu}&=(\alpha_{\nu}-\frac{\alpha}{\varepsilon})(\gamma-2\alpha_{\nu})+
\alpha_{\nu}(\varepsilon-\delta)-q, \\
P_{\nu}&=\alpha_{\nu}[(\alpha_{\nu}+\delta)-\frac{\alpha}{\varepsilon}].
\end{align}
The initial conditions for left-hand side termination of the derived series at
$\nu=0$ are $a_{-2}=a_{-1}=0$. As a result $R_0=0$. This is possible if
$\alpha_0=\gamma$ or $\alpha_0=\frac{\alpha}{\varepsilon}$. Then the final expression is written explicitly as follows:
\begin{equation}
\fbox{$\displaystyle w=\sum_{\nu=0}^{\infty}a_{\nu}F(\alpha_0+\nu,\gamma,-\varepsilon z)
$}
\label{expanKummer}
\end{equation}
and the coefficients of the recurrence take the form
\begin{align}
R_{\mu}&=(\alpha_0+\mu-\gamma)(\alpha_0+\mu-\alpha/\varepsilon) \\
Q_{\mu}&=(\alpha_0+\mu-\alpha/\varepsilon)(\gamma-2\alpha_0-2\mu)+(\alpha_0+\mu)(\varepsilon-\delta)-q \\
P_{\mu}&=(\alpha_0+\mu)((\alpha_0+\mu+\delta)-\alpha/\varepsilon)
\end{align}
The expansion is valid if $\varepsilon \not =0$ and $\gamma \not\in \mathbb{Z}^{-}$.
The expansion (\ref{expanKummer}) is right-hand terminated for some $\nu=N$ if
$a_N\not =0$ and $a_{N+1}=a_{N+2}=0.$ Then, it should be $P_N=0$. If $\alpha_0=\alpha/\varepsilon$, the condition $P_N=0$ is satisfied if
\begin{equation}
\alpha_0=-N\Rightarrow \alpha/\varepsilon=-N,\; {\rm or} \;\;\delta=-N
\end{equation}
If $\alpha_0=\gamma$, the only possibility since $\gamma\in\mathbb{Z}^{+}$, is:
\begin{equation}
\gamma+\delta+\left(-\frac{\alpha}{\varepsilon}\right)=-N.
\end{equation}
For each of these cases, there are $N+1$ values of $q$ for which the termination occurs. Indeed, for $N=1$ we get the relations
\begin{equation}
R_1a_1+Q_0a_0=0\Rightarrow a_1=\frac{-Q_0a_0}{R_1},
\end{equation}
and $Q_0=\alpha_0(\varepsilon-\delta)-q, \;R_1=(1+\alpha_0-\gamma)(1+\alpha_0-\alpha/\varepsilon)=1+\alpha_0-\gamma$. In this case these values are determined from the condition
\begin{equation}
Q_1a_1+P_0a_0=0, \;\;(a_2=0),
\end{equation}
where $P_0=\alpha_0\delta,Q_1=(\alpha_1-\frac{\alpha}{\varepsilon})(\gamma-2\alpha_1)+
\alpha_1(\varepsilon-\delta)-q$, consequently
\begin{equation}
\left[(\alpha_1-\frac{\alpha}{\varepsilon})(\gamma-2\alpha_1)+
\alpha_1(\varepsilon-\delta)-q\right]\left[-\frac{(\alpha_0(\varepsilon-\delta)-q)a_0}{1+\alpha_0-\gamma}\right]+
\alpha_0\delta a_0=0
\end{equation}
while for $N=2$ these values are determined from the condition:
\begin{equation}
Q_2a_2+P_1a_1=0,\;\;(a_3=0)
\end{equation}
while from the equation $R_2a_2+Q_1a_1+P_0a_0=0$ we solve for $a_2$
\begin{align}
&R_2a_2+Q_1a_1+P_0a_0=0\Rightarrow a_2=\frac{-Q_1a_1-P_0a_0}{R_2} \nonumber \\
&=-\frac{\left[(\alpha_1-\frac{\alpha}{\varepsilon})(\gamma-2\alpha_1)+
\alpha_1(\varepsilon-\delta)-q\right]\left[-\frac{(\alpha_0(\varepsilon-\delta)-q)a_0}
{1+\alpha_0-\gamma}\right]-P_0a_0}{R_2}
\end{align}
and
\begin{equation}
Q_2=2(\gamma-2\alpha_0-4)+(\alpha_0+2)(\varepsilon-\delta)-q
\end{equation}
\subsection{Exact solution of the radial equation for a massive scalar particle in the Kerr-Newman spacetime \label{AKTINAKFGMASSIVE}}
\label{aktinamazikoubatmotoul0}
\subsubsection{case I: Neutral massive scalar particle }

In this subsection we will derive the exact solution of the radial part of the KGF equation for a massive particle in the Kerr-Newman spacetime.
The radial equation in this case is given by:
\begin{align}
& \Delta^{KN}\frac{{\rm d}}{{\rm d}r}\left(\Delta^{KN}\frac{{\rm d}R}{{\rm d}r}\right)+\Biggl[ \omega^2(r^2+a^2)^2-4Ma\omega m r+2e^2 a\omega m-\mu^2 r^2 \Delta^{KN} \nonumber \\
&+ m^2 a^2-(\omega^2 a^2+\mathcal{K}_{lm})\Delta^{KN}-2eqr[(r^2+a^2)\omega-am]+e^2q^2r^2\Biggr]R=0
\label{RKNKGFM}
\end{align}
where $\Delta^{KN}$ is obtained by setting $\Lambda=0$ in equation (\ref{DiscrimiL}).
Assume initially an electrically neutral particle ($q=0$). Introducing a new independent variable $x$ through:
\begin{equation}
M\chi=r-r_{+},\;\;\;r_{\pm}=M\pm Md,
\end{equation}
the radial equation takes the form:

\begin{align}
&\frac{{\rm d}}{{\rm d}\chi}\left[\chi(\chi+2d)\frac{{\rm d}R}{{\rm d}\chi}\right]+
\Biggl[\frac{\omega^2}{M^2 \chi(\chi+2d)}\{M^2[(\chi+d+1)^2-(d^2-1)]-e^2\}^2 +\frac{2e^2a\omega m}{M^2 \chi(\chi+2d)} \nonumber \\
&-\frac{4a\omega m(\chi+d+1)}{\chi(\chi+2d)}
-\mu^2 M^2 (\chi+d+1)^2+\frac{m^2 a^2}{M^2 \chi(\chi+2d)}-(\omega^2 a^2+\mathcal{K}_{lm})\Biggr]R=0,
\end{align}
where the following relations are also valid:
\begin{align}
\Delta^{KN}=M^2 \chi(\chi+2d),\;\Delta^{KN}+2Mr=M^2(\chi+d+1)^2-M^2(d^2-1)
\end{align}
Using the change of variables
\begin{equation}
R(\chi)=Z(\chi)(\chi(\chi+2d))^{-1/2},
\end{equation}
the radial equation acquires the form:
\begin{align}
&\frac{{\rm d}^2 Z}{{\rm d}\chi^2}+Z\Biggl\{ (\omega^2-\mu^2)M^2
+\frac{1}{M^2 \chi^2(\chi+2d)^2}((\omega^2[M^4 4(\chi+d+1)^2+4M^4(\chi+d+1)\chi(\chi+2d)\nonumber \\
&-2e^2M^2[\chi(\chi+2d)+2(\chi+d+1)]+e^4] \nonumber \\
&-4a\omega m M^2(\chi+d+1)+2e^2a\omega m-\mu^2 M^4[2\chi+(d+1)^2]\chi(\chi+2d) \nonumber \\
&+m^2 a^2-(\omega^2a^2+\mathcal{K}_{lm})M^2 \chi(\chi+2d)+d^2M^2))
\Biggr\}=0.
\end{align}
Using the partial fractions technique the previous differential equation is written:
\begin{equation}
\frac{{\rm d}^2Z}{{\rm d}\chi^2}+\Biggl[M^2(\omega^2-\mu^2)+\frac{1}{M^2}\Biggl\{\frac{A}{\chi^2}+\frac{B}{\chi}
+\frac{C}{(\chi+2d)^2}+\frac{D}{\chi+2d}\Biggr\}\Biggr]Z=0,
\label{INITIARADIAL}
\end{equation}
where the coefficients are calculated to be:
\begin{align}
A&=\frac{d^2 M^2+(am+(-2(1+d)M^2+e^2)\omega)^2}{4d^2},\\
B&=\frac{1}{4d^3}(-a^2m^2+d^2M^2(-1-2\mathcal{K}_{lm}-2(1+d)^2M^2\mu^2)+2 am(2M^2-e^2) \omega \nonumber \\
&-(2a^2 d^2 M^2-4(1+d)^2(-1+2d)M^4+4(-1+d^2)M^2e^2+e^4)\omega^2)\\
C&=\frac{d^2M^2+(a m+(2(-1+d)M^2+e^2)\omega)^2}{4d^2} \\
D&=\frac{1}{4d^3}(d^2M^2(1+2\mathcal{K}_{lm}+2(-1+d)^2 M^2 \mu^2)+2 a m(-2M^2+e^2)\omega \nonumber \\
&+(4(-1+d)^2(1+2d)M^4+4(-1+d^2)M^2 e^2+e^4)\omega^2+a^2 (m^2+2d^2 M^2\omega^2))
\end{align}
Using a change in the independent variable
\begin{equation}
\zeta=-\frac{\chi}{2d},
\label{fundachangevariablez}
\end{equation}
equation (\ref{INITIARADIAL}) reduces to a \textit{normal form} of the confluent Heun equation \cite{RONVEAUX}:
\begin{equation}
\frac{{\rm d}^2 Z}{{\rm d}\zeta^2}+\left[4d^2M^2(\omega^2-\mu^2)+
\frac{1}{M^2}\left(\frac{A}{\zeta^2}+\frac{-2d B}{\zeta}+\frac{C}{(\zeta-1)^2}+\frac{-2d D}{\zeta-1}\right)\right]Z=0
\end{equation}

We have arrived in the equation:
\begin{align}
&\frac{{\rm d}^2w}{{\rm d}\zeta^2}+\frac{{\rm d}w}{{\rm d}\zeta}\left[\frac{1}{\zeta}+\frac{1}{\zeta-1}\right]+w(\zeta)\Biggl\{\left(\frac{A}{M^2}-\frac{1}{4}\right)\frac{1}{\zeta^2}+
\left(\frac{C}{M^2}-\frac{1}{4}\right)\frac{1}{(\zeta-1)^2} \nonumber \\
&+\left(\frac{-2dB}{M^2}-\frac{1}{2}\right)\frac{1}{\zeta}+\left(\frac{-2dD}{M^2}+\frac{1}{2}\right)\frac{1}{\zeta-1}+
4d^2M^2(\omega^2-\mu^2)\Biggr\}=0
\end{align}
The indicial equation for the exponentials of the singular points at $\zeta=0$ and $\zeta=1$ is
\begin{equation}
r(r-1)+r+B^{\prime}_i=0, i=1,2
\end{equation}
with the roots $\mu^{(1,2)}_{i}=\pm i\sqrt{B^{\prime}_i}$ or
\begin{align}
2\mu_1^{(1,2)}&=\pm 2 i\sqrt{B^{\prime}_1}=\pm \frac{i}{M}\sqrt{4A-M^2}, \\
2\mu_2^{(1,2)}&=\pm 2 i\sqrt{B^{\prime}_2}=\pm \frac{i}{M} \sqrt{4C-M^2}
\end{align}
We now apply the homotopic transformation of the dependent variable
\begin{align}
w(\zeta)&=e^{\nu \zeta}\prod_{i=1}^{2}(\zeta-\zeta_i)^{\mu_i}Y(\zeta) =e^{\pm 2i \zeta d M\sqrt{\omega^2-\mu^2}}\zeta^{\mu_1}(\zeta-1)^{\mu_2}Y(\zeta) \nonumber \\
&=e^{\pm 2i \zeta d M\sqrt{\omega^2-\mu^2}}\zeta^{\frac{\pm i}{2M}\sqrt{4A-M^2}}(\zeta-1)^{\frac{\pm i}{2M}\sqrt{4C-M^2}}Y(\zeta)
\end{align}
where  $\nu_{\pm}:=\pm 2idM\sqrt{\omega^2-\mu^2}$,
which yields the confluent Heun equation
\begin{equation}
Y^{\prime\prime}(\zeta)+\left(\alpha+\frac{\gamma}{\zeta}+\frac{\delta}{\zeta-1}\right)Y^{\prime}(\zeta)
+\frac{w\zeta-\sigma}{\zeta(\zeta-1)}Y(\zeta)=0
\label{exactnessKGF}
\end{equation}
The parameters of the confluent equation are given by
\begin{align}
\alpha_{\pm}&=2\nu_{\pm}=\pm 4idM\sqrt{\omega^2-\mu^2}, \gamma_{\pm}=1\pm\frac{i}{M}\sqrt{4A-M^2},\delta_{\pm}=1\pm \frac{i}{M}\sqrt{4C-M^2},\label{alphagammadeltaexactKFG}\\
\sigma_{\pm}&=\left(\frac{-2dB}{M^2}-\frac{1}{2}\right)+\frac{1}{2}\pm\frac{4idM\sqrt{\omega^2-\mu^2}}{2}
\left(1\pm\frac{i}{M}\sqrt{4A-M^2}\right) \nonumber \\
&-\frac{1}{2}\left(1\pm\frac{i}{M}\sqrt{4A-M^2}\right)\left(1\pm\frac{i}{M}\sqrt{4C-M^2}\right) \label{sigmaexactKFG} \\
w_{\pm}&=\frac{-2d}{M^2}(B+D)\pm4idM\sqrt{\omega^2-\mu^2}\pm\frac{4idM\sqrt{\omega^2-\mu^2}}{2}\left[
\pm\frac{i}{M}\sqrt{4A-M^2}\pm\frac{i}{M}\sqrt{4C-M^2}\right]
\label{wexactKFG}
\end{align}
Summarising an exact solution of the radial KGF equation for  a massive neutral particle in the KN black hole spacetime is the following:
\begin{equation}
\fbox{$
R(\zeta)=\frac{M}{\sqrt{\Delta^{KN}}}e^{-2idM\sqrt{\omega^2-\mu^2}\zeta}\zeta^{\frac{1}{2}-\frac{i}{2M}\sqrt{4A-M^2}}
(\zeta-1)^{\frac{1}{2}-\frac{i}{2M}\sqrt{4C-M^2}} H_c(\alpha_{-},w_{-},\gamma_{-},\delta_{-},\sigma_{-},\zeta)$}.
\label{GVKKGFEXACTKNA}
\end{equation}
The parameters of the confluent Heun function $H_c(\alpha_{-},w_{-},\gamma_{-},\delta_{-},\sigma_{-},\zeta)$ are given in
(\ref{alphagammadeltaexactKFG})-(\ref{wexactKFG}), while $\Delta^{KN}$ denotes the radial quartic polynomial $\Delta_r^{KN}$ for $\Lambda=0$.
If we want to apply the theory of the previous section in our exact solution, as expressed in equations (\ref{exactnessKGF}) and (\ref{GVKKGFEXACTKNA}), and wish to constrain the parameters of the theory so that the solution can be simplified and written in terms of the confluent Kummer hypergeometric functions we derive that the series expansion of confluent Heun functions  in terms of $F(\alpha_0+\mu,\gamma_0,s_0\zeta)$ is right hand terminated if
\begin{equation}
\delta_{\pm}=-N \;{\rm or}\; \frac{w_{\pm}}{\alpha_{\pm}}=-N
\end{equation}
Also if $\alpha_0=\gamma$ the series is right hand terminated if
\begin{equation}
\gamma_{\pm}+\delta_{\pm}+\left(-\frac{w_{\pm}}{\alpha_{\pm}}\right)=-N
\end{equation}

\subsubsection{Asymptotic solutions at infinity}
Using expansions for large $\chi, \chi\rightarrow \infty$ we obtain:
\begin{eqnarray}
&\frac{C}{(\chi+2d)^2}=C\left[\frac{1}{\chi^2}-\frac{4d}{\chi^3}+\cdots\right], \\
&\frac{D}{\chi+2d}=D\left[\frac{1}{\chi}-\frac{2d}{\chi^2}+\cdots\right]
\\
&\frac{{\rm d}^2Z}{{\rm d}\chi^2}=\left[M^2(\mu^2-\omega^2)-\frac{1}{M^2}\left(\frac{A+C-2dD}{\chi^2}+\frac{B+D}{\chi}\right)+\mathcal{O}\left(\frac{1}{\chi^3}\right)\right]Z
\label{farouthorizon}
\end{eqnarray}
Introducing the variable $\xi=2M(\mu^2-\omega^2)^{1/2}\chi$ in the large $\chi$ limit the radial differential equation reduces to the Whittaker differential equation \cite{EdmundWHITTAKER}:
\begin{equation}
\frac{{\rm d}^2Z}{{\rm d}\xi^2}=\left(\frac{1}{4}-\frac{k}{\xi}+\frac{m^2-\frac{1}{4}}{\xi^2}\right)Z,
\label{WhittakerE}
\end{equation}
with the parameters of the Whittaker equation determined in terms of the physical and separation parameters as follows:
\begin{equation}
k=\frac{B+D}{2M^3\sqrt{\mu^2-\omega^2}},\;\;\;\frac{1}{4}-m^2=\frac{1}{M^2}[A+C-2dD].
\end{equation}
Standard solutions of (\ref{WhittakerE}) are
\begin{eqnarray}
M_{k,m}(\xi)=e^{-\xi/2}\xi^{m+1/2}M(m-k+\frac{1}{2},2m+1,\xi),\\
W_{k,m}(\xi)=e^{-\xi/2}\xi^{m+1/2}U(m-k+\frac{1}{2},2m+1,\xi).
\end{eqnarray}
The large $\chi$ limit is characterized by:
\begin{equation}
W_{k,m}(\xi)\sim e^{-\xi/2}\xi^{k}\;\;\; (\xi\rightarrow\infty,|{\rm Arg}(\xi)|\leq\frac{3}{2}\pi-\delta_1)
\end{equation}
Thus as $\chi\rightarrow \infty$ so $r\rightarrow \infty$ and asymptotically the solutions of the radial KGF equation are
\begin{eqnarray}
&R(r)\sim\frac{M}{\sqrt{\Delta^{KN}}}e^{-\xi/2}\xi^{k}\Rightarrow \nonumber \\
&R(r)\sim\frac{M}{\sqrt{\Delta^{KN}}}e^{-\sqrt{\mu^2-\omega^2}(r-r_{+})}[2(\mu^2-\omega^2)^{1/2}(r-r_{+})]^{\frac{B+D}{2M^3 \sqrt{\mu^2-\omega^2}}}.
\end{eqnarray}
\subsubsection{The limit $\chi\rightarrow 0\equiv r\rightarrow r_{+}$}
In this case expanding the $C$ and $D$ terms in (\ref{INITIARADIAL}) for small $\chi$ and neglecting terms of $\mathcal{O}(\chi)$ one obtains the Whittaker differential equation:
\begin{equation}
\frac{{\rm d}^2Z}{{\rm d}\eta^2}=\left(\frac{1}{4}-\frac{k_h}{\xi}+\frac{m_h^2-\frac{1}{4}}{\eta^2}\right)Z,
\label{WhittakerE2}
\end{equation}
with
\begin{equation}
k_h=\frac{B}{2M^2\underset{=:\sqrt{\mathcal{F}}}{\underbrace{\sqrt{M^2(\mu^2-\omega^2)-\frac{C}{4M^2d^2}-\frac{D}{2M^2 d}}}}},\;
m^2_h=\frac{1}{4}-\frac{A}{M^2}
\end{equation}
In this case the first solution of (\ref{WhittakerE2}) is written in terms of the confluent Kummer hypergeometric function
\begin{equation}
M_{k_h,m_h}(\eta)=e^{-\frac{1}{2}\eta}\eta^{\frac{1}{2}+m_h}F(m_h+\frac{1}{2}-k_h,2m_h+1,\eta)
\end{equation}
where $\eta=2\sqrt{\mathcal{F}}\chi$.
Near the event horizon limit
\begin{equation}
M_{k_h,m_h}(\eta)\sim \eta^{\frac{1}{2}+m_h}, \eta\rightarrow 0
\end{equation}
or in terms of the original variables and expanding the Kummer hypergeometric function \cite{ROWANSTEPHEN}
\begin{equation}
R(r)\sim \frac{M}{\sqrt{\Delta^{KN}}}e^{-\sqrt{\mathcal{F}}\frac{(r-r_{+})}{M}}
\left(2\sqrt{\mathcal{F}}\frac{(r-r_{+})}{M}\right)^{\frac{1}{2}+m_h},\;\;r\rightarrow r_{+}
\end{equation}

\subsubsection{Case II: Charged massive scalar particle}

In this subsection we derive the exact solution of Eqn.(\ref{RKNKGFM}) in the Kerr-Newman spacetime for $q\not =0$, i.e for a charged massive particle. Following similar steps with previous sections we arrive at the equation:
\begin{align}
&\frac{{\rm d}^2 Z}{{\rm d}\chi^2}+Z\Biggl\{ (\omega^2-\mu^2)M^2
+\frac{1}{M^2 \chi^2(\chi+2d)^2}((\omega^2[M^4 4(\chi+d+1)^2+4M^4(\chi+d+1)\chi(\chi+2d)\nonumber \\
&-2e^2M^2[\chi(\chi+2d)+2(\chi+d+1)]+e^4] \nonumber \\
&-4a\omega m M^2(\chi+d+1)+2e^2a\omega m-\mu^2 M^4[2\chi+(d+1)^2]\chi(\chi+2d) \nonumber \\
&+m^2 a^2-(\omega^2a^2+\mathcal{K}_{lm})M^2 \chi(\chi+2d)+d^2M^2 \nonumber \\
&-2eqM^3\chi(\chi+2d)(\chi+d+1)\omega-4eqM^3 (\chi+d+1)^2 \omega+2e^3qM(\chi+d+1)\omega+ \nonumber \\
&+ 2eqM(\chi+d+1)am+e^2q^2M^2[\chi(\chi+2d)+2\chi+(d+1)^2])
\Biggr\}=0.
\end{align}
Using the partial fractions technique the previous differential equation is written:
\begin{equation}
\frac{{\rm d}^2Z}{{\rm d}\chi^2}+\Biggl[M^2(\omega^2-\mu^2)+\frac{1}{M^2}\Biggl\{\frac{A^{\prime}}{\chi^2}+\frac{B^{\prime}}{\chi}
+\frac{C^{\prime}}{(\chi+2d)^2}+\frac{D^{\prime}}{\chi+2d}\Biggr\}\Biggr]Z=0,
\label{CHARGEDINITIARADIAL}
\end{equation}

\begin{align}
A^{\prime}&=A-\frac{1}{4d^2}\left(-e^2q^2M^2(1+d)^2+4eM^3q\omega(1+d)^2-2e^3qM\omega(d+1)\right)\nonumber \\
B^{\prime}&=B-\frac{1}{4d^3}\left(2aemMq+e^2M^2q^2(1-d^2)+2d^2M^4\mu^2(1+d)^2+4eM^3q\omega(d^3+2d^2-1)+2e^3qM\omega\right)
\nonumber \\
C^{\prime}&=C-\frac{1}{4d^2}\left(2aeqmM(d-1)-e^2q^2M^2(1-d)^2+4eqM^3\omega(d-1)^2+2e^3qM\omega(d-1)\right) \nonumber \\
D^{\prime}&=D-\frac{1}{4d^3}(-2aeqmM-e^2q^2M^2(1-d^2)-2d^2M^4\mu^2(d-1)^2+4eq\omega M^3(1-2d^2+d^3)-2e^3qM\omega)
\label{chargedcoeffi}
\end{align}
Following similar steps as in the previous section the exact solution of the radial part of the KGF differential equation for a massive charged particle in the KN black hole spacetime will involve the confluent Heun function.
A particular exact solution is
\begin{equation}
\fbox{$
R(\zeta)=\frac{M}{\sqrt{\Delta^{KN}}}e^{-2idM\sqrt{\omega^2-\mu^2}\zeta}\zeta^{\frac{1}{2}-\frac{i}{2M}\sqrt{4A^{\prime}-M^2}}
(\zeta-1)^{\frac{1}{2}-\frac{i}{2M}\sqrt{4C^{\prime}-M^2}} H_c(\alpha^{\prime}_{-},w^{\prime}_{-},\gamma^{\prime}_{-},\delta^{\prime}_{-},\sigma^{\prime}_{-},\zeta).$}
\label{GVKKGFEXACTKNCr}
\end{equation}
where
\begin{align}
\alpha^{\prime}_{\pm}&=\pm4idM\sqrt{\omega^2-\mu^2},  \;\;\gamma^{\prime}_{\pm}=1\pm\frac{i}{M}\sqrt{4A^{\prime}-M^2}, \;\;\delta^{\prime}_{\pm}=1\pm\frac{i}{M}\sqrt{4C^{\prime}-M^2},
\label{alphagammadeltaexactKFG13C}\\
\sigma^{\prime}_{\pm}&=\left(\frac{-2dB^{\prime}}{M^2}-\frac{1}{2}\right)+\frac{1}{2}+\frac{\pm 4idM\sqrt{\omega^2-\mu^2}}{2}
\left(1\pm\frac{i}{M}\sqrt{4A^{\prime}-M^2}\right) \nonumber \\
&-\frac{1}{2}\left(1\pm\frac{i}{M}\sqrt{4A^{\prime}-M^2}\right)\left(1\pm\frac{i}{M}\sqrt{4C^{\prime}-M^2}\right) \label{sigmaexactKFG13C} \\
w^{\prime}_{\pm}&=\frac{-2d}{M^2}(B^{\prime}+D^{\prime})\pm 4idM\sqrt{\omega^2-\mu^2}\pm\frac{4idM\sqrt{\omega^2-\mu^2}}{2}\left[
\pm\frac{i}{M}\sqrt{4A^{\prime}-M^2}\pm\frac{i}{M}\sqrt{4C^{\prime}-M^2}\right],
\label{wexactKFGRADCHARHE}
\end{align}
the variable $\zeta$ is given in (\ref{fundachangevariablez}) and  we can write the exact solution also in terms of the confluent Heun  function, ${\rm HeunC}(\alpha_{\rm M},\beta_{\rm M},\gamma_{\rm M},\delta_{\rm M},\eta_{\rm M},\zeta)$, defined in Maple. The correspondence among the parameters of the two functions $H_c$ and HeunC is:
\begin{align}
&Hc(\alpha^{\prime}_{-},w^{\prime}_{-},\gamma^{\prime}_{-},\delta^{\prime}_{-},\sigma^{\prime}_{-},\zeta)\rightarrow{\rm HeunC}\left( \alpha_{\rm M},\beta_{\rm M},\gamma_{\rm M},\delta_{\rm M},
\eta_{\rm M},\zeta\right)\nonumber \\
&\alpha_{-}^{\prime}=\alpha_{\rm M}, \gamma^{\prime}_{-}=1+\beta_{\rm M},\;\delta^{\prime}_{-}=1+\gamma_{\rm M},\;\delta_{\rm M}=-\frac{2d}{M^2}
(B^{\prime}+D^{\prime}),\;\eta_{\rm M}=\frac{1}{2}+\frac{2dB^{\prime}}{M^2}\nonumber \\
&\sigma^{\prime}_{-}=-\eta_{\rm M}+\frac{1}{2}+\frac{\alpha_{\rm M}}{2}(1+\beta_{\rm M})-\frac{1}{2}(1+\beta_{\rm M})(1+\gamma_{\rm M})\nonumber \\
&w^{\prime}_{-}=\delta_{\rm M}+\alpha_{\rm M}+\frac{\alpha_{\rm M}}{2}(\beta_{\rm M}+\gamma_{\rm M})
\label{surreousaHEUNC}
\end{align}
Using the Maple implemented Heun function, HeunC, a general solution for the radial equation of a charged massive particle in KN spacetime over the range $0\leq\zeta<\infty$ can be written:
\begin{align}
R(\zeta)&=\frac{M}{\sqrt{\Delta^{KN}}}e^{\frac{1}{2}\alpha_{\rm M}\zeta}\zeta^{\frac{1}{2}(1+\beta_{\rm M})}(\zeta-1)^{\frac{1}{2}(1+\gamma_{\rm M})}\{c_1 {\rm HeunC}(\alpha_{\rm M},\beta_{\rm M},\gamma_{\rm M},\delta_{\rm M},\eta_{\rm M},\zeta)\nonumber \\
&+c_2\zeta ^{-\beta_{\rm M}}{\rm HeunC}(\alpha_{\rm M},-\beta_{\rm M},\gamma_{\rm M},\delta_{\rm M},\eta_{\rm M},\zeta)\}
\label{FORTIORADEXACTHCG}
\end{align}
with $c_1,c_2$ constants. For zero electric charge (i.e. $q=0$) of the massive particle, (\ref{FORTIORADEXACTHCG}) reduces correctly to the result in \cite{BEZERRA}.

Constraining the parameters of the theory so that the solution when expanded in terms of the confluent Kummer hypergeometric functions is right hand terminated we derive the conditions:
\begin{align}
\delta^{\prime}_{\pm}&=1\pm\frac{i}{M}\sqrt{4C^{\prime}-M^2}=-N\; {\rm or} \\
\frac{w^{\prime}_{\pm}}{\alpha^{\prime}_{\pm}}&=\frac{\frac{-2d}{M^2}(B^{\prime}+D^{\prime})\pm 4idM\sqrt{\omega^2-\mu^2}\pm\frac{4idM\sqrt{\omega^2-\mu^2}}{2}\left[
\pm\frac{i}{M}\sqrt{4A^{\prime}-M^2}\pm\frac{i}{M}\sqrt{4C^{\prime}-M^2}\right]}{\pm 4idM\sqrt{\omega^2-\mu^2}}=-N
\end{align}
Also if $\alpha_0=\gamma^{\prime}$ the series is right hand terminated if
\begin{equation}
\gamma^{\prime}_{\pm}+\delta^{\prime}_{\pm}+\left(-\frac{w^{\prime}_{\pm}}{\alpha^{\prime}_{\pm}}\right)=-N
\end{equation}

\subsubsection{Asymptotic solutions at infinity-$r\rightarrow\infty$}
\label{farmakriaeventhorizon}
Using similar expansions as in (\ref{farouthorizon}), for large $\chi$, we bring the radial equation (\ref{CHARGEDINITIARADIAL}) into the Whittaker's form. The solutions then far from the event horizon will involve the Kummer and Tricomi confluent hypergeometric functions with parameters modified by the electric charge of the scalar particle.
Using the important asymptotic series of the Tricomi function ($b=1+a-c$) \cite{OLVER}:
\begin{equation}
U(a,c,\xi)\sim \xi^{-a}\left[1-\frac{ab}{\xi}+\frac{a(a+1)b(b+1)}{2!\xi^2}-\cdots\right]=
\frac{1}{\xi^a}\sum_{\nu=0}\frac{(a)_{\nu}(b)_{\nu}}{(1)_{\nu}}
\left(\frac{-1}{\xi}\right)^{\nu}\;\;\xi\; {\rm large}
\end{equation}
yields
\begin{equation}
W_{k,m}(\xi)\sim e^{-\xi/2}\xi^{k}\left[1
+\frac{m^2-\left(k-\frac{1}{2}\right)^2}{\xi}+\frac{(m^2-\left(k-\frac{1}{2}\right)^2)(m^2-\left(k-\frac{3}{2}\right)^2)}{2! \xi^2}+\cdots \right]
\end{equation}

Thus we find as $\chi\rightarrow \infty\Leftrightarrow r\rightarrow \infty$ the solutions of the radial KGF equation for a massive charged particle are:
\begin{equation}
R(r)\sim\frac{M}{\sqrt{\Delta^{KN}}}e^{-\sqrt{\mu^2-\omega^2}(r-r_{+})}[2(\mu^2-\omega^2)^{1/2}(r-r_{+})]^{\frac{B^{\prime}+D^{\prime}}{2M^3\sqrt{\mu^2-\omega^2}}}
\label{CMFarhorizonWhitt1}
\end{equation}
where the particles electric charge contribution is through the factors $B^{\prime},D^{\prime}$.
Since $W_{-k,m}(-\xi)$ forms another independent solution of the Whittaker equation we also have that as $r\rightarrow \infty$
\begin{equation}
R(r)\sim \frac{M}{\sqrt{\Delta^{KN}}}e^{+\sqrt{\mu^2-\omega^2}(r-r_{+})}[-2(\mu^2-\omega^2)^{1/2}(r-r_{+})]^{-\frac{B^{\prime}+D^{\prime}}{2M^3\sqrt{\mu^2-\omega^2}}}
\label{CMFarhorizonWhit2}
\end{equation}

We can also obtain the far horizon limit of our closed form analytic radial solutions as follows. The CHE (\ref{exactnessKGF}) is a differential equation with an irregular singularity at infinity. Following the discovery of Thom\'{e} that such a differential equation can be satisfied in the neighbourhood of an irregular singularity by a series of the form \cite{OLVER}
\begin{equation}
Y=e^{\lambda \zeta}\zeta^{\mu}\sum_{s=0}^{\infty}\frac{a_s}{\zeta^s}
\end{equation}
we determine the exponential parameters $\lambda,\mu$ for the case of CHE to be:
\begin{align}
\lambda_1&=0,\;\;\;\;\;\;\;\;\;\;\mu_1=-\frac{w_{-}^{\prime}}{\alpha^{\prime}_{-}}=
-\left[\frac{\delta_{\rm M}}{\alpha_{\rm M}}+\frac{1}{2}(2+\beta_{\rm M}+\gamma_{\rm M})\right] \\
\lambda_2&=-\alpha_{-}^{\prime}=-\alpha_{\rm M},\;\;\mu_2=\left[-(\gamma_{-}^{\prime}+\delta_{-}^{\prime})+\frac{w^{\prime}_{-}}{\alpha_{-}^{\prime}}\right]=
\frac{\delta_{\rm M}}{\alpha_{\rm M}}-\frac{1}{2}(2+\beta_{\rm M}+\gamma_{\rm M})
\end{align}
Thus for $r\rightarrow \infty$
\[H_c(\alpha^{\prime}_{-},w^{\prime}_{-},\gamma^{\prime}_{-},\delta^{\prime}_{-},\sigma^{\prime}_{-},\zeta)
\sim\left\{\begin{array}{c}
\zeta^{-\frac{w_{-}^{\prime}}{\alpha_{-}^{\prime}}}\\
e^{-\alpha_{-}^{\prime}\zeta}\zeta^{-(\gamma_{-}^{\prime}+\delta_{-}^{\prime})+\frac{w^{\prime}_{-}}{\alpha_{-}^{\prime}}}\end{array}\Leftrightarrow\right.
\]
\[{\rm HeunC}(\alpha_{\rm M},\beta_{\rm M},\gamma_{\rm M},\delta_{\rm M},\eta_{\rm M},\zeta)\sim\left\{\begin{array}{c}
\zeta^{-\left[\frac{\delta_{\rm M}}{\alpha_{\rm M}}+\frac{1}{2}(2+\beta_{\rm M}+\gamma_{\rm M})\right]}\\
e^{-\alpha_{\rm M}\zeta}\zeta^{\frac{\delta_{\rm M}}{\alpha_{\rm M}}-\frac{1}{2}(2+\beta_{\rm M}+\gamma_{\rm M})}\end{array}\right.\]
Thus in terms of the original variables we find:
\begin{equation}
R(r)\sim\left\{\begin{array}c
\frac{M}{\sqrt{\Delta^{KN}}}e^{-\sqrt{\mu^2-\omega^2}(r-r_{+})}\left(
-\frac{r-r_{+}}{2Md}\right)^{\frac{(B^{\prime}+D^{\prime})}{2M^3\sqrt{\mu^2-\omega^2}}}\\
\frac{M}{\sqrt{\Delta^{KN}}}e^{+\sqrt{\mu^2-\omega^2}(r-r_{+})}\left(
-\frac{r-r_{+}}{2Md}\right)^{-\frac{(B^{\prime}+D^{\prime})}{2M^3\sqrt{\mu^2-\omega^2}}}
\end{array}\right.
\label{MKGFCFHORIZONTHOME}
\end{equation}
Our results in (\ref{MKGFCFHORIZONTHOME}) for $q=0$ (i.e. neutral massive scalar field) reduce correctly to the result in \cite{BEZERRA}.
When we compare the results in (\ref{MKGFCFHORIZONTHOME}) with the results in (\ref{CMFarhorizonWhitt1}),(\ref{CMFarhorizonWhit2}) we see they differ slightly. The are equivalent, except for a multiplicative constant.

\subsubsection{The near event horizon limit-$\chi\rightarrow 0\Leftrightarrow r\rightarrow r_{+}$}\label{kontastonorizonta}

By expanding the $C^{\prime},D^{\prime}$ in (\ref{CHARGEDINITIARADIAL}) we derive for small $\chi$-and neglecting terms of $\mathcal{O}(\chi)$-a Whittaker's differential equation:
\begin{equation}
\frac{{\rm d}^2 Z}{{\rm d}\eta^{\prime 2}}
=\left[\frac{1}{4}
-\frac{1}{M^2}\left(\frac{A^{\prime}}
{\eta^{\prime 2}}+\frac{B^{\prime}}{2\sqrt{\mathcal{F}^{\prime}}}\frac{1}{\eta^{\prime}}\right)\right]Z
\end{equation}
where
\begin{equation}
\mathcal{F}^{\prime}=M^2(\mu^2-\omega^2)-\frac{C^{\prime}}{4M^2 d^2}-\frac{D^{\prime}}{2M^2 d},\;\;\eta^{\prime}:=2\sqrt{\mathcal{F}^{\prime}}\chi
\end{equation}
and with the parameters $k^{\prime}:=\frac{B^{\prime}}{2M^2\sqrt{\mathcal{F}^{\prime}}},\;m_h^{\prime 2}:=1/4-\frac{A^{\prime}}{M^2}$.
Near the event horizon limit and expanding Kummer's confluent hypergeometric function which is involved in the solution of Whittaker's equation we find:
\begin{equation}
M_{k^{\prime},m_{h}^{\prime}}(\eta^{\prime})=e^{-\frac{\eta^{\prime}}{2}}{\eta^{\prime}}^{\frac{1}{2}+m_h^{\prime}}
F(m_h^{\prime}+\frac{1}{2}-k^{\prime},2m_h^{\prime}+1,\eta^{\prime})
=e^{-\frac{\eta^{\prime}}{2}}{\eta^{\prime}}^{\frac{1}{2}+m_h^{\prime}}
\sum_{\nu=0}^{\infty}\frac{(m_h^{\prime}+\frac{1}{2}-k^{\prime})_{\nu}}{(2m_h^{\prime}+1)_{\nu}} \frac{\eta^{\prime \nu}}{\nu!}
\end{equation}
\begin{equation}
R(r)\sim \left\{\begin{array}{l} \frac{M}{\sqrt{\Delta^{KN}}}e^{-\sqrt{\mathcal{F}^{\prime}}\frac{(r-r_{+})}{M}}
\left(2\sqrt{\mathcal{F}^{\prime}}\frac{(r-r_{+})}{M}\right)^{\frac{1}{2}+m_h^{\prime}},\;\;\;r\rightarrow r_{+}\\
\frac{M}{\sqrt{\Delta^{KN}}}e^{-\sqrt{\mathcal{F}^{\prime}}\frac{(r-r_{+})}{M}}
\left(2\sqrt{\mathcal{F}^{\prime}}\frac{(r-r_{+})}{M}\right)^{\frac{1}{2}-m_h^{\prime}},\;\;\;r\rightarrow r_{+}
\end{array}\right.
\label{OrizontasGegonotonWHIT}
\end{equation}

We can also use the convergent power series for the confluent function in the vicinity of $\zeta=0$ to derive the near event horizon limit:
\begin{align}
H_c(\alpha^{\prime}_{-},w^{\prime}_{-},\gamma^{\prime}_{-},\delta^{\prime}_{-},\sigma^{\prime}_{-},\zeta)=\sum_{k=0}^{\infty}c_k\zeta^k
=1+\frac{\sigma^{\prime}_{-}}{-\gamma^{\prime}_{-}}\zeta
+\frac{-(-\alpha^{\prime}_{-}+\gamma^{\prime}_{-}+\delta^{\prime}_{-})\sigma^{\prime}_{-}+\sigma^{\prime 2}_{-}+w^{\prime}_{-}\gamma^{\prime}_{-}}{2\gamma^{\prime}_{-}(1+\gamma^{\prime}_{-})}\zeta^2+\cdots
\end{align}
\begin{equation}
R(r)\sim \left\{\begin{array}{l}
\frac{M}{\sqrt{\Delta^{KN}}}\left[-\frac{(r-r_{+})}{2dM}\right]^{\frac{1}{2}-
\frac{i}{2M}\sqrt{4A^{\prime}-M^2}}\\
\frac{M}{\sqrt{\Delta^{KN}}}\left[-\frac{(r-r_{+})}{2dM}\right]^{\frac{1}{2}+
\frac{i}{2M}\sqrt{4A^{\prime}-M^2}}\end{array}\right.
\label{EventHorizonPOWERSERIES}
\end{equation}
Our results in (\ref{EventHorizonPOWERSERIES}) for $q=0$ (i.e. neutral massive scalar field) reduce correctly to the result in \cite{BEZERRA}.
When we compare our results in (\ref{EventHorizonPOWERSERIES}) with
the result in (\ref{OrizontasGegonotonWHIT}) and by expanding the latter equation up to the first order in $\chi=(r-r_{+})/M$, we see that the two results agree, except for a multiplicative constant.

\section{Conclusions\label{symperasma}}

In this work we have derived exact analytic solutions of the KGF equation for a massive charged scalar in the Kerr-Newman-de Sitter  and Kerr-Newman black hole spacetimes. We first derived the radial and angular Fuchsian differential equations that result by separating variables in the general relativistic massive KGF equation in the KN-(a)dS black hole spacetime.

The exact solutions for a massive neutral and a massive charged scalar particle in the KN black hole spacetime are expressed in terms of confluent Heun functions. We derived conditions in the parameters of confluent Heun functions such as the solutions can be written in terms of confluent Kummer hypergeometric functions. Under certain conditions on the parameters they reduce to a sum-with finite number of terms-of confluent Kummer hypergeometric functions.

In the general case in which the cosmological constant is present the resulting radial and angular equations are Fuchsian differential equations with more than four regular singularities, thereby they constitute generalisation of the Heun equation with four regular singularities. As a result the solutions will generalise the Heun functions and local solutions.
For some particular values of the scalar mass in terms of the cosmological constant $\Lambda$ the solutions can be expressed in terms of Heun functions. For some other values of the scalar mass the extra singular points become false or apparent singular points and again these can lead in principle to analytic solutions in terms of Heun functions. We have derived the conditions in the parameters of the theory such that an extra singularity with exponents $(0,2)$ becomes a false singularity.
In the case of a massive scalar particle in the KNdS black hole spacetime we have derived the elliptic function representation for those values of the parameters for which the Fuchsian equation becomes a Heun equation. This in principle can be generalised to the case of a resulting Fuchsian equation with additional false singular points besides the four regular singular points of a Heun equation.

Following recent work in the mathematical literature \cite{TAYLORAPPELL} and starting from the equation obeyed by the derivative, Eqn.(\ref{funfverzweipunkt}) we constructed several expansions of the solutions of the general Heun equation in terms of the Lauricella $F_D$ and the Appell $F_1$ generalised hypergeometric functions of three and two variables respectively.
We expect this to be of relevance also on the isomonodromy problem of the generalised Fuchsian equations with more than four regular singularities that appeared in this work. Such an analysis is beyond the scope of this paper and it will be a subject of a future publication.

As we mentioned in the introduction a possible application of our work will be the computation of gravitational radiation from a hypothetical axion cloud around a KNdS black hole. Indeed a  superradiant instability \cite{zeldovich} effectively takes place if the Compton wavelength of the axion mass $\mu$ has the order of the gravitational radius of a black hole. Thus an interesting application of our exact analytic solutions of the KGF equation in the curved spacetime of a KNdS black hole derived in this work will be the investigation of superradiant instabilities in such gravitational backgrounds that can be used to constrain the mass of ultralight axionic degrees of freedon- especially when combined with precision measurements of the relativistic effects for the galactic centre SgrA* black hole which will determine its fundamental parameters $M,a,e,\Lambda$.

Another interesting research avenue is the following. There is a deep connection between  a Fuchsian equation with false singular points and \textit{finite-gap} elliptic Schr\"{o}dinger equation. It is  worth exploring further generalisations of this connection from closed form solutions of massive KGF equation in curved BH backgrounds with false singular point(s).

We have entered a very exciting era of general relativity and the theory of spacetime.

\section*{Acknowledgements}
This research has been co-financed by the European Union (European Social Fund-ESF) and Greek national funds through the Operational Program ``Education and Lifelong Learning" of the National Strategic Reference Framework (NSRF)-
Research Funding Program: THALIS-Investing in the society of knowledge through the European Social Fund, project: ``Beyond the Standard Model: Theoretical Physics of Elementary Particles and Cosmology under the light of LHC", MIS 375734, code $E\Lambda KE$ 80803. The author is grateful to K. Tamvakis for reading early versions of the manuscript. He also thanks  A. Ishkhanyan, K. S. Virbhadra, R. Konoplya C. Kostoulas and M. G. Fern\'{a}ndez for discussions and useful correspondence and the referees for their constructive comments.

\appendix{}
\section{Heun's differential equation and its elliptic function representation\label{HeunMun}}

The German mathematician Karl Heun generalised in 1888 the work of Riemann on Gau\ss\; hypergeometric function. He obtained a second-order differential equation with variable coefficients with four regular singularities. Namely, he discovered the following differential equation which bears his name in its canonical form \cite{KARLHEUNmunich}:
\begin{equation}
\frac{\mathrm{d}^2 y}{\mathrm{d}z^2}+\left(\frac{\gamma}{z}+\frac{\delta}{z-1}+\frac{\varepsilon}{z-a}\right)
\frac{\mathrm{d}y}{\mathrm{d}z}+\frac{\alpha\beta z-q}{z(z-1)(z-a)}y=0
\label{KarlHeunI}
\end{equation}
In (\ref{KarlHeunI}), $y$ and $z$ are regarded as complex variables and $\alpha,\beta,\gamma,\delta,\varepsilon,q,a$ are parameters, generally complex and arbitrary, except that $a\in\mathbb{C}\setminus
\{0,1\}$. The first five parameters are linked by the equation
\begin{equation}
\gamma+\delta+\varepsilon=\alpha+\beta+1
\label{Fuchs}
\end{equation}
Heun's equation is thus of Fuchsian type with regular singularities at the points $z=0,1,a,\infty$. The exponents at these singularities are computed through the indicial equation to be:$\{0,1-\gamma\}$;$\{0,1-\delta\}$;$\{0,1-\varepsilon\}$;$\{\alpha,\beta\}$. The sum
of these exponents must take the value $2$, according to the general theory of Fuchsian equations. It is this fact that yielded equation $(\ref{Fuchs})$.
Its Klein-B\^ocher-Ince formula is [0,4,0]\cite{RONVEAUX} .
The Heun equation includes an \textit{accessory} or \textit{auxiliary} parameter, namely the quantity $q\in\mathbb{C}$, which in many applications appears as a spectral parameter.
\subsubsection{The set of local solutions and Heun functions}\label{localHEUN}

Following \cite{RONVEAUX} we adopt the symbol '$Hl$', standing for 'Heun-local' to represent the series $y=\sum_{r=0}^{\infty}c_r z^r, (c_0\not =0)$ with the normalization $c_0=1$  as follows:
\begin{equation}
Hl(a,q;\alpha,\beta,\gamma,\delta;z)
\label{TOPIKESHEUNSYNARTISEIS}
\end{equation}
It should be noted that the parameter $\varepsilon$ does not appear explicitly in this notation, so that the Fuchs relation for Heun's equation $\varepsilon=\alpha+\beta-\gamma-\delta+1$, must be kept in mind.
Naturally, this function $Hl$ is defined in the first instance, for $|z|<1$; its analytic continuation is an aspect of the \textit{connection problem} \cite{SCHAFKE} that we will discuss further in the main text.
In the first place, there are eight local solutions of Heun's equation, one corresponding to each of the exponents at each of the four singularities.
Next, there are 24 mappings which take threee of the four points $\{0,1,a,\infty\}$ into $\{0,1,\infty\}$ so there are 192 solutions of the
Heun equation \cite{MAIER}. In \cite{MAIER} the \textit{group structure} was determined, of the set of transformations that can be applied to any normalised Fuchsian equation on $\mathbb{P}^1(\mathbb{C})$ with $n$ singular points. This automorphsim group has order $2^{n-1}n!$ and acts on the parameter space of the equation. It is isomorphic to the Coxeter group $\mathcal{D}_n$ the group of even-signed permutations of an $n$-set. Thus each of the $192$ local solutions $Hl$ of the Heun equation is labeled by an element of $\mathcal{D}_4$.

A  Heun function $Hf$ is, by definition, a solution of Heun's equation which is valid in a region containing two singularities $s_1,s_2$ of the equation in the sense that it is simultaneously a Frobenius solution about $s_1$ corresponding to one of the exponents there, and also a Frobenius solution about $s_2$ corresponding to one of the exponents there. For such a solution it is customary the term \textit{Heun function} relative to $\{s_1,s_2\}$ \cite{RONVEAUX}.

\subsection{The Confluent Heun Equation (CHE)}\label{confluentheunfuchs}

This is obtained by merging the singularity at $z=a$ of Heun's equation with that at $z=\infty$, resulting in an equation still having regular singularities at $z=0$ and $z=1$, and an irregular singularity of rank $1$ at $z=\infty$ \cite{RONVEAUX}.
Indeed, dividing (\ref{KarlHeunI}) by $a$ we derive:
\begin{align}
& z(z-1)\left(\frac{z}{a}-1\right)y^{\prime\prime}(z)+\left[\gamma(z-1)\left(\frac{z}{a}-1\right)
+\delta z \left(\frac{z}{a}-1\right)+\frac{\varepsilon}{a}z(z-1)\right]y^{\prime}(z) \nonumber \\
& +\left(\alpha\frac{\beta}{a}z-\frac{q}{a}\right)y(z)=0.
\end{align}
We let $a\rightarrow \infty$ and simultaneously let $\beta,\varepsilon,q\rightarrow \infty$ in such a way that
\begin{equation}
\frac{\beta}{a}\rightarrow \frac{\varepsilon}{a}\rightarrow -\nu,\;\;\frac{q}{a}\rightarrow -\sigma,
\end{equation}
which yields
\begin{equation}
\frac{{\rm d}^2y}{{\rm d}z^2}+\left[\frac{\gamma}{z}+\frac{\delta}{z-1}+\nu\right]\frac{{\rm d}y}{{\rm d}z}+\left[\frac{\alpha\nu z-\sigma}{z(z-1)}\right]y(z)=0,
\end{equation}
in which $\gamma,\delta,\alpha$ are the same parameters as in the original equation (\ref{KarlHeunI}) while $\nu,\sigma$ are new.

If one, following Darboux \cite{MGDARBOUX}, applies the transformation
\begin{equation}
z={\rm sn}^2(u,k),\;z_3=a=k^{-2},
\end{equation}
obtains the elliptic function representation of Heun's equation (\ref{KarlHeunI}), namely
\begin{align}
& \frac{{\rm d}^2 y}{{\rm d}u^2}+\Biggl[(2\gamma-1)\frac{{\rm cn}u{\rm dn}u}{{\rm sn}u}-(2\delta-1)\frac{{\rm sn}u{\rm dn}u}{{\rm cn}u}-k^2(2\varepsilon-1)\frac{{\rm sn}u{\rm cn}u}{{\rm dn}u}\Biggr]\frac{{\rm d}y}{{\rm d}u} \nonumber \\
&+(4\alpha\beta k^2 {\rm sn}^2 u-4k^2 q)y=0,
\end{align}
where ${\rm sn}u,{\rm dn}u,{\rm cn}u$ are the Jacobian elliptic functions with two periods $4mK,4niK^{\prime},m,n\in \mathbb{Z}$:
\begin{align}
{\rm sn}(u+4mK+4niK^{\prime})&={\rm sn}u,\\
{\rm cn}(u+4mK+4niK^{\prime})&={\rm cn}u,\\
{\rm dn}(u+4mK+4niK^{\prime})&={\rm dn}u,
\end{align}
where
\begin{equation}
K=\int_0^1 \frac{{\rm d}t}{\sqrt{(1-t^2)(1-k^2t^2)}}, K^{\prime}=\int_1^{1/k} \frac{{\rm d}x}{\sqrt{(x^2-1)(1-k^2 x^2)}}=\int_0^{\cos^{-1}k}\frac{{\rm d}\theta}{\sqrt{\cos^2\theta-k^2}}
\end{equation}
We also used in the calculation of the elliptic representation fundamental properties of the derivatives of the Jacobian elliptic functions:
\begin{align}
\frac{{\rm d}z}{{\rm d}u}&=\frac{\rm d}{{\rm d}u}{\rm sn}^2(u,k)=2{\rm sn}(u,k)\frac{{\rm d sn}u}{{\rm d}u}=2{\rm sn}(u,k){\rm cn}(u,k){\rm dn}(u,k),\\
 \frac{{\rm d}y}{{\rm d}z}&=\frac{{\rm d}y}{{\rm d}u}\frac{{\rm d}u}{{\rm d}z} =\frac{1}{2{\rm sn}u\;{\rm cn}u\;{\rm dn}u}\frac{{\rm d}y}{{\rm d}u}
\end{align}
and
\begin{align}
\frac{{\rm d}^2 y}{{\rm d}z^2}&=\frac{1}{2{\rm sn}u\; {\rm cn}u\; {\rm dn}u}\frac{{\rm d}u}{{\rm d}z}\frac{{\rm d}^2 y}{{\rm d}u^2} \nonumber \\
&-\frac{1}{2{\rm sn}^2 u \;{\rm cn}^2 u\; {\rm dn}^2u}\left({\rm cn^2}u\;{\rm dn}^2 u-{\rm sn}^2 u\;{\rm dn}^2 u-k^2{\rm sn}^2 u \;{\rm cn}^2u\right)\frac{{\rm d}u}{{\rm d}z}
\frac{{\rm d}y}{{\rm d}u},
\end{align}
\begin{align}
4z(1-z)(1-k^2z)\frac{{\rm d}^2y}{{\rm d}z^2}&=\frac{{\rm d}^2y}{{\rm d}u^2}
-\frac{({\rm cn^2}u\;{\rm dn}^2u-{\rm sn}^2 u\;{\rm dn}^2 u-k^2{\rm sn}^2 u\; {\rm cn}^2 u)}{{\rm sn}u\; {\rm cn} u\; {\rm dn} u}\frac{{\rm d}y}{{\rm d}u}
\end{align}
The connection of Jacobian elliptic functions to the Weierstra\ss elliptic functions can be obtained as follows. Let us suppose for instance, $X=4(x-e_1)(x-e_2)(x-e_3)$, with $e_1>e_2>e_3$
\begin{align}
u&=\int_x^{\infty}\frac{{\rm d}x}{\sqrt{4x^3-g_2 x-g_3}}=\int_x^{\infty}
\frac{{\rm d}x}{\sqrt{4(x-e_1)(x-e_2)(x-e_3)}}=\wp^{-1}x \nonumber \\
&=\frac{1}{\sqrt{e_1-e_3}}{\rm sn}^{-1}\sqrt{\frac{e_1-e_3}{x-e_3}} \nonumber \\
&=\frac{1}{\sqrt{e_1-e_3}}{\rm cn}^{-1}\sqrt{\frac{x-e_1}{x-e_3}}=\frac{1}{\sqrt{e_1-e_3}}{\rm dn}^{-1}\sqrt{\frac{x-e_2}{x-e_3}},
\label{KarlWeierstrass}
\end{align}
from which we deduce:
\begin{align}
\frac{e_1-e_3}{\wp(u)-e_3}&={\rm sn}^2(u\sqrt{e_1-e_3}),\frac{\wp(u)-e_2}{\wp(u)-e_3}&={\rm dn}^2(u\sqrt{e_1-e_3}),\frac{\wp(u)-e_1}{\wp(u)-e_3}&={\rm cn}^2(u\sqrt{e_1-e_3})
\end{align}

\section{Exact solution of Heun's differential equation with a false singular point}

Consider the Fuchsian Heun equation with a false singular point:
\begin{equation}
\frac{{\rm d}^2Y}{{\rm d}\zeta^2}+\left(\frac{\gamma}{\zeta}+\frac{\delta}{\zeta-1}+\frac{-1}{\zeta-a}\right)\frac{{\rm d}Y}{{\rm d}\zeta}+\frac{(\alpha\beta\zeta-q)Y)}{\zeta(\zeta-1)(\zeta-a)}=0,
\label{apparentHeun}
\end{equation}
the point $\zeta=a$ is the false singularity. The exponents at this point are equal to $0$ and $2$ and thus $\varepsilon=-1$. Using the Fuchs relation that the sum of all exponents depend only on the number of singular points we now have that $\delta=2-\gamma+\beta+\alpha$.

The differential equation (\ref{apparentHeun}), as was first claimed in \cite{Craster}, and we prove in detail in this appendix,  has the exact solution in terms of Gau$\ss$ hypergeometric function:
\begin{equation}
Y(\zeta)=(1-a)(\gamma-1)F(\alpha,\beta,\gamma-1,\zeta)+(q-a(1+\alpha+\beta+\alpha\beta-\gamma))
F(\alpha,\beta,\gamma,\zeta)
\label{GAUSSHEUNFEXACT}
\end{equation}
Indeed, Heun's equation with a false singularity at $\zeta=a$ is written:
\begin{equation}
\zeta(\zeta-1)(\zeta-a)\frac{{\rm d}^2Y}{{\rm d}\zeta^2}+\left\{\gamma (\zeta-1)(\zeta-a)+\delta\zeta(\zeta-a)-\zeta(\zeta-1)\right\}\frac{{\rm d}Y}{{\rm d}\zeta}+(\alpha\beta\zeta-q)Y=0
\label{falsesingHeun}
\end{equation}
The derivatives can be written
\begin{equation}
\frac{{\rm d}Y}{{\rm d}\zeta}=(1-a)\alpha\beta F(\alpha+1,\beta+1,\gamma,\zeta)+
(q-a(1+\alpha+\beta+\alpha\beta-\gamma))\frac{\alpha\beta}{\gamma}F(\alpha+1,\beta+1,\gamma+1,\zeta),
\end{equation}
\begin{align}
\frac{{\rm d}^2Y}{{\rm
d}\zeta^2}&=(1-a)\alpha\beta\frac{(\alpha+1)(\beta+1)}{\gamma}F(\alpha+2,\beta+2,\gamma+1,\zeta) \nonumber \\
&+(q-a(1+\alpha+\beta+\alpha\beta-\gamma))\frac{\alpha\beta}{\gamma}\frac{(\alpha+1)(\beta+1)}{\gamma+1}
F(\alpha+2,\beta+2,\gamma+2,\zeta),
\end{align}
where we used the fundamental property of Gau$\ss$ hypergeometric function:
\begin{equation}
\frac{{\rm d}^m}{{\rm d}x^m}F(\alpha,\beta,\gamma,x)=\frac{\Gamma(\alpha+m)\Gamma(\beta+m)}{\Gamma(\gamma+m)}\frac{\Gamma(\gamma)}{\Gamma(\alpha)\Gamma(\beta)}
F(\alpha+m,\beta+m,\gamma+m,x)
\end{equation}
Now the second derivative is written:
\begin{align}
\frac{{\rm d}^2 Y}{{\rm d}\zeta^2}&=\frac{(1-a)\alpha\beta(\alpha+1)(\beta+1)}{\gamma}\Biggl\{
\frac{-\gamma(\gamma-1)}{(\alpha+1)(\beta+1)\zeta}\Biggl[ F(\alpha+1,\beta+1,\gamma,\zeta)\nonumber \\
&-
\frac{(\alpha+1-\gamma)(\beta+1-\gamma)\zeta}{\gamma(\gamma-1)(1-\zeta)}F(\alpha+1,\beta+1,\gamma+1,\zeta) \nonumber \\
&+\frac{\gamma(1-\gamma-(\alpha+\beta+3-2\gamma)\zeta)}{\gamma(\gamma-1)(1-\zeta)}F(\alpha+1,\beta+1,\gamma,\zeta)
\Biggr]\Biggr\} \nonumber \\
&+(q-a(1+\alpha+\beta+\alpha\beta-\gamma))\frac{\alpha\beta(\alpha+1)(\beta+1)}{\gamma(\gamma+1)}
\nonumber \\
&\times
\Biggl\{\frac{-\gamma(\gamma+1)}{(\alpha+1)(\beta+1)}\Biggl[
\frac{F(\alpha+1,\beta+1,\gamma+1,\zeta)-F(\alpha+1,\beta+1,\gamma,\zeta)}{\zeta}
\Biggr] \Biggr\}
\end{align}

In producing the last equation we used the formulae:
\begin{equation}
F(\alpha+2,\beta+2,\gamma+1,\zeta)=\frac{-\gamma(\gamma-1)}{(\alpha+1)(\beta+1)\zeta}
[F(\alpha+1,\beta+1,\gamma,\zeta)-F(\alpha+1,\beta+1,\gamma-1,\zeta)],
\end{equation}
and
\begin{align}
-F(\alpha+1,\beta+1,\gamma-1,\zeta)&=\frac{-(\alpha+1-\gamma)(\beta+1-\gamma)}{\gamma(\gamma-1)(1-\zeta)}
F(\alpha+1,\beta+1,\gamma+1,\zeta)\zeta\nonumber\\
&+\frac{\gamma(1-\gamma-(\alpha+\beta+3-2\gamma)\zeta)}{\gamma(\gamma-1)(1-\zeta)}
F(\alpha+1,\beta+1,\gamma,\zeta)
\end{align}

The terms involving the first derivative can be written:

\begin{align}
&(\gamma(\zeta-1)(\zeta-a)+\delta\zeta(\zeta-a)-\zeta(\zeta-1))\frac{{\rm d}Y}{{\rm d}\zeta} \nonumber \\ &=[\gamma(\zeta-1)(\zeta-a)+[2-\gamma+\beta+\alpha](\zeta-a)\zeta-\zeta(\zeta-1)](1-a)\alpha\beta F(\alpha,\beta,\gamma,\zeta) \nonumber \\
&+ [\gamma(\zeta-1)(\zeta-a)+[2-\gamma+\beta+\alpha](\zeta-a)\zeta-\zeta(\zeta-1)](1-a)(-\alpha(\gamma-1))
F(\alpha,\beta,\gamma,\zeta) \nonumber \\
&+ [\gamma(\zeta-1)(\zeta-a)+[2-\gamma+\beta+\alpha](\zeta-a)\zeta-\zeta(\zeta-1)](1-a)\alpha(\gamma-1)
F(\alpha,\beta,\gamma-1,\zeta) \nonumber \\
&+[\gamma(\zeta-1)(\zeta-a)+[2-\gamma+\beta+\alpha](\zeta-a)\zeta-\zeta(\zeta-1)](1-a)
\frac{\alpha\beta(\beta+1)\zeta}{\gamma}F(\alpha+1,\beta+2,\gamma+1,\zeta) \nonumber \\
&+[\gamma(\zeta-1)(\zeta-a)+[2-\gamma+\beta+\alpha](\zeta-a)\zeta-\zeta(\zeta-1)][q-a(1+\alpha+\beta+\alpha\beta-\gamma)]
\frac{1-\gamma}{\zeta}F(\alpha,\beta,\gamma,\zeta) \nonumber \\
&+[\gamma(\zeta-1)(\zeta-a)+[2-\gamma+\beta+\alpha](\zeta-a)\zeta-\zeta(\zeta-1)][q-a(1+\alpha+\beta+\alpha\beta-\gamma)]
\frac{\gamma-1}{\zeta}F(\alpha,\beta,\gamma-1,\zeta),
\end{align}
where we used the formula:
\begin{equation}
F(\alpha+1,\beta+1,\gamma,\zeta)=F(\alpha,\beta,\gamma,\zeta)+\frac{\zeta\alpha}{\gamma}F(\alpha+1,\beta+1,\gamma+1,\zeta)
+\frac{(\beta+1)\zeta}{\gamma}F(\alpha+1,\beta+2,\gamma+1,\zeta)
\end{equation}

We further simplify matters using the formula
\begin{align}
F(\alpha+1,\beta+2,\gamma+1,\zeta)&=\frac{\gamma-\beta-1-\alpha\zeta}{(\beta+1)(\zeta-1)}
\left(\frac{-\gamma(\gamma-1)}{\alpha\beta\zeta}\right)\left[F(\alpha,\beta,\gamma,\zeta)-F(\alpha,\beta,\gamma-1,\zeta)\right] \nonumber \\
&+ \frac{1}{\alpha(\beta+1)(\zeta-1)}(\gamma(\beta-\gamma)+(\gamma-\alpha-\beta)\gamma)F(\alpha,\beta,\gamma,\zeta)
\end{align}

Plugging the previous formulae for the Gau$\ss$ hypergeometric function in the differential equation (\ref{falsesingHeun}) we arrive at the result:

\begin{equation}
\left\{-q^2+q[a(\alpha+\beta+2 \alpha\beta)+1-\gamma]-a\alpha\beta
[a(1+\alpha+\beta+\alpha\beta)-\gamma]\right\}F(\alpha,\beta,\gamma,\zeta)=0
\end{equation}
The last equality is true since this is equivalent to the condition that the logarithmic terms are absent which ensures that the point $z=a$ is a false singularity. Thus we proved that the exact solution of Heun equation with a false singularity, equation (\ref{apparentHeun}), is given in terms of Gau$\ss$ hypergeometric function by equation (\ref{GAUSSHEUNFEXACT}).

\section{Solutions of the general Heun equation and Appell-Lauricella hypergeometric expansions}\label{LauricellaDivision}

The  solution of Heun equation (\ref{KarlHeunI}) is written as we mentioned in section \ref{localHEUN} as $y=Hl(a,q;\alpha,\beta,\gamma,\delta;z)$ assuming the value of $\varepsilon$ is determined by the Fuchsian relation.

As was first discussed in \cite{TAYLORAPPELL} the function:
\begin{equation}
v=z^{\gamma}(z-1)^{\delta}(z-a)^{\varepsilon}\frac{{\rm d}y}{{\rm d}z}
\label{substitute}
\end{equation}
obeys the differential equation:
\begin{align}
&\frac{{\rm d}^2v}{{\rm d}z^2}+\left(\frac{1-\gamma}{z}+\frac{1-\delta}{z-1}+\frac{1-\varepsilon}{z-a}-
\frac{\alpha\beta}{\alpha\beta z-q}\right)\frac{{\rm d}v}{{\rm d}z}+\frac{\alpha\beta z-q}{z(z-1)(z-a)}v=0
\label{funfverzweipunkt}
\end{align}
Indeed the function $v$ satisfies:
\begin{equation}
\frac{{\rm d}v}{{\rm d}z}=-(\alpha\beta z-q)z^{\gamma-1}(z-1)^{\delta-1}(z-a)^{\varepsilon-1}y,
\end{equation}
\begin{align}
&\frac{{\rm d}^2 v}{{\rm d}z^2}=-\frac{(\alpha\beta z-q)}{z(z-1)(z-a)}v -\alpha\beta z^{\gamma-1}(z-1)^{\delta-1}(z-a)^{\varepsilon-1}y \nonumber -(\gamma-1)(\alpha\beta z-q)z^{\gamma-2}(z-1)^{\delta-1}(z-a)^{\varepsilon-1}y \\ &-(\delta-1)(\alpha\beta z-q)z^{\gamma-1}(z-1)^{\delta-2}(z-a)^{\varepsilon-1}y -(\varepsilon-1)(\alpha\beta z-q)z^{\gamma-1}(z-1)^{\delta-1}(z-a)^{\varepsilon-2}y
\end{align}
In general eqn.(\ref{funfverzweipunkt}) is a Fuchsian equation having five regular singular points. The additional singularity is located at the point
$z=q/(\alpha\beta)$ with exponents $(0,2)$. It can be seen at once that in four particular cases, namely, if $q=0,q=\alpha\beta,q=a\alpha\beta$ and $\alpha\beta=0$, the point $q/(\alpha\beta)$ coincides with one of already existing singular points. Thus, in these four cases the number of singularities remain four and Eqn.(\ref{funfverzweipunkt}) represents a general Heun equation with altered parameters as compared to eqn.(\ref{KarlHeunI}).
If we do not restrict ourselves in the above four cases, let us consider now a power-series expansion of a point $z_0$ of complex plane, ordinary or singular, finite or infinite. For instance, let $z_0$ be a finite point and consider the following expansion:
\begin{equation}
v=(z-z_0)^{\mu}\sum_{\nu=0}^{\infty}a_{\nu}(z-z_0)^{\nu},
\label{anaptygma}
\end{equation}
Substituting the series into Eqn.(\ref{substitute}) and integrating term by term, we arrive at the expansion:
\begin{equation}
y=C_0+\sum_{\nu=0}^{\infty}a_{\nu}(\int z^{-\gamma}(z-1)^{-\delta}(z-a)^{-\varepsilon}(z-z_0)^{\mu+\nu}{\rm d}z),
\label{verzweipunktINT}
\end{equation}
where $C_0$ is a constant.
In many cases the integrals involved in this sum are expressed in terms of the Appell hypergeometric function of two variables or Gau$\ss$ hypergeometric function \cite{TAYLORAPPELL}. As we shall also show for the first time in some cases the integrals are expressed in closed analytic form in terms of the Lauricella's fourth hypergeometric function $F_D$ of three variables.

The first hypergeometric function of  Appell $F_1(\alpha,\beta,\beta^{\prime},\gamma,x,y)$, is  a two variable
hypergeometric function with variables $x,y$ and parameters $\alpha,\beta,\beta^{\prime},\gamma$
that admits the integral
representation:
\begin{equation}
\int_0^1 u^{\alpha-1}(1-u)^{\gamma-\alpha-1}(1-u x)^{-\beta}(1-u
y)^{-\beta^{\prime}}{\rm
d}u=\frac{\Gamma(\alpha)\Gamma(\gamma-\alpha)}{\Gamma(\gamma)}
F_1(\alpha,\beta,\beta^{\prime},\gamma,x,y)
\end{equation}

Let us embark on the proof of the above statements. The most obvious situations where the integrals are computed in closed form in terms of the Appell's hypergeometric function $F_1$ is by choosing $z_0$ as a singular point of the Heun equation (\ref{KarlHeunI}), that is if $z_0=0,1,a,\infty$. Indeed let us work with $z_0=0$, so that the expansion (\ref{anaptygma}) represents a Frobenius solution of Eqn.(\ref{funfverzweipunkt}) in the neighbourhood of its singular point $z=0$:
\begin{equation}
v=z^{\mu}\sum_{\nu=0}^{+\infty}a_{\nu}^{(1)}z^{\nu}, \mu=0,\gamma
\end{equation}
In this case the expansion reads:
\begin{equation}
u=C_0+\sum_{\nu=0}^{\infty}\left(\int z^{-\gamma}(z-1)^{-\delta}(z-a)^{-\varepsilon}z^{\mu+\nu}{\rm d}z\right)
\end{equation}
Let us calculate the complex integral. We choose a parametrisation $z(u)=uz, u\in[0,1]$ so that $\int f(z){\rm d}z=\int f(z(u))z^{\prime}(u){\rm d}u$. Then we compute:
 \begin{align}
y_{\nu}&=\int z^{-\gamma}(z-1)^{-\delta}(z-a)^{-\varepsilon}z^{\mu+\nu}{\rm d}z \nonumber \\
&=\frac{(-1)^{(-\delta)}}{(-a)^{\varepsilon}}z^{1-\gamma+\mu+\nu}\int_0^1u^{-\gamma+\mu+\nu}
(1-uz)^{-\delta}\left(1-u\frac{z}{a}\right)^{-\varepsilon}{\rm d}u \nonumber \\
&=\frac{(-1)^{(-\delta)}}{(-a)^{\varepsilon}}z^{\gamma_0+\nu}
\frac{\Gamma(\gamma_0+\nu)}{\Gamma(\gamma_0+\nu+1)}F_1\left(\gamma_0+\nu,\delta,\varepsilon,
1+\gamma_0+\nu,z,\frac{z}{a}\right)
\end{align}
where we suppose $|z|<1<|a|$ and we defined $\gamma_0:=1-\gamma+\mu$.
Thus we get the series:
\begin{align}
&y=C_0+\sum_{\nu=0}^{\infty}a_{\nu}^{(1)}y_{\nu} =C_0+\sum_{\nu=0}^{\infty}a_{\nu}^{(1)}\frac{(-1)^{(-\delta)}}{(-a)^{\varepsilon}}z^{\gamma_0+\nu}
\frac{1}{\gamma_0+\nu}F_1\left(\gamma_0+\nu,\delta,\varepsilon,
1+\gamma_0+\nu,z,\frac{z}{a}\right)\Rightarrow \nonumber \\
&Hl(a,q;\alpha,\beta,\gamma,\delta;z)=C_0+\sum_{\nu=0}^{\infty}a_{\nu}^{(1)}\frac{(-1)^{-\delta}}{(-a)^{\varepsilon}}
\frac{z^{\gamma_0+\nu}}{\gamma_0+\nu}F_1\left(\gamma_0+\nu,\delta,\varepsilon,
1+\gamma_0+\nu,z,\frac{z}{a}\right), \mu=0,\gamma
\end{align}
Similarly, choosing $z_0=1$ we compute the integral in terms of Appell's function $F_1$ as follows:
\begin{align}
y&=C_0+\sum_{\nu=0}^{\infty}a_{\nu}^{(2)}\left(\int z^{-\gamma}(z-1)^{-\delta+\mu+\nu}(z-a)^{-\varepsilon}{\rm d}z\right)\Rightarrow \nonumber \\
y&=C_0+\frac{z^{1-\gamma}(-1)^{-\delta+\mu+\nu}}{(-a)^{\varepsilon}}\sum_{\nu=0}^{\infty}
a_{\nu}^{(2)}\frac{\Gamma(1-\gamma)}{\Gamma(2-\gamma)}F_1\left(1-\gamma,\delta-\mu-\nu,\varepsilon,2-\gamma,z,\frac{z}{a}\right),\mu=0,\delta
\end{align}
Similarly choosing $z_0=a$ reads:
\begin{align}
y&=C_0+\sum_{\nu=0}^{\infty}a_{\nu}^{(3)}\left(\int z^{-\gamma}(z-1)^{-\delta}(z-a)^{-\varepsilon+\mu+\mu}{\rm d}z\right) \nonumber \\
&=C_0+\frac{(-1)^{-\delta}}{(-a)^{\varepsilon-\mu-\nu}}z^{1-\gamma}
\sum_{\nu=0}^{\infty}a_{\nu}^{(3)}\frac{\Gamma(1-\gamma)}{\Gamma(2-\gamma)}
F_1\left(1-\gamma,\delta,\varepsilon-\mu-\nu,2-\gamma,z,\frac{z}{a}\right), \mu=0,\varepsilon
\end{align}
If $z_0=\frac{q}{\alpha\beta}$ or $z_0$ is an ordinary point the integrals involved in (\ref{verzweipunktINT}) in general are not expressed in terms of the Appell function. Under circumstances they do as we shall show below.
If we consider the non-logarithmic solution in (\ref{funfverzweipunkt}) in the neighbourhood of the singular point $z_0=\frac{q}{\alpha\beta}$ with exponent $\mu=2 $ we have:
\begin{equation}
y_{\nu}=\int z^{-\gamma}(z-1)^{-\delta}(z-a)^{-\varepsilon}(z-z_0)^{2+\nu}{\rm d}z\Rightarrow
\end{equation}
\begin{align}
y_{\nu}&=\int_0^1 u^{-\gamma}z^{-\gamma}(-1)^{-\delta}(1-uz)^{-\delta}(-a)^{-\varepsilon}
\left(1-\frac{zu}{a}\right)^{-\varepsilon}(-z_0)^{2+\nu}\left[1-\frac{uz}{z_0}\right]
^{2+\nu}z{\rm d}u \nonumber \\
&=\frac{(-1)^{-\delta}}{(-a)^{\varepsilon}}(-z_0)^{2+\nu}z^{1-\gamma}
\int_0^1 u^{-\gamma}(1-uz)^{-\delta}\left(1-\frac{uz}{a}\right)^{-\varepsilon}
\left(1-\frac{zu}{z_0}\right)^{2+\nu}{\rm d}u \nonumber \\
&=\frac{(-1)^{-\delta}}{(-a)^{\varepsilon}}(-z_0)^{2+\nu}z^{1-\gamma}
\frac{\Gamma(1-\gamma)}{\Gamma(2-\gamma)}F_D\left(1-\gamma,\delta,\varepsilon,-2-\nu,2-\gamma,
z,\frac{z}{a},\frac{z}{z_0}\right),
\label{HEUNLAURICELLA}
\end{align}
where $F_D$ denotes the fourth hypergeometric function of Lauricella with three variables.

For $\varepsilon=0$ our exact analytic result in Eqn.(\ref{HEUNLAURICELLA}) reduces to the result in terms of Appell's $F_1$ obtained in \cite{TAYLORAPPELL} for this particular value of this parameter:
\begin{equation}
y_{\nu}=\frac{(-1)^{-\delta}}{(-a)^{\varepsilon}}\frac{z^{1-\gamma}}{(-z_0)^{-2-\nu}}
\frac{1}{1-\gamma}F_1\left(1-\gamma,\delta,-2-\nu,2-\gamma,z,\frac{z}{z_0}\right)
\end{equation}
For the definition of
Lauricella's $4^{th}$
hypergeometric function $F_D$ of $m$-variables and its integral
representation we refer the reader to Appendix A, Eqns (A.1)-(A.7) of \cite{CQGKraniotis}.

\section{Isomonodromic mappings in Fuchs spaces}

In this subsection, we will review the space of Fuchsian spaces and the isomonodromic mappings in such spaces. It will also be discussed how one can reduce the number of false singular point using the concept of isomonodromy along the lines of \cite{Craster}.
\begin{definition}
Let $\rm{U}$, a linear space of functions. If the following conditions hold:
(a) the space has dimension two over $\mathbb{C}$, i.e. any three elements of $\rm{U}$ are linearly dependent with complex coefficients.

(b) $\forall\; U\in{\rm U}$ is valid that every element $U$ is a regular function of the complex variable $\zeta$ everywhere except a finite set of singular points $a_1,\ldots a_r$.

(c) for every element of this set, let us say $a_j$, two fundamental functions
$U_{j,1}$ and $U_{j,2}$ can be selected among the elements of the space $\rm{U}$, such that they obey the ansatz:
\begin{align}
Y_{j,1}(\zeta)&=(\zeta-a_j)^{\alpha_j}y_1(\zeta), \;Y_{j,2}(\zeta)=(\zeta-a_j)^{\beta_j}y_2(\zeta)\;{\rm or} \\
Y_{j,1}(\zeta)&=(\zeta-a_j)^{\alpha_j}y_1(\zeta)+\log(\zeta-a_j)Y_{j,2}(\zeta),\;Y_{j,2}(\zeta)
=(\zeta-a_j)^{\beta_j}y_2(\zeta)
\label{NEPER}
\end{align}
or the ansatz for a singular point at infinity in which the previous expressions acquire the form
\begin{align}
Y_{j,1}(\zeta)&=\zeta^{-\alpha_j}y_1(1/\zeta),\;Y_{j,2}(\zeta)=\zeta^{-\beta_j}y_2(1/\zeta)
\label{infinityP1}\\
Y_{j,1}(\zeta)&=\zeta^{-\alpha_j}y_1(1/\zeta)+\log(\zeta)Y_{j,2}(\zeta),\;Y_{j,2}=\zeta^{-\beta_j}y_2(1/\zeta),
\label{InfinityP2}
\end{align}
then we call $\rm U$ a $\bf{Fuchs\; space}$. Also $y_i(\zeta),i=1,2$ are regular and non-zero at $a_j$, while in (\ref{NEPER}) the exponents obey the condition $\beta_j-\alpha_j\in\mathbb{Z}^+$. Likewise the functions $y_1(\zeta)$ and $y_2(\zeta)$ in (\ref{infinityP1}),(\ref{InfinityP2}) are regular and non-zero at $\zeta=0$ \cite{Craster}. If all the conditions are satisfied, except that at infinity, there are fundamental functions of the form \cite{RONVEAUX},\cite{Craster}:
\begin{equation}
Y_1(\zeta)=e^{\lambda_1 \zeta}\zeta^{-\mu_1}\sum_{\nu=0}^{\infty}c_{\nu}\zeta^{-\nu},\;Y_2(\zeta)=e^{\lambda_2 \zeta}\zeta^{-\mu_2}\sum_{\nu=0}^{\infty}d_{\nu}\zeta^{-\nu},
\end{equation}
and then we talk for a \textbf{confluent\;Fuchsian} space. The solutions of each (confluent) Fuchsian equation form a (confluent) Fuchsian linear space.
Given some point in the complex plane, it falls into one of two categories: either it belongs to the set $\{a_j\}$, in which case it is singular, or it is a regular point. Only strong singular points of the equation become singular points of the space of its solutions. The false singular points of the equation are actually regular points of the space.
\end{definition}

\begin{definition}

Let us consider two distinct Fuchsian, or confluent Fuchsian spaces $\rm U$ and $\rm V$. The invertible linear mapping:
\begin{equation}
{\rm U}\xrightarrow{\varphi}{\rm V}
\end{equation}
is an \textbf{isomonodromy} if

(a) the set of non-trivial singular points $\{a_1,\ldots,a_r\}$ for both spaces coincide.

(b) The exponents $\alpha_j^{*},\beta_j^*$ of $\rm U$ and the exponents $\alpha_j^{**},\beta_j^{**}$ of $\rm V$ are such that $\alpha_j^*-\alpha_j^{**}\in\mathbb{Z}$ and $\beta_j^*-\beta_j^{**}\in\mathbb{Z}$. If infinity is the irregular singular point, then $\lambda_1^{*}=\lambda_1^{**},\lambda_2^{*}=\lambda_2^{**}$ and
$\mu_1^*-\mu_1^{**}\in \mathbb{Z},\mu_2^{*}-\mu_2^{**}\in\mathbb{Z}$.

(c) For every singular point $a_j$, the image of the fundamental function $U_{j,1}$ with exponent $\alpha_j^*$ is the fundamental function $V_{j,1}$ with exponent $\alpha_j^{**}$.
\end{definition}

\subsection{Explicit isomonodromy mappings and connection between Fuchsian spaces and Fuchsian equations}
In \cite{Craster} the following important theorem was proved:
\begin{theorem}
Let $\rm U$ be a Fuchsian, or confluent Fuchsian space with basis $(U_1,U_2)$ and let ${\rm U}\rightarrow \varphi({\rm U})$ and ${\rm U}\rightarrow \psi({\rm U})$  be some isomonodromy mappings. Assuming the determinant
\begin{equation}
D(\zeta)=\left|\begin{array}{cc}
\varphi(U_1) & \psi (U_1) \\
\varphi(U_2) & \psi (U_2)
\end{array}
 \right|
\end{equation}
is not identically zero, then there exist two rational functions $R_1(\zeta)$ and $R_2(\zeta)$, such that for any $U\in{\rm U}$
\begin{equation}
U=R_1\varphi(U)+R_2\psi(U).
\end{equation}
\end{theorem}
An important corollary of the theorem is that for the isomonodromy mappings
${\rm U^{\prime\prime}}\rightarrow {\rm U^{\prime}}$ and ${\rm U}^{\prime\prime}\rightarrow {\rm U}$ if
\begin{equation}
D(\zeta)=\left|\begin{array}{cc}
U_1^{\prime}&U_1 \\
U_2^{\prime}&U_2
\end{array}
\right|
\end{equation}
is not identically zero there exist two rational functions $f(\zeta)$ and $g(\zeta)$, such that for any $U\in {\rm U}$ the relation
\begin{equation}
U^{\prime\prime}(\zeta)=f(\zeta)U^{\prime}(\zeta)+g(\zeta)U(\zeta)
\label{COEFFICIEFUCHS}
\end{equation}
is valid. This means that the Fuchsian (or confluent Fuchsian) space is the space of solutions for a Fuchsian (or a confluent Fuchsian) differential equation. The explicit form of the coefficients is determined from the system of equations
\begin{align}
U_1^{\prime\prime}&=f(\zeta)U_1^{\prime}+g(\zeta)U_1(\zeta), \\
U_2^{\prime\prime}&=f(\zeta)U_2^{\prime}+g(\zeta)U_2(\zeta),
\end{align}
\begin{align}
f(\zeta)&=\frac{\left|\begin{array}{cc}
U_2^{\prime\prime} & U_2 \\
U_1^{\prime\prime} &U_1
\end{array}\right|}{\left|\begin{array}{cc}
U_2^{\prime} & U_2 \\
U_1^{\prime} &U_1
\end{array}\right|}=\frac{(-)\left|\begin{array}{cc}
U_1^{\prime\prime} & U_1 \\
U_2^{\prime\prime} &U_2
\end{array}\right|}{(-)\left|\begin{array}{cc}
U_1^{\prime} & U_1 \\
U_2^{\prime} &U_2
\end{array}\right|}= \frac{\left|\begin{array}{cc}
U_1^{\prime\prime} & U_1 \\
U_2^{\prime\prime} &U_2
\end{array}\right|}{D(\zeta)}\\
g(\zeta)&=\frac{\left|\begin{array}{cc}
U_1^{\prime} & U_1^{\prime\prime}\\
U_2^{\prime} & U_2^{\prime\prime}
\end{array}\right|}{D(\zeta)}
\end{align}
At this point we observe that the false or apparent singular points are the roots of the determinant
\begin{align}
D(\zeta)=\left|\begin{array}{cc}
U_1^{\prime} & U_1 \\
U_2^{\prime} & U_2
\end{array}\right|
\end{align}
Indeed, the following statements are equivalent: 1) A point $\zeta_0$ is an apparent singularity of the Fuchsian equation:
\begin{equation}
\frac{{\rm d}^2 u}{{\rm d}\zeta^2}+p(\zeta)\frac{{\rm d}u}{{\rm d}\zeta}+q(\zeta)u=0,
\label{FUCHSDIFFERENTIAL}
\end{equation}
2) There exists two linearly independent holomorphic solutions of (\ref{FUCHSDIFFERENTIAL}) around $\zeta_0$ whose Wronskian vanishes at $\zeta_0$ \cite{Yoshida}.
Under the condition $D\not=0$ identically, for any isomonodromy mapping \footnote{We mention that the theory of isomonodromic deformation of a differential equation has its roots to the Riemann-Hilbert problem \cite{Yoshida}.}
$\varphi:{\rm V}\rightarrow {\rm U}$ there exist functions $J(\zeta),H(\zeta)$ such that:
\begin{equation}
V(\zeta)=J(\zeta)U(\zeta)+H(\zeta)U^{\prime}(\zeta)
\label{ISOMFUCH}
\end{equation}
Assuming that the Fuchsian equation for ${\rm V}$ has the general form
\begin{equation}
V^{\prime\prime}(\zeta)=f^{*}(\zeta)V^{\prime}+g^{*}(\zeta)V(\zeta)
\label{WICHTIGFUCHS}
\end{equation}
one can determine the rational functions $f^*(\zeta),g^*(\zeta)$ as follows:
Let the coefficients $f$ and $g$ in (\ref{COEFFICIEFUCHS}) for ${\rm U}$ are known then $\forall U\in{\rm U}$ and $V=\varphi(U)$,
\begin{equation}
V^{\prime}(\zeta)=I(\zeta)U(\zeta)+Z(\zeta)U^{\prime}(\zeta),
\label{FUCCOEF}
\end{equation}
where
\begin{equation}
I(\zeta):=J^{\prime}(\zeta)+H(\zeta)g(\zeta),\;\;Z(\zeta):=J(\zeta)+H(\zeta)f(\zeta)+H^{\prime}(\zeta)
\end{equation}
Differentiating (\ref{FUCCOEF}) we obtain:
\begin{equation}
V^{\prime\prime}(\zeta)=[I^{\prime}(\zeta)+Z(\zeta)g(\zeta)]U(\zeta)+
[I(\zeta)+Z^{\prime}(\zeta)+Z(\zeta)f(\zeta)]U^{\prime}(\zeta)
\end{equation}
Using (\ref{ISOMFUCH}) and (\ref{WICHTIGFUCHS}) one obtains:
\begin{align}
f^{*}(\zeta)& =\frac{\left|\begin{array}{cc}
I^{\prime}(\zeta)+Z(\zeta)g(\zeta) & J(\zeta) \\
I(\zeta)+Z^{\prime}(\zeta)+Z(\zeta)f(\zeta) & H(\zeta)
\end{array}\right|}{\left|\begin{array}{cc}
I(\zeta) & J(\zeta) \\
Z(\zeta) & H(\zeta)\end{array} \right|} \\
g^{*}(\zeta)&=\frac{\left|\begin{array}{cc}
I(\zeta) & I^{\prime}(\zeta)+Z(\zeta)g(\zeta) \\
Z(\zeta) & I(\zeta)+Z^{\prime}(\zeta)+Z(\zeta)f(\zeta)
\end{array}\right|}{-J(\zeta)Z(\zeta)+I(\zeta)H(\zeta)}.
\end{align}
We thus can write
\begin{equation}
\left[\begin{array}{c}
V \\
V^{\prime}\end{array}\right]=\left[\begin{array}{cc}
J & H \\
I & Z\end{array}\right]\left[\begin{array}{c}
U \\
U^{\prime}\end{array}\right]
\end{equation}
while the inverse transformation reads as follows:
\begin{equation}
\left[\begin{array}{c}
U \\
U^{\prime}\end{array}\right]=\frac{1}{JZ-IH}\left[\begin{array}{cc}
Z & -H \\
-I & J\end{array}\right]\left[\begin{array}{c}V\\
V^{\prime}\end{array}\right]
\end{equation}
This framework has been advocated in \cite{Craster} in using isomonodromy for reducing the number of false singular points. By considering Fuchsian equations that correspond to different Fuchsian spaces each forming an isomonodromy to each other, one can investigate the possibility in which while the number of nontrivial strong singular points of all equations coincide, the number of trivial singular points and false singular points can be different, leading to equations which are easier to solve. In \cite{Craster}, a form of $D(\zeta)$
was given, under the simplifying assumption that infinity is a regular point,
\begin{equation}
D(\zeta)=P_{\epsilon}(\zeta)\prod_{i=1}^{r}(\zeta-a_1)^{\alpha_i+\beta_i-1}, \;\;\;\epsilon=-\sum_{i=1}^{r}(\alpha_i+\beta_i-1)-2,
\label{APPARENTNUMBER}
\end{equation}
where $P_{\epsilon}$ is a polynomial of order $\epsilon$, the set $\{a_1\cdots a_r\}$ denotes the singular points with corresponding pairs of the exponents $(\alpha_i,\beta_i),\;i=1 \cdots r$.
The number of false singular points is defined by $\epsilon$, as one can proves using the Fuchs relation. These singular points have exponents $(0,2)$ and correspond to the case of simple roots of $P_{\epsilon}$.  If infinity is a regular singular point of ${\rm U}$, then the number of false singular points is still defined in (\ref{APPARENTNUMBER}) \cite{Craster}. In \cite{Craster} this theory of monodromy was applied to the case of a Heun equation with a false singular point. Using the ansatz given in their Eqn.(4.3) they showed that the solution is given by hypergeometric functions of Gau$\ss$ \cite{Craster}. In Appendix B we prove analytically in detail for the first time this assertion using properties of the hypergeometric functions of Gau$\ss$. Therefore we expect that in the case of Fuchsian equation with 5 singularities as it is the case for the radial and angular differential equations for a massive charged scalar particle in the KNdS black hole spacetime for most of the parameter space, that if one of the singularities is false, the solution will be expressed in terms of Heun functions. However, the verification of the conjecture is beyond the scope of the current paper and it will be a subject of a future publication.


\begin{thebibliography}{99}

\bibitem{KLEIN} O. Klein, \textit{Quantumtheorie und f\"unfdimensionale Relativit\"atstheorie}, Z.Phys.\textbf{37} (1926) 895-906
\bibitem{GORDON} W. Gordon, \textit{Der Comptoneffect nach der Schr\"odingerschen Theorie}, Z. Phys.\textbf{40} (1926),117-133
\bibitem{FOCK} V. Fock, \textit{Zur Schr\"odingerschen Wellenmechanik}, Z. Phys. \textbf{38} (1926), 242-250; V. Fock, \textit{\"Uber die invariante Form der Wellen- und der Bewegungsgleichungen f\"ur einen geladenen Massenpunkt}, Z. Phys.\textbf{39} (1926), 226-232

\bibitem {GhezA}A. M. Ghez \textit{et al, Measuring distance and properties of
the Milky Way's central supermassive black hole with stellar
orbits, }Astrophys. J. \textbf{689},(2008)1044, (arXiv:0808.2870),
L. Meyer \textit{et al, The Shortest-Known-Period Star
Orbiting Our Galaxy's Supermassive Black Hole, }Science \textbf{338 }(2012)84


\bibitem {GenzelETAL}R. Genzel \textit{et al} 2010, \textit{Rev.Mod. Phys.
}\textbf{82} 3121-95, R. Sch\"{o}del \textit{et al}, \textit{The nuclear cluster of Milky Way: our primary testbed for the interaction of a dense star cluster with a massive black hole} Class. Quantum Grav. \textbf{31} (2014) 244007


\bibitem{GRGKRANIOTIS} G. V. Kraniotis, \textit{Gravitational lensing and frame dragging of light in the Kerr-Newman and the Kerr-Newman-(anti) de Sitter black hole spacetimes}, Gen. Rel. Grav. {\bf 46} (2014) 1818 [arXiv:1401.7118]
\bibitem{CQGKraniotis} G. V. Kraniotis \textit{Precise analytic treatment of Kerr and Kerr-(anti) de Sitter black holes as gravitational lenses}, Class. Quant.Grav. {\bf 28} (2011) 085021

\bibitem{SHADOW} A. Abdujabbarov, B. Toshmatov, Z. Stuchl\'{\i}k, B. Ahmedov,
\textit{Shadow of the rotating black hole with quintessential energy in the presence of the plasma}, arXiv:1512.05206, E.F. Eiroa, C.M. Sendra, \textit{Strong deflection lensing by charged black holes in scalar-tensor gravity}, Eur.Phys.J.C74 (2014) 11, 3171, V. Bozza, \textit{Gravitational Lensing by Black Holes }, Gen.Rel.Grav.42 (2010) 2269-2300,  A. de Vries, Class.Quantum Grav. \textbf{17} (2000) 123-144, K.S. Virbhadra and G.F.R. Ellis, Phys.Rev. D62 (2000) 084003, H. C. Ohanian, Am. J. Phys. \textbf{55} (1987) 428-432


\bibitem {GeorgeVKraniotis} G. V. Kraniotis, \textit{Periapsis and
gravitomagnetic precessions of stellar orbits in Kerr and Kerr-de Sitter black
hole spacetimes,}Class. Quantum Grav. \textbf{24 }(2007) 1775-1808;
C. M. Will, ApJ, 674 (2008) L25,
D. Merritt, T. Alexander, S. Mikkola and C. M. Will, Phys. Rev.D \textbf{81} (2010)
062002, L. Iorio,arXiv:1008.1720v4[gr-qc],Mon.Not.R.Astron.Soc.(2011)\textbf{411},453-63, also: Jaroszy\'{n}ski M. Acta
Astronomica (1998) \textbf{48}, 653, G. F. Rubilar and A. Eckart (2001) A\&A \textbf{374}, 95,
P.C. Fragile and G. J. Mathews 2000, ApJ \textbf{542}, 328, N. N. Weinberg, M.
Milosavljevi\'{c} and A. M. Ghez, (2005) ApJ \textbf{622}, 878,
G. V. Kraniotis, \textit{Gravitational lensing and frame dragging of light in the Kerr-Newman and the Kerr-Newman-(anti) de Sitter black hole spacetimes}, Gen. Rel. Grav. {\bf 46} (2014) 1818 [arXiv:1401.7118],
G. V. Kraniotis, \textit{Frame dragging and bending
of light in Kerr and Kerr-(anti) de Sitter spacetimes,} Class. Quantum Grav.
\textbf{22} (2005) 4391-4424


\bibitem{LINHE} G. He, W. Lin, \textit{Second order Kerr-Newman time delay} Phys.Rev.D \textbf{93} (2016) 023005

\bibitem{HERRERA} A. Herrera-Aguilar, U. Nucamendi, \textit{Kerr black hole parameters in terms of the redshift/blueshift of photons emitted by geodesic particles} Phys.Rev.D92 (2015) 045024

\bibitem{PRETOSAHA}Preto M. and P. Saha (2009) ApJ \textbf{703}, 1743


\bibitem{PHIGGS}P.W.Higgs, \textit{Broken symmetries and the masses of gauge bosons}, Phys.Rev.Lett.\textbf{13} (1964)508-509;P.W.Higgs, Phys.Rev.145 (1966),1156-1163; F. Englert and R. Brout, Phys.Rev.Lett.13 (1964)321-323

\bibitem{CMS}CMS collaboration, \textit{Observation of a new boson at a mass of 125 GeV with the CMS experiment at the LHC}, Phys.Lett.B. 716 (2012)30-61

\bibitem{ATLAS} ATLAS collaboration, \textit{Observation of a new particle in the search for the Standard Model Higgs boson with the ATLAS detector at the LHC}, Phys.Lett.B.716 (2012),1-29

\bibitem{GW150914} B. P. Abbott \textit{et al},\textit{Observation of Gravitational Waves from a Binary Black Hole Merger} Phys.Rev.Lett.\textbf{116}, 061102 (2016)
\bibitem{GW151226} B. P. Abbott \textit{et al},\textit{GW151226: Observation of Gravitational Waves from a 22-Solar-Mass Binary} Phys.Rev.Lett.\textbf{116}, 241103 (2016)

\bibitem{DIMO} A. Arvanitaki, S. Dimopoulos, S. Dubovsky, N. Kaloper and J. March-Russell, Phys.Rev. D \textbf{81} 123530 (2010)

\bibitem{AXIONCLOUD} H. Yoshino and H. Kodama, \textit{Gravitational radiation from an axion cloud around a black hole: Superradiant phase} Prog. Theor. Exp. Phys. (2014) 043E02

\bibitem{diameaxion} T.C. Bachlechner, C. Long, L. McAllister, JHEP12(2015)042

\bibitem {Newman}E. T. Newman, E. Couch, K. Chinnapared, A. Exton, A. Prakash
and R. Torrence, \textit{Metric of a Rotating, Charged Mass,} Journal of
Mathematical Physics \textbf{6, }918 (1965)

\bibitem{HansOhanian} H. Ohanian and R. Ruffini 1994, \textit{Gravitation and
Spacetime} (New York: Norton and Company)

\bibitem {KerrR}R P Kerr, \textit{Gravitational field of a spinning mass as an
example of algebraically special metrics, }Phys. Re. Lett. \textbf{11 }(1963) 237
\bibitem {Appell}G. Lauricella \textit{Sulle funzioni ipergeometriche a
pi\`{u} variabili, }Rend.Circ.Mat. Palermo \textbf{7 (}1893) pp 111-158; P.
Appell \textit{Sur les fonctions hyperg\'{e}ometriques de deux variables, }J.
Math.Pure Appl.\textbf{8} (1882) 173-216

\bibitem {Ruffini}M. Johnston and R. Ruffini, \textit{Generalized Wilkins
effect and selected orbits in a Kerr-Newman geometry},
Phys.Rev.D.\textbf{10} (1974) 2324-2329

\bibitem {EVAh}E. Hackmann and H. Xu, \textit{Charged particle motion in
Kerr-Newmann spacetimes, }Phys.Rev.D. \textbf{87} (2013),124030

\bibitem {RufII}D. Pugliese, H. Quevedo and R. Ruffini, \textit{Equatorial
circular orbits of neutral test particles in the Kerr-Newman spacetime, }Phys.
Rev. D\textbf{88 }(2013) 024042


\bibitem{CHANDRA} S. Chandrasekhar, \textit{The Mathematical Theory of Black Holes} Oxford University Press (1998)


\bibitem{STEUKOLSKY} S. A. Teukolsky, \textit{Rotating Black Holes: Separable Wave Equations for Gravitational and Electromagnetic  Perturbations}, Phys.Rev. Lett. \textbf{29} (1972) 1114-1118, A.A. Starobinsk\u{i} and S.M. Churilov, \textit{Amplification of electromagnetic and gravitational waves scattered by a rotating black hole} Sov.Phys.JETP, Vol.\textbf{38} (1974) 1-5

\bibitem{page} D. N. Page, \textit{Dirac equation around a charged, rotating black hole}, Phys. Rev. D14 (1976) 1509-1510; C. H. Lee, Phys.Lett.B 68 (1977)152

\bibitem{ROWANSTEPHEN} D. J. Rowan and G. Stephenson, \textit{The Klein-Gordon equation in a Kerr-Newman background}, J.Phys.A.Math.Gen.\textbf{10} (1977) 15-23
\bibitem{Blandin} J. Blandin, R. Pons and G. Marcilhacy, Lett Al. Nuovo Cimento, \textbf{38} (1983) 561-567

\bibitem{WU} S. Wu and X. Cai, J. Math. Phys. \textbf{40} (1999),4538, P. P. Fiziev, Class.Quantum Grav. \textbf{27} (2010) 135001

\bibitem{Batic} D. Batic and H. Schmidt, \textit{The Dirac propagator in the extreme Kerr metric} Journal of Phys.A (2007) \textbf{40} 13443-13451

\bibitem{HODs} S. Hod, \textit{Rotating black holes can have short bristles}, Phys.Lett.B \textbf{739} (2014) 196-200

\bibitem{BEZERRA} V.B. Bezerra, H. S. Vieira and A. A. Costa, \textit{The Klein-Gordon equation in the spacetime of a charged and rotating black hole} Clas.Quantum Grav. \textbf{31} (2014) 045003
\bibitem{ProfDrKRANIOTIS} G. V. Kraniotis, Work in Progress


\bibitem{KARLHEUNmunich} K. Heun, \textit{Zur Theorie der Riemann'schen Functionen zweiter Ordnung mit vier Verzweipunkten}, Mathematische Annalen 33, (1889) pp 161-179
\bibitem{RONVEAUX} A. Ronveaux. \textit{ed} \textit{Heun's Differential Equations} Oxford Science Publications, OUP (1995)
\bibitem{Svartholm} N. Svartholm,\textit{Die L\"osung der Fuchssehen Differentialgleichung zweiter Ordnung durch hypergeometrische Polynome} Mathematische Annalen \textbf{116} (1939) 413-421
    \bibitem{Kristensson} G. Kristensson, \textit{Second Order Differential Equations} Springer 2010
\bibitem{Erdelyi} A. Erd\'{e}lyi, \textit{Certain expansions of solutions of the Heun equation}, Q. J. Math.(Oxford), \textbf{15} (1944), pp. 62-69
\bibitem{EdmundWHITTAKER} E. T. Whittaker and G. N. Watson \textit{A course of modern analysis} (1927), 4th ed. Cambridge Univ.Press

\bibitem{INOZEMTSEV} V. I. Inozemtsev, \textit{Lax representation with Spectral Parameter on a torus for Integrable Particle Systems} Letters in Mathematical Physics {\bf 17} (1989) 11-17
\bibitem{MGDARBOUX} G. Darboux, \textit{Sur une \'{e}quation lin\'{e}aire},
Comptes Rendus Acad. Sci. Paris {\bf 44} (1882), 1645-1648

\bibitem{SMIRNOV} A. O. Smirnov, \textit{Finite-gap solutions of the Fuchsian Equations}, Letters in Math.Physics (2006) 76:297-316


\bibitem{Craster} A. V. Shanin and R. V. Craster, \textit{Removing false singular points as a method of solving ordinary differential equations} Euro Jnl of Applied Mathematics (2002) Vol. 13, pp 617-639

\bibitem {Suzuki} H. Suzuki, E. Takasugi and H. Umetsu\textit{Perturbations of Kerr-de Sitter Black holes and Heun's equations, }Prog. of Theor. Phys.Vol \textbf{100} (1998) 491-505; Prog.of Theor.Phys. Vol.102 (1999) 253-272

\bibitem{RK} R.A. Konoplya and A. Zhidenko, arXiv:0707.1890

\bibitem{ISHKHANYAN} T. A. Ishkhanyan and A. M. Ishkhanyan \textit{Expansions of the solutions to the confluent Heun equation in terms of the Kummer confluent hypergeometric functions} AIP Advances \textbf{4}, 087132 (2014)


\bibitem{SCHAFKE} R. Sch\"{a}fke, \textit{The connection problem for two neighboring regular singular points of general linear complex ordinary differential equations}, Siam J. Math.Anal.Vol 11 (1980) 863-875; R. Sch\"{a}fke, \textit{A connection problem for a regular and an irregular singular point of complex ordinary differential equations}, Siam J. Math.Anal. Vol 15 (1984) 253-271

\bibitem{MAIER} R.S. Maier, \textit{The 192 solutions of the Heun equation} Mathematics of Computation, Vol.76 (2007), 811-843

\bibitem {Stuchlik1}Z. Stuchl\'{\i}k, G. Bao, E. \O stgaard and S.
Hled\'{\i}k,\textit{\ Kerr-Newman-de Sitter black holes with a restricted
repulsive barrier of equatorial photon motion, }Phys. Rev. D. \textbf{58}
(1998) 084003

\bibitem {GrifPod}J. B. Griffiths and Ji\v{r}\'{\i} Podolsk\'{y},
\textit{Exact spacetimes in Einstein's General Relativity, }Cambridge
Monographs in Mathematical Physics, Cambirdge University Press (2009)

\bibitem {BCAR}B. Carter, \textit{Global structure of the Kerr family of
gravitational fields }Phys.Rev.\textbf{174 }(1968)1559-71


\bibitem{ZST} Z. Stuchl\'{\i}k, \textit{The motion of test particles in black-hole backgrounds
with non-zero cosmological constant}, Bull. of the Astronomical
Institute of Chechoslovakia \textbf{34} (1983) 129-149

\bibitem{STATICRADIUS}Z. Stuchl\'{\i}k and S. Hled\'{\i}k,\textit{Some properties of the Schwarzschild-de Sitter and Schwarzschild-anti-de Sitter spacetimes} Phys.Rev.D\textbf{60} (1999) 044006
    \bibitem{RADIUSBEFOREL} Z. Stuchl\'{\i}k, \textit{Influence of the relict cosmological constant on accretion discs} Mod.Phys.Lett.A \textbf{20} (2005) 561-575

\bibitem{Yoshida} M. Yoshida, \textit{Fuchsian differential equations}, Aspects of Mathematics, Springer (1987)


\bibitem{TAYLORAPPELL} C. Leroy and A.M. Ishkhanyan, \textit{Expansions of the solutions of the confluent Heun equation in terms of the incomplete Beta and the Appell generalized functions} Integral Transforms and Special Functions (2015) 1-9,  A. M. Ishkhanyan, \textit{The Appell hypergeometric expansions of the general Heun Equation}, arXiv:1405.2871

\bibitem{Bicak} J. Bi\u{c}ak, Z. Stuchl\'{\i}k and V. Balek, \textit{The motion of charged particles in the field of rotating
charged black holes and naked singularities}, Bull. of the
Astronomical Institute of Chechoslovakia \textbf{40} (1989), 65-92

\bibitem{ZSTUCH} Z. Stuchl\'{\i}k, \textit{Equatorial circular orbits and the motion of the shell of dust in the field of a rotating naked singularity} Astronomical Institute of Chechoslovakia \textbf{31} (1980) pp 129-144
\bibitem{NakedZSTUCH} Z. Stuchl\'{\i}k and Jan Schee, \textit{Ultra-high energy collisions in the superspinning Kerr geometry}, Clas. Quantum Grav. \textbf{30} 2013, 075012






\bibitem {Calvani}M. Calvani and R. Turolla, \textit{Complete description of
photon trajectories in the Kerr-Newman space-time, }J. Phys. A.
Math.Gen.\textbf{14} (1981),1931-1942

\bibitem{ZdeStu} Z. Stuchl\'{\i}k and S.Hled\'{\i}k,
\textit{Equatorial photon motion in the Kerr-Newman spacetimes
with a non-zero cosmological constant}, Class. Quantum Grav.
\textbf{17} (2000) 4541-4576

\bibitem{zeldovich} Ya B. Zel'dovich, \textit{Amplification of Cylindrical Electromagnetic Waves Reflected from a Rotating Body}, Soviet Physics JETP Vol35 (1972) pp 1085-1087; R. Brito, V. Cardoso and P. Pani, Springer, LNP Vol 906 (2015)






\bibitem{OLVER} F. W. J. Olver, \textit{Asymptotics and special functions}
Academic Press,  Editor W. Rheinbolt (1974)

\end{thebibliography}
\end{document}